# Bayesian Learning in Structural Dynamics: A Comprehensive Review and Emerging Trends

Wang-Ji Yan[1,2], Lin-Feng Mei[1], Yuan-Wei Yin[1], Jiang Mo[1], Costas Papadimitriou[3], Ka-Veng Yuen[1,2], Michael Beer[4,5,6]

[1]*State Key Laboratory of Internet of Things for Smart City and Department of Civil and Environmental Engineering, University of Macau, China*

[2]*Guangdong-Hong Kong-Macau Joint Laboratory for Smart Cities, China*

[3]*Department of Mechanical Engineering, University of Thessaly, Volos, Greece*

[4]*Leibniz Universität Hannover, Institute for Risk and Reliability, Hannover, Germany*

[5]*Department of Civil and Environmental Engineering, University of Liverpool, Liverpool L69 3BX, United Kingdom*

[6]*International Joint Research Center for Resilient Infrastructure & International Joint Research Center for Engineering Reliability and Stochastic Mechanics, Tongji University, Shanghai 200092, PR China*

**Abstract:** Bayesian learning has emerged as a compelling and vital research direction in the field of structural dynamics, offering a probabilistic lens to understand and refine the analysis of complex dynamical systems. This state-of-the-art review meticulously traces the three-decade evolution of Bayesian learning in structural dynamics, illuminating core principles, groundbreaking methodologies, and diverse applications that have significantly influenced the field. The narrative commences by delving into the basics of Bayesian theory, clarifying essential concepts, and introducing primary methods for deriving posterior distributions, with an in-depth exploration of three types: Laplace approximation, stochastic sampling, and variational inference. Subsequently, the text explores the strategies and implementation of two types of Bayesian learning in structural dynamics: physical model learning and data-centric statistical model learning. Physical model learning emphasizes inferring physical model parameters directly within a general Bayesian framework for system identification and prediction, with extensions that include a thorough exploration of three types: sparse, hierarchical, and approximate Bayesian learning. On the other hand, statistical model learning integrates Bayesian learning methodologies into data-centric statistical modeling within the domain of probabilistic machine learning, encompassing four major categories: Bayesian nonparametric clustering, Gaussian process, Bayesian dynamic linear model, and Bayesian neural network. Inferences for both physical and data-centric statistical models resonate across various applications, including modal analysis, model updating, damage detection, reliability updating, response prediction, and model selection, highlighting their pivotal role in enhancing comprehension of dynamical systems and decision-making processes. The paper navigates obstacles by proposing ways to enhance Bayesian inference efficiency through techniques such as surrogate models and parallel computation. Furthermore, it explores advanced methods for estimating posterior distributions, addresses non-Gaussian and non-stationary modeling errors, and envisions the forefront of online Bayesian learning amid evolving trends in machine learning. Distinguished from previous research, this study offers a thorough examination of both traditional and cutting-edge Bayesian methods. It not only underscores the transformative influence of Bayesian approaches but also serves as a beacon, guiding researchers towards fresh advancements and opportunities while aiding in the judicious selection and refinement of suitable methods for various challenges in structural dynamics.

*Corresponding author.
E-mail address: wangjiyan@um.edu.mo (W.J. Yan); yc17409@um.edu.mo (L.F. Mei)

**List of Abbreviations**

| | |
|---|---|
| ABC | approximate Bayesian computation |
| ABC-NS | ABC-nested sampling |
| ABC-PMC | ABC-population Monte Carlo |
| ABC-SMC | ABC-sequential Monte Carlo |
| ABC-SubSim | ABC-subset simulation |
| AIES | affine-invariant ensemble sampler |
| ARD | automatic relevance determination |
| BASIS | Bayesian annealed sequential importance sampling |
| BCNN | Bayesian convolutional neural network |
| BDLM | Bayesian dynamic linear model |
| BFFTA | Bayesian fast Fourier transform approach |
| BLRM | Bayesian linear regression model |
| BNN | Bayesian neural network |
| BNPMM | Bayesian nonparametric mixture model |
| BOMA | Bayesian operational modal analysis |
| BRNN | Bayesian recurrent neural network |
| BSDA | Bayesian spectral density approach |
| BTDA | Bayesian time domain approach |
| BUS | Bayesian updating with structural reliability methods |
| CNN | convolutional neural network |
| c.o.v. | coefficient of variation |
| DKF | dual Kalman filter |
| DL | deep learning |
| DNN | deep neural network |
| DOF | degree-of-freedom |
| DP | Dirichlet process |
| DPMM | Dirichlet process mixture model |
| DRAM | delayed rejection adaptive Metropolis |
| DREAM | differential evolution adaptive Metropolis |
| EKF | extended Kalman filter |
| EK-PK | extended Kalman particle filter |
| ELBO | evidence lower bound |
| EM | expectation-maximization |
| EOV | environmental and operational variability |
| FBG | fiber Bragg grating |
| FBSDA | fast Bayesian spectral density approach |
| FBSTA | fast Bayesian spectral trace approach |
| FCBNN | fully connected Bayesian neural network |
| FE | finite element |
| FFT | fast Fourier transform |
| FIM | Fisher information matrix |
| FRF | frequency response function |
| GP | Gaussian process |
| GPR | Gaussian process regression |
| GRU | gated recurrent unit |
| HMC | Hamiltonian Monte Carlo |
| HPM | hierarchical prior model |
| HSM | hierarchical stochastic model |
| iOSE | improved orthogonal series expansion |
| KF | Kalman filter |
| KL | Kullback-Leibler |
| KPCA | kernel principal component analysis |
| LSTM | long-short term memory |
| MAC | modal assurance criteria |
| MAL | Metropolis-adjusted-Langevin |
| MAP | maximum a posteriori |
| MCMC | Markov chain Monte Carlo |
| MDS | multi-dimensional scaling |
| MFVI | mean-field variational inference |
| MH | Metropolis-Hastings |
| ML | machine learning |
| MMD | maximum mean discrepancy |
| MPV | most probable value |
| MSD | Mahalanobis squared distance |
| NLLF | negative log-likelihood function |
| NNM | nonlinear normal mode |
| OMA | operational modal analysis |
| PCE | polynomial chaos expansion |
| PDF | probability density function |
| PF | particle filter |
| PIML | physics-informed machine learning |
| PINN | physics-informed neural network |
| PSD | power spectral density |
| RNN | recurrent neural network |
| RUL | residual useful life |
| SBL | sparse Bayesian learning |
| SFE | spectral finite element |
| SGD | stochastic gradient descent |
| SHMC | shadow HMC |
| S2HMC | separable shadow HMC |
| SI | system identification |
| SRS | stochastic response surface |
| SVI | stochastic variational inference |
| TEMCMC | transitional ensemble MCMC |
| TMCMC | transitional MCMC |
| ToF | time of flight |
| UGW | ultrasonic guided wave |
| UKF | unscented Kalman filter |
| VBMC | variational Bayesian Monte Carlo |
| VI | variational inference |
| VIV | vortex-induced vibration |
| WFE | wave and finite element |
| WIM | weigh-in-motion |

# 1 Introduction

As the bedrock of our society, engineering structures carry the weighty responsibility of serving humanity for generations to come. However, their durability can be significantly compromised by harsh environmental conditions, rigorous loading, and inadequate maintenance practices. Consequently, the realm of structural dynamics has emerged as a crucial discipline for safeguarding the safety and reliability of these vital structures. This encompasses not only pinpointing the pivotal parameters within structural systems and monitoring their operational states through dynamic response measurements [1, 2], but also evaluating the overarching dynamic behaviors in the realms of civil, mechanical, and aerospace engineering. Over the years, numerous researchers have devoted their efforts to demonstrating the applicability of the forward and inverse structural dynamic analysis in real-world problems, achieving substantial progress that has been comprehensively reviewed in many previous studies [3-9]. Despite these advances and achievements, the development of structural dynamic analysis has not yet been full-fledged, with one of the most pressing issues regarding the uninsured robustness of its results [2, 10], especially in the inverse problem or system identification, as conventional methods are typically restricted to deterministic optimal values while ignoring the parametric uncertainties. As a result, there is still a demand for more robust algorithms with a systematic framework to address uncertainties in structural dynamic analysis, which has attacked increasing attention in recent years [2, 11-13].

Generally, there are various kinds of uncertainties involved in dynamic problems, such as uncertainties stemming from incomplete information, errors due to imperfect modeling of physical phenomena, as well as measurement noise, etc. [14]. Disregarding the diversity of uncertainty sources, the uncertainties in fields within the model universe can be categorized into two types, namely the aleatoric uncertainty and epistemic uncertainty (Fig. 1) [15]:

- **Aleatoric uncertainty:** This type of uncertainty arises from the inherent randomness or noise of data [16], which is irreducible even if more data are collected to refine the model and includes the homoscedastic and heteroscedastic parts [17]. The former remains constant for different inputs, while the latter depends on the inputs as some inputs potentially have more noisy outputs than others. Aleatoric uncertainty embodies the inherently stochastic nature of inputs, outputs, and the dependency between them, which, in the context of structural dynamics, can source from various factors such as variability of material properties and external loads, measurement noise of excitation and dynamic response, environmental and operational variabilities (EOVs), manufacturing variabilities, inappropriate data pre-processing methods, etc. Moreover, aleatoric uncertainty can also accumulate from multiple sources and propagate into the model.
- **Epistemic uncertainty:** This type of uncertainty stems from a lack of knowledge about which model is best suited to describe the given data [16], which can be theoretically explained out given enough data to refine the model. The sources of epistemic uncertainty are similar within engineering problems, which can be categorized as uncertainty from model parameters and uncertainty from model structure [15]. For example, adopting a linear assumption to model the dynamic response of a nonlinear structural system inherently introduces epistemic uncertainties, stemming from incomplete knowledge about the system's behavior.

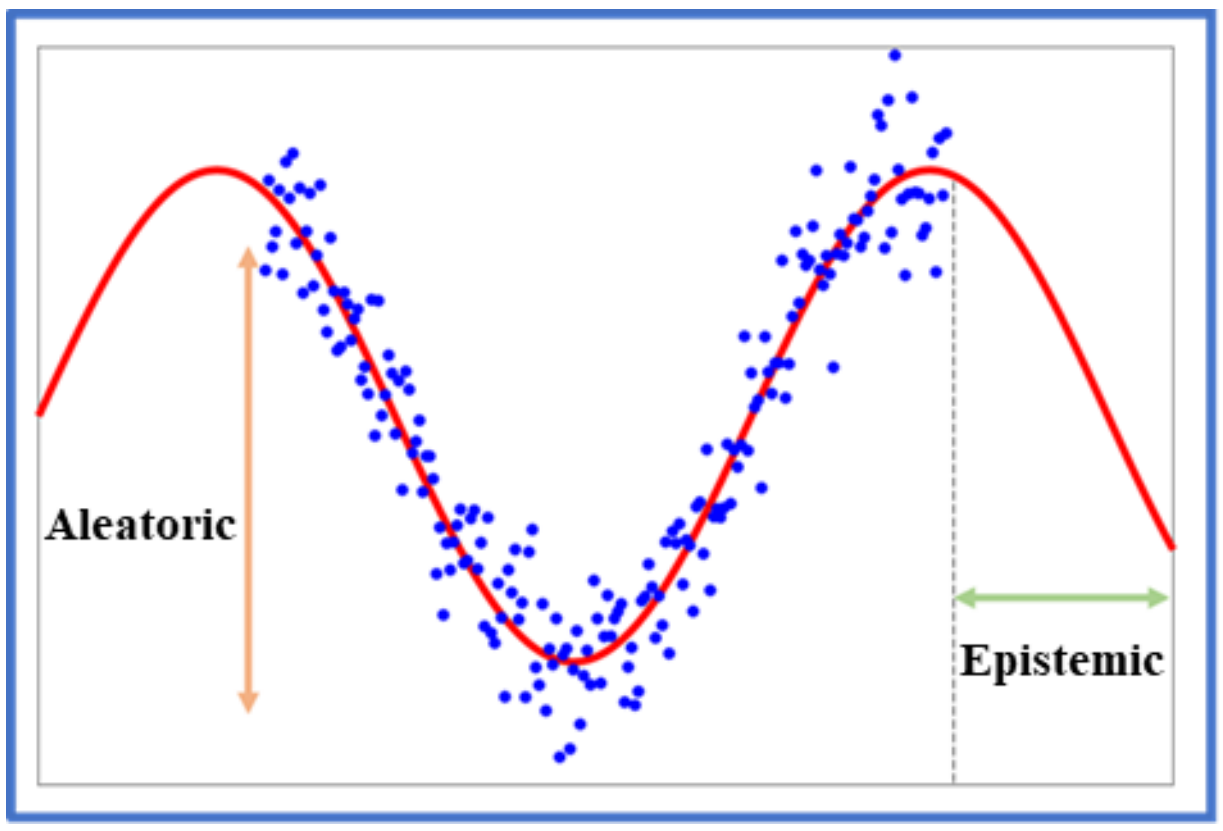

Figure 1. A schematic view of the aleatoric and epistemic uncertainties (reproduced from [18]).

To address the ubiquitous uncertainties in structural dynamics, an intuitive and commonly used approach is applying statistical inference to consider the probability distribution of each variable. There are two main broad approaches to statistical inference, namely the 'frequentist' and the Bayesian approach [19]. The definition of probability in 'frequentist' statistics is a long-run limiting relative frequency [20] while in Bayesian statistics, on the other hand, probability is viewed as a measure of the plausibility of a proposition conditional on incomplete information, with the true or false status of the proposition remaining unknown [21]. From the Bayesian perspective, the probability of an event can be updated based on observed evidence [22]. Specifically, the Bayesian theory explains the probability of an event given that another event has been observed, which, in contrast to 'frequentist' statistics, allows for reversely estimating the probability given an observation. Moreover, both the data and model parameters are assigned probabilities within the Bayesian perspective, leading to a systematic framework for addressing the epistemic uncertainty through the probability distributions of model parameters. However, despite the advantage of Bayesian statistics in handling uncertainty, it was not widely embraced until the late 20th century because of a prevalent belief that probability can only be applied to aleatoric uncertainty, rather than epistemic uncertainty. Since the 1950s, efforts by Cox [23, 24] and Jaynes [25, 26] to establish a rigorous logic foundation for the Bayesian approach have led to its popularity as a mathematical approach for statistical inference and uncertainty quantification, which has since been applied and investigated in various fields.

On the other hand, there are two main approaches for structural dynamic analysis, namely physical model-based and data-centric statistical model-based approaches [1]. Conventionally, structural dynamics analysis has relied on physics-based or law-based models, such as finite element models, of the investigated structural system, as illustrated in Fig. 2. Subsequent analysis tasks are then performed using a detailed physical description of the model in combination with measured data from the real structure [1]. Bayesian inference was introduced into this framework in the 1990s by Beck and Katafygiotis [27, 28]. This approach offers a way to handle the uncertainties conditioned on the non-repeatable nature of many engineering problems, which limits statistical inference from the ‘frequentist’ perspective. Another advantage of this approach is that the true values of the model parameters are commonly absent as estimation is often not unique, which makes it more reasonable to interpret probability as a multivalued logic for plausible reasoning conditioned on incomplete information [29]. Following this pioneering insight of employing Bayesian inference for physical models in structural dynamics, there have been considerable developments in this field over the years for various purposes, including propagating uncertainties for robust response and reliability predictions [30, 31], updating the parameters of finite element models [32, 33], selecting the model class from a family of competitive model classes [14, 34], and conducting uncertainty quantification for more reliable structural damage diagnosis results [35]. However, building accurate physical models can sometimes be challenging due to incomplete knowledge of physical mechanisms as well as difficulties in specifying complex system components like joints and bonds. This limitation has led to the rise of data-centric statistical model-based approaches in structural dynamics [1], which directly construct data-centric statistical models using measured data to represent input-output mapping relationships for various tasks. The advent of advanced sensing networks and machine learning (ML) techniques has further fueled the development of these statistical model-based approaches. Meanwhile, Bayesian inference has also been emphasized within these approaches, as ML models are prone to overconfident but inaccurate predictions when uncertainties are not properly recognized or handled [2]. Bayesian inference addresses these shortcomings by providing a more robust framework for uncertainty estimation.

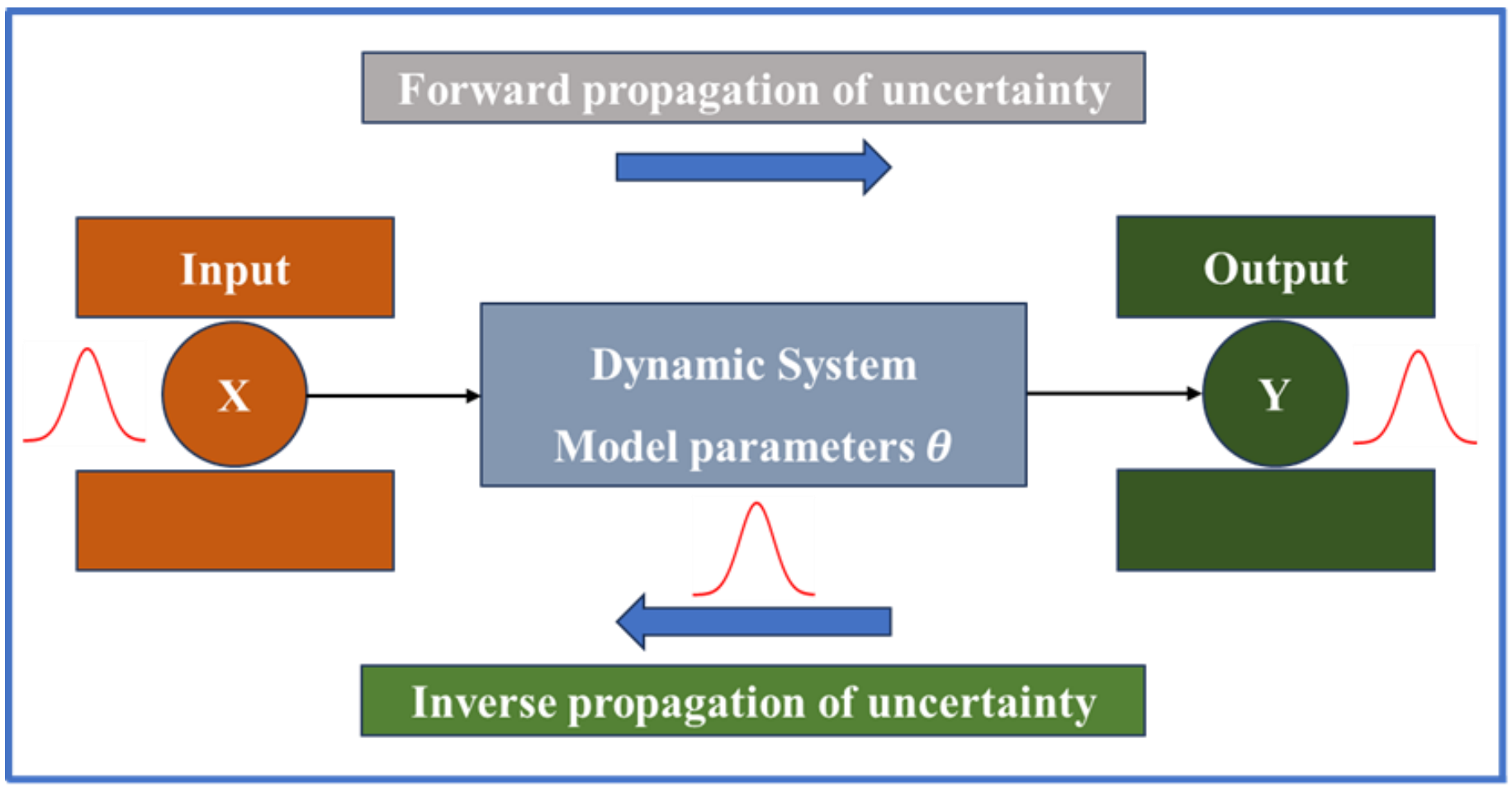

Figure 2. A schematic overview of Bayesian statistics for structural dynamics, where both the data and model parameters are viewed as uncertain variables.

Despite the growing prevalence and achievements of Bayesian approaches in structural dynamics, to the best of the authors' knowledge, there has yet to be a comprehensive review that categorizes these Bayesian approaches and outlines their specific application scenarios, particularly from the perspective of Bayesian inference for physical models and data-centric statistical models. Existing review works primarily focused on specific techniques or application scenarios in structural dynamics [32, 33, 36-39]. To fill this gap, this work provides a systematic review of traditional and state-of-the-art Bayesian methods for both physical models and data-centric statistical models, along with their applications across various tasks in structural dynamics. Specifically, we offer a detailed introduction to the Bayes' theorem, exploring three mainstream approaches for estimating the posterior distribution. Furthermore, we categorize Bayesian approaches in structural dynamics into two main types: Bayesian inference for physical models and for data-centric statistical models. We then delve into the applications of these approaches in some essential tasks of structural dynamics, including system identification (SI), model updating, damage identification, and reliability updating. Specific Bayesian approaches designed for each task are also elaborately discussed, focusing on the selection of priors, formulation of likelihood functions, and estimation of posterior distributions. A schematic diagram illustrating the Bayesian approaches for structural dynamics covered in this review is presented in Fig. 3a, while Fig. 3b presents a roadmap outlining the structure of the paper.

The remaining parts of this review are organized as follows: Section 2 provides the theoretical background of the Bayesian inference framework, along with three mainstream approaches to estimating the posterior distribution. On this basis, Section 3 delves into traditional and state-of-the-art methodologies of Bayesian inference for physical models and their applications in structural dynamics. Section 4 examines four approaches to Bayesian inference for statistical models that are prevalent in structural dynamics, along with their applications in this area. Section 5 discusses some research challenges that require future efforts, as well as some potential directions for further exploration. Section 6 concludes this review by highlighting the virtues, potentialities, and open challenges of Bayesian approaches in structural dynamics.

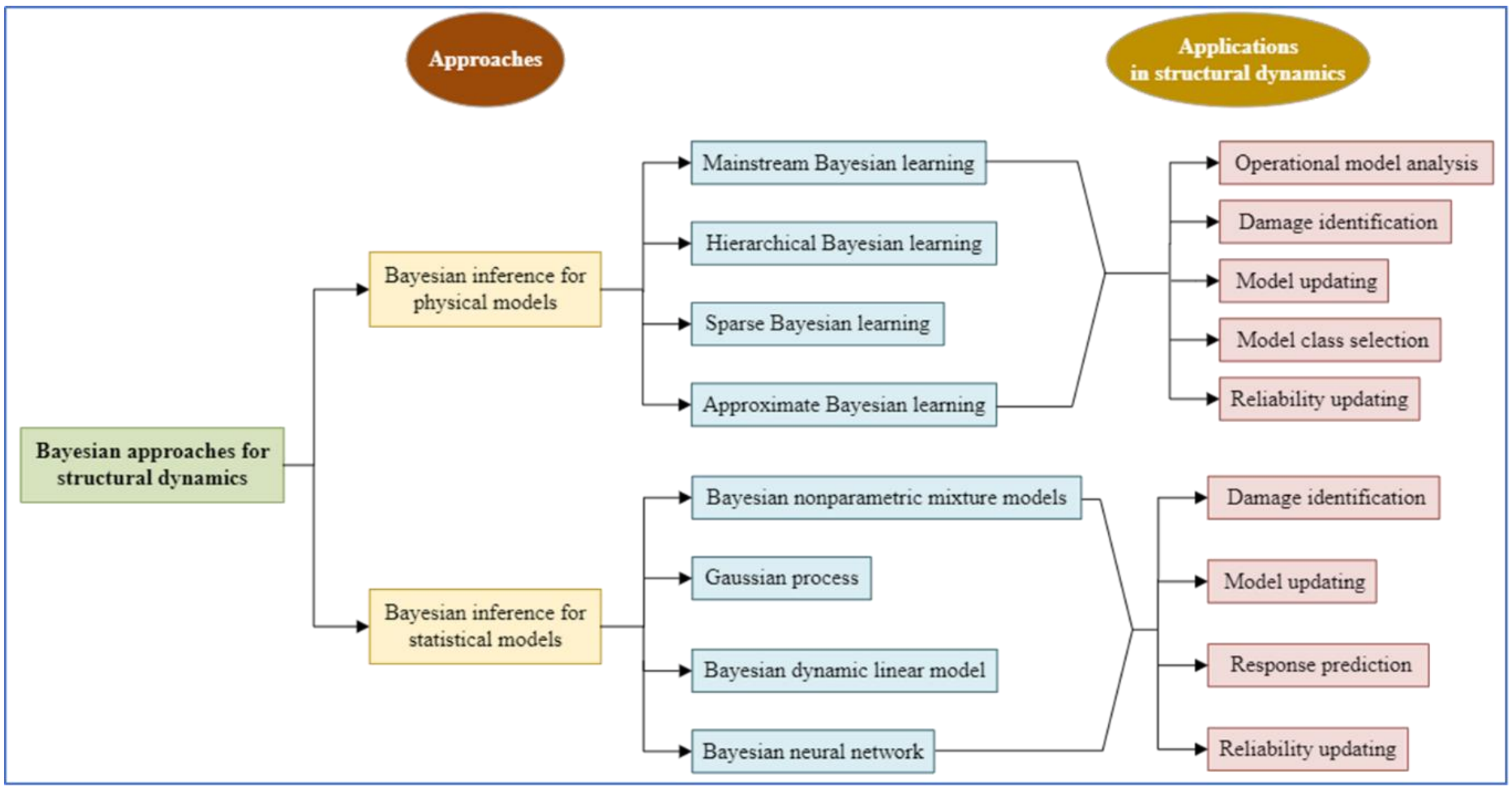

(a)

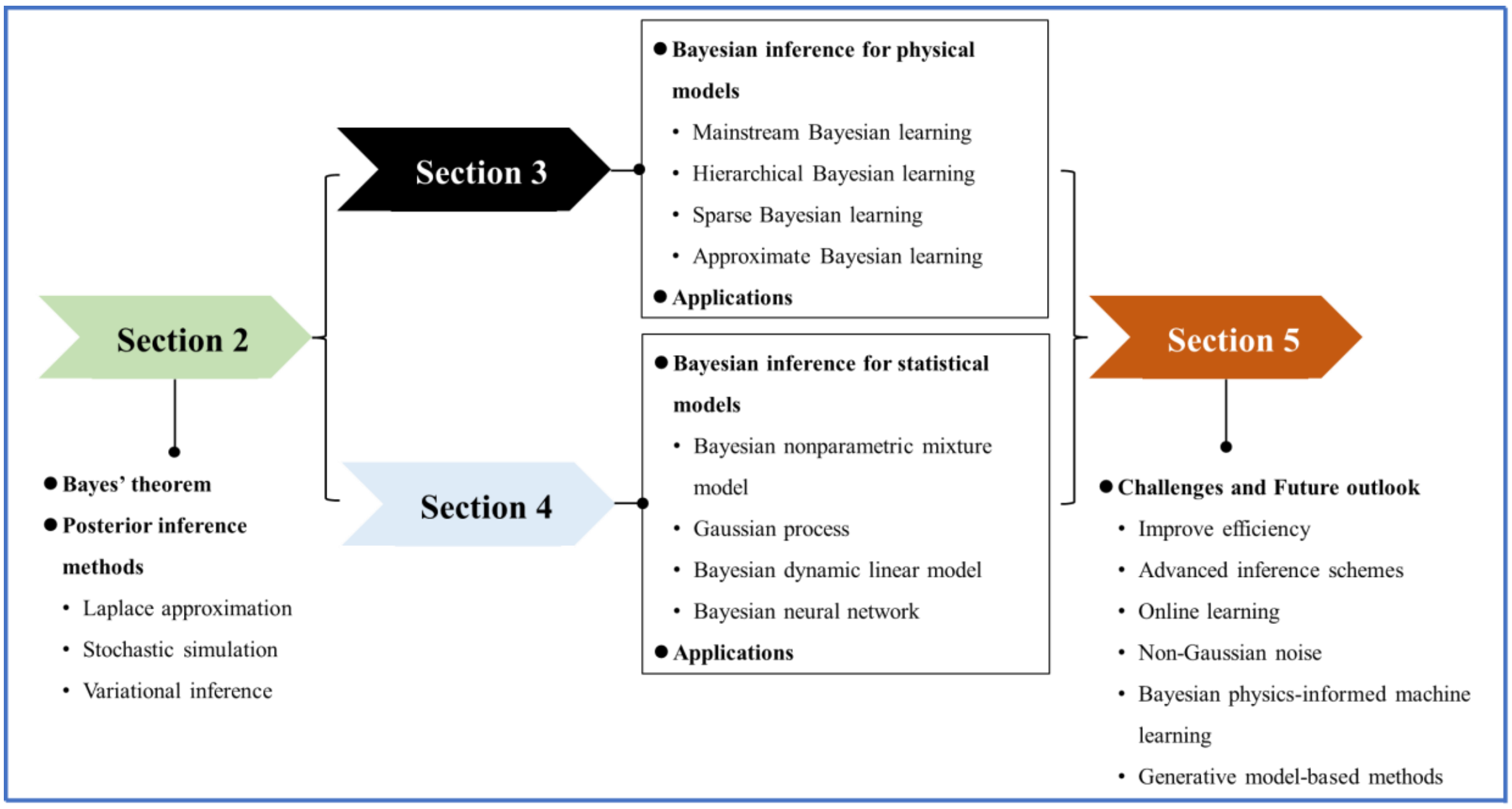

(b)

Figure 3. (a) A schematic diagram of the contents included in this review; (b) Roadmap of the review structure.

## 2 Fundamentals of Bayesian Inference

### 2.1 Bayes' theorem

Named after Thomas Bayes, the Bayes' theorem stands as a potent instrument for recalibrating probability distributions in light of fresh evidence. With a storied past and broad-ranging applications across various disciplines, it stands as a fundamental pillar in contemporary statistics and probability theory. If we denote two events as $\mathcal{A}$ and $\mathcal{B}$, the Bayes' theorem can be derived as follows:

$$p(\mathcal{A}|\mathcal{B}) = \frac{p(\mathcal{B}|\mathcal{A}) \cdot p(\mathcal{A})}{p(\mathcal{B})} \tag{1}$$

The Bayes' theorem offers a method for deducing uncertain models of systems from their measurements. Rooted in probability logic, the Bayesian perspective posits that probability signifies a level of confidence in a statement.

Within the realm of engineering applications, the Bayes' theorem plays a pivotal role in enriching our comprehension of a system's behavior through the assimilation of fresh observational data. This is accomplished via parametric identification, which entails assigning values of unknown parameters to a series of mathematical models that depict a specific physical phenomenon or system, along with selecting a suitable class of mathematical models for parametric identification. In the domains of physical model learning or data-centric statistical model learning in structural dynamics, Bayesian inference can be written as [27]:

$$p(\varpi|\mathfrak{M},\mathcal{D})=\frac{p(\mathcal{D}|\mathfrak{M},\varpi)\cdot p(\varpi|\mathfrak{M})}{p(\mathcal{D}|\mathfrak{M})}=\frac{p(\mathcal{D}|\mathfrak{M},\varpi)\cdot p(\varpi|\mathfrak{M})}{\int_{\Theta}p(\mathcal{D}|\mathfrak{M},\varpi)\cdot p(\varpi|\mathfrak{M})\cdot d\varpi} \tag{2}$$

where $\varpi$ denotes the physical parameters of the structural dynamical system or the hyperparameters of data-centric statistical models to be inferred; $\mathcal{D}$ represents the available measurements or training data; $\mathfrak{M}$ is the model class. $p(\varpi|\mathfrak{M},\mathcal{D})$ , $p(\varpi|\mathfrak{M})$ and $p(\mathcal{D}|\mathfrak{M},\varpi)$ represent the posterior distribution, the prior distribution and the likelihood function, respectively. In Eq. (2), the likelihood function $p(\mathcal{D}|\mathfrak{M},\varpi)$ as a function of $\varpi$ expresses the probability of getting data $\mathcal{D}$ from the model class $\mathfrak{M}$ given a value of the parameter vector $\varpi$; the prior distribution $p(\varpi|\mathfrak{M})$ is specified by $\mathfrak{M}$ which is chosen to quantify the initial plausibility of each possible value of the parameter vector $\varpi$; $p(\mathcal{D}|\mathfrak{M})$ is called the evidence for the model class given by data $\mathcal{D}$. Although $p(\mathcal{D}|\mathfrak{M})$ is a normalizing constant in Eq. (2) and so it does not affect the shape of the posterior distribution, it has great importance in model class selection and averaging.

Leveraging Bayes' Theorem, this approach establishes a fundamental framework for data processing, facilitating the derivation of conclusions regarding parameters that align with modeling assumptions and the principles of probability. The Bayesian methodology presents notable advantages over non-Bayesian techniques in a spectrum of scenarios [40] where developing estimators for crucial parameters poses challenges, resolving significant uncertainties in identification that require quantification, and managing cases where identification outcomes are intricately linked to underlying assumptions, demanding the quantification of effects or the choice of better assumptions. By integrating Bayesian inference into structural dynamics analysis, engineers and researchers can enhance decision-making processes, accommodating uncertainties, and refining the accuracy of structural assessments based on observed data.

## 2.2 Strategies for posterior inference

In Bayesian learning, the posterior distribution plays a pivotal role in updating our beliefs about the parameter values of a model based on observed data. This distribution captures the updated uncertainty about these parameters after taking into account both the prior beliefs and the information contained in the data. However, directly computing the posterior distribution can be challenging due to the intractable nature of the evidence $p(\mathcal{D}|\mathfrak{M})$. To address this challenge, various sophisticated methods have been developed to approximate the posterior distribution effectively over time. These methods are crucial for making Bayesian inference practical in a wide range of applications. Some of the primary approaches include Laplace approximation, stochastic approximation and variational inference. These methods represent the most fundamental strategies employed to tackle the challenge of inferring the posterior distribution in Bayesian learning. By approximating the posterior distribution, researchers are able to leverage the power of Bayesian inference for uncertainty quantification and decision-making in a wide array of complex problems.

### 2.2.1 Laplace approximation

Laplace approximation is a classical method used in Bayesian statistics to approximate a posterior distribution with a Gaussian distribution, particularly beneficial when the posterior distribution is intricate and challenging to compute directly. The fundamental concept behind Laplace approximation involves approximating the posterior distribution's shape near its peak (mode) with a multivariate Gaussian distribution. This entails identifying the mode of the posterior distribution, often called the maximum a posteriori (MAP) estimate, and then conducting a second-order Taylor expansion of the log posterior around this mode. The resulting Gaussian approximation simplifies the form of the distribution for easier computational manipulation.

To obtain an approximate posterior around the MAP estimation of the parameters $\varpi$ through a Gaussian distribution [30], the second-order Taylor expansion to the log posterior PDF $p\left(\varpi|\mathfrak{M},\mathcal{D}\right)$ at $\varpi=\hat{\varpi}$ can be conducted as follows:

$$\log p\left(\varpi|\mathfrak{M},\mathcal{D}\right)\approx\log p\left(\hat{\varpi}|\mathfrak{M},\mathcal{D}\right)-\frac{1}{2}(\varpi-\hat{\varpi})^T\left[\boldsymbol{H}(\hat{\varpi})\right](\varpi-\hat{\varpi}) \tag{3}$$

where $\hat{\varpi}$ is the MAP estimate, i.e. $\hat{\varpi}=\arg\max_{\varpi}\left[\log p\left(\varpi|\mathfrak{M},\mathcal{D}\right)\right]=\arg\max_{\varpi}\left[\log p\left(\varpi,\mathcal{D}|\mathfrak{M}\right)\right]$ (notice that the evidence $p\left(\mathcal{D}|\mathfrak{M}\right)$ is independent of $\varpi$); $\boldsymbol{H}(\hat{\varpi})$ is the Hessian matrix of $-\log p\left(\varpi,\mathcal{D}|\mathfrak{M}\right)$ evaluated at $\hat{\varpi}$. As the first term in Eq. (3) is independent of $\varpi$, the posterior PDF simplifies to:

$$p\left(\varpi|\mathfrak{M},\mathcal{D}\right)\propto\exp\left(-\frac{1}{2}(\varpi-\hat{\varpi})^T\left[\boldsymbol{S}^{-1}(\hat{\varpi})\right](\varpi-\hat{\varpi})\right) \tag{4}$$

where $\boldsymbol{S}(\hat{\varpi})=\boldsymbol{H}^{-1}(\hat{\varpi})$. Therefore, the posterior PDF of the parameters $\varpi$ can be well-approximated in the neighborhood sufficiently close to $\hat{\varpi}$ by a Gaussian PDF with mean $\hat{\varpi}$ and covariance matrix $\boldsymbol{S}(\hat{\varpi})$.

In practical applications, Laplace approximation is frequently employed as a swift and computationally efficient method for approximating the posterior distribution, making Bayesian inference feasible in instances where exact computation proves challenging. This technique shines particularly when dealing with unimodal and symmetric posterior distributions, where the Gaussian approximation can effectively encapsulate crucial distribution characteristics near its peak. However, the application of Laplace approximation encounters hurdles in scenarios with limited data or with sufficiently large the number of model parameters or when working with unidentifiable model classes. Moreover, it may struggle with highly skewed or multimodal distributions. This approximation necessitates non-convex optimization in a high-dimensional parametric space, presenting computational challenges, especially in cases where the model class lacks global identifiability and multiple local maxima exist.

### 2.2.2 Stochastic simulation

Due to the complexities involved in computing the high-dimensional integrals required for Bayesian updating, alternative approaches have been developed that focus on approximation through stochastic sampling, in which samples consistent with the posterior PDF $p(\varpi|\mathfrak{M},\mathcal{D})$ are generated. Currently, the most prevalent and widely used stochastic approximation approaches are Markov chain Monte Carlo (MCMC) methods, which construct an ergodic Markov chain whose stationary distribution, says $\pi$, is the posterior $p(\varpi|\mathfrak{M},\mathcal{D})$, and thus allow for generating samples from the stationary distribution. The main building block of this class of algorithms is the Metropolis-Hastings (MH) algorithm [41, 42]. It requires the definition of a family of proposal distributions $\{q(\varpi^*|\varpi_i)\}$ whose role is to generate possible transitions for the Markov chain, say from current sample $\varpi_i$ to the next candidate sample $\varpi^*$ of the chain. The candidate sample is accepted with probability $\alpha$ defined as:

$$\alpha = \min\left\{1, \frac{\pi(\varpi^*)q(\varpi_i|\varpi^*)}{\pi(\varpi_i)q(\varpi^*|\varpi_i)}\right\} \tag{5}$$

This type of approximation approach can handle more general cases than Laplace approximation as the samples are ensured to be drawn from the true posterior [43], but difficulty arises from ensuring the stationarity of the Markov chain. Moreover, MCMC methods are often inefficient for practical Bayesian inference. For example, the MH algorithm may be not efficient when the uncertain variables are highly correlated conditioning on the data and when the posterior PDF is very peaked [44].

Many other stochastic simulation methods have been proposed to address accuracy issues and computational challenges by reducing the number of model runs, accelerating convergence. Due to the limit of the length of this survey, we only outline briefly some of the typical methods.

- The Transitional MCMC (TMCMC) algorithm [45] aims to tackle the challenge of sampling from complex target probability density functions (PDFs) by instead sampling from a succession of intermediate PDFs that gradually approach the target PDF and are simpler to sample from. Compared with the previous approaches, TMCMC has several advantages: This method efficiently generates samples asymptotically distributed as the posterior PDF, which can be applied to a wide range of cases including multimodal distributions, peaked probability density functions, and probability density functions with flat regions over a lower-dimensional manifold of the parameter space. Furthermore, as a by-product of the procedure it can estimate the evidence without additional computational burden, which is important for Bayesian model class selection [45].
- Hamiltonian Monte Carlo (HMC) [46], also referred to as hybrid Monte Carlo, is a more advanced MCMC algorithm designed to handle high-dimensional problems more efficiently. It uses gradient information of the target distribution to guide the proposal steps, resulting in faster mixing and more efficient sampling, especially in high-dimensional spaces. Instead of making random-walk-like proposals (as in MH algorithm), HMC simulates a physical system with momentum and uses Hamiltonian dynamics to make proposals that follow the shape of the target distribution more closely.

- More recently, another approach, referred to as BUS (Bayesian Updating with Structural reliability methods) [47, 48], has been developed for effectively updating mechanical and other computational models, which opens up the possibility of efficient solution by subset simulation [49] as an alternative to MCMC. The BUS formulation integrates a rejection sampling strategy with structural reliability methods. Its main advantages lie in its simplicity and the capability to leverage the complete range of structural reliability techniques and related software for Bayesian updating.

### 2.2.3 Variational inference

The field of Bayesian statistics often deals with probabilistic models where the posterior distribution of latent variables is difficult or impossible to compute directly due to high-dimensional integrals or complex dependencies. Traditional sampling-based methods like MCMC, while accurate, can be computationally expensive and slow, especially for large datasets and high-dimensional models. In response to these challenges, variational inference (VI) arose as a more scalable, deterministic alternative to such sampling-based methods [50, 51]. VI is a deterministic approximation approach that employs a variational distribution to approximate the true posterior distribution by minimizing the Kullback-Leibler (KL) divergence between the approximate variational distribution and the true posterior [43]. Specifically, assuming the variational distribution is denoted by $q_{\phi}(\varpi)$ with a set of free parameters $\phi$, the KL-divergence between the variational distribution and the true posterior is:

$$
\begin{aligned}
D_{KL}\left(q_\phi(\varpi)\,||\,p(\varpi|\mathcal{D})\right) &= \int_\varpi q_\phi(\varpi)\log\frac{q_\phi(\varpi)}{p(\varpi|\mathcal{D})}\mathrm{d}\varpi \\
&= \int_\varpi q_\phi(\varpi)\log\frac{q_\phi(\varpi)p(\mathcal{D})}{p(\varpi,\mathcal{D})}\mathrm{d}\varpi \\
&= \log p(\mathcal{D}) - \left(\mathbb{E}_q\left[\log p(\varpi,\mathcal{D})\right] - \mathbb{E}_q\left[\log q_\phi(\varpi)\right]\right)
\end{aligned} \tag{6}
$$

where $\mathbb{E}_q[\bullet]$ denotes the expectation with respect to $q_\phi(\varpi)$. As $\log p(\mathcal{D})$ is irrelevant to the variational parameters $\phi$, minimizing the KL-divergence is equivalent to maximizing the remaining part of the right side in Eq. (6), which is referred to as the evidence lower bound (ELBO) as shown in Fig. 4 and denoted by:

$$
\text{ELBO} = \mathbb{E}_q\left[\log p(\varpi,\mathcal{D})\right] - \mathbb{E}_q\left[\log q_\phi(\varpi)\right] \tag{7}
$$

Hence, VI effectively transforms the objective of inferring the posterior distribution into an optimization task, which is actually to find a tradeoff between prediction accuracy and model complexity [52].

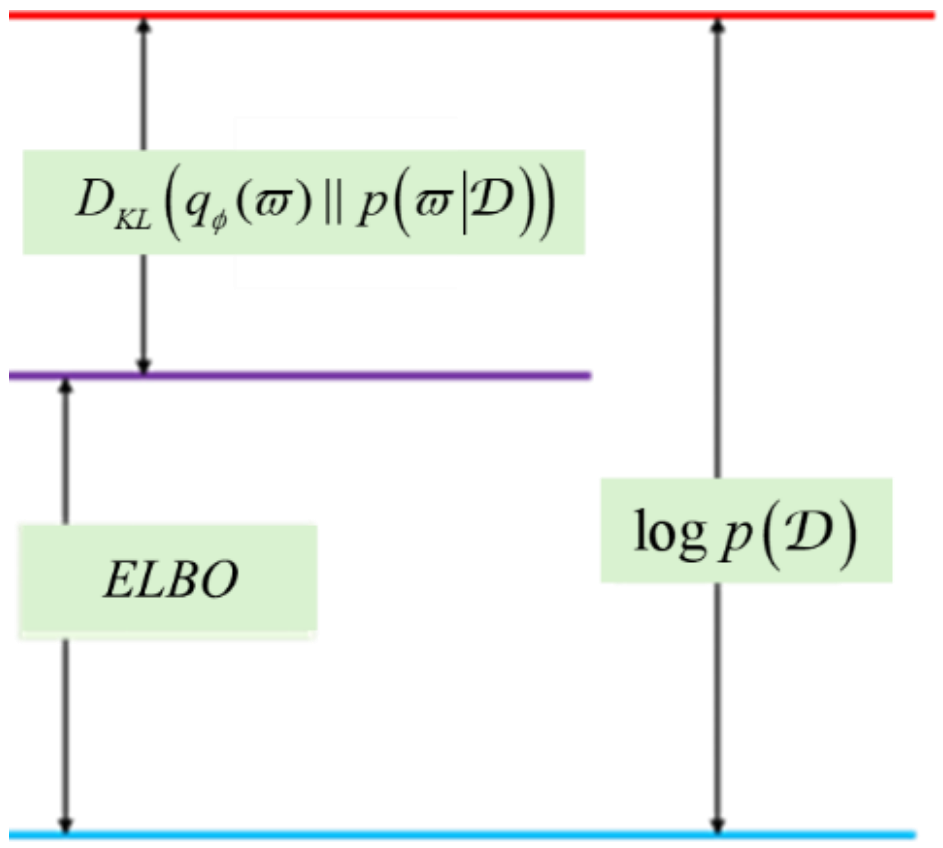

Figure 4. The relation between the KL-divergence, ELBO and log-evidence.

Mean-field VI (MFVI) [43] is one of the most basic and widely used forms of VI. The idea is to assume that the variational distribution factorizes over groups of latent variables, making the approximation more tractable. Each latent variable or group of variables is treated as independent. A fully factorized variational distribution based on the mean-field assumption takes the form of $q_{\phi}(\varpi)=\prod_{i=1}^{M} q_{\phi_i}(\varpi_i)$, which allows for analytical derivation of the optimal distribution of each factor [43] in the form:

$$q_{\phi_i}^{*}(\varpi_i)=\frac{\exp\left\{\mathbb{E}_{i\neq j}\left[\ln p(\mathcal{D},\varpi)\right]\right\}}{\int \exp\left\{\mathbb{E}_{i\neq j}\left[\ln p(\mathcal{D},\varpi)\right]\right\} d\varpi_j} \tag{8}$$

where the symbol $\mathbb{E}_{i\neq j}[\cdots]$ denotes expectation with respect to all variables $\varpi_j$, excluding $\varpi_i$. According to (8), the estimation of each one of $q_{\phi_i}^{*}(\varpi_i)$ requires knowledge of $q_{\phi_j}^{*}(\varpi_j)$ for all $j\neq i$. Subsequently, the optimal distribution $q_{\phi}^{*}(\varpi)$ can be approximated by updating each factorized distribution $q_{\phi_i}^{*}(\varpi_i)$ iteratively until the ELBO converges, which is referred to as coordinate ascent algorithm [53]. MFVI simplifies the optimization problem, making it computationally efficient. However, the independence assumption can be too simple and may fail to capture complex dependencies between variables.

Stochastic variational inference (SVI) is another prevalent kind of method for estimating the posterior [51]. The "stochastic" aspect of SVI comes from the use of stochastic optimization methods, such as stochastic gradient descent (SGD) in maximizing the ELBO. In many real-world applications, datasets are too large to fit into memory or to compute exact gradients over the entire dataset. Instead, SVI operates on mini-batches (small, random subsets of the data) to estimate noisy but efficient gradients, allowing it to scale to massive datasets. SVI can be applied to a wide variety of probabilistic models, including complex hierarchical Bayesian models, where exact inference would be intractable. Some remarkable extensions to SVI include the natural gradient [51], the Bayes by Backprop [54] and the local reparameterization trick [55]. Conventionally, these approaches also use factorized variational distributions, which leads to a straightforward ELBO, but the approximation capability is limited due to the ignorance of posterior correlations among variational parameters.

VI provides a framework that can be adapted to a wide range of models, including non-conjugate models (where the posterior distribution does not have a closed form). By utilizing various models such as parametric distributions and neural networks, VI allows flexibility in the choice of approximate posterior distributions, making it applicable to a broad set of problems. Compared to other approximation methods, VI stands out for its flexibility, scalability, and computational efficiency, but there is still a demand to improve its approximation capability without compromising efficiency.

# 3 Bayesian Inference for Physical Models and Applications to Structural Dynamics

## 3.1 Bayesian Inference Methodologies for Physical Models

### 3.1.1 Mainstream Bayesian system identification paradigm

The introduction of Bayesian inference methods in structural dynamics can be attributed to around three decades ago [56]. The adoption of Bayesian methods in structural dynamics has undergone substantial growth since 1998, marked by the seminal work of Beck and Katafygiotis, who introduced a systematic framework for Bayesian inference [27, 28]. This framework has since become a cornerstone of Bayesian learning paradigms in the field, embraced by a multitude of research groups and practitioners.

In structural dynamics, Bayesian inference plays a crucial role in incorporating the predictions of structural models into a probabilistic framework. Suppose system data $\mathcal{D} = \{y^{mea}, \hat{u}\}$ is available that consists of measured output $y^{mea}$ of the system and possibly the corresponding system input $\hat{u}$. This approach enables the decomposition of predictions from structural models into two key components: a deterministic part representing the expected behavior based on the model, and a random part that accounts for uncertainties and variability in the system. Through introducing the prediction-error parameter $\epsilon$, the measured responses $y^{mea}$ can be connected with the model parameters $\theta$ [29]:

$$y^{mea} = x(u,\theta) + \epsilon \tag{9}$$

where $x(u,\theta)$ denotes the model prediction given the model parameters $\theta$ and system input $u$; $\epsilon$ is an additive noise variable representing the measurement noise and model error, which provides a bridge between the behaviors of the deterministic model and the real system. The maximum entropy PDF for the prediction error $\epsilon$ over an unrestricted range is discrete-time Gaussian white noise. Therefore, the predictive PDF for the system output $y_i^{mea}$ at discrete time $t_i$, conditional on the parameter vector $\theta$, is given by the following Gaussian PDF with the mean equal to the model output $x_i(u,\theta)$ and with a parameterized covariance matrix $\mathbf{\Sigma}(\theta)$:

$$p\left(y_i^{mea}|u,\mathfrak{M},\varpi\right) = \frac{1}{(2\pi)^{n_o/2}|\det\mathbf{\Sigma}|^{\frac{1}{2}}}\exp\left(-\frac{1}{2}\left(y_i^{mea}-x_i(u,\theta)\right)^T\mathbf{\Sigma}^{-1}\left(y_i^{mea}-x_i(u,\theta)\right)\right) \tag{10}$$

Conditioned on a set of measured quantities $\mathcal{D}=\left\{y_i^{mea}\right\}_{i=1}^{N}$, the likelihood function $p\left(\mathcal{D}|\mathfrak{M},\varpi\right)$ can be expressed as:

$$p\left(\mathcal{D}|\mathfrak{M},\varpi\right)=\prod_{i=1}^{N}\left(y_i^{mea}|u,\mathfrak{M},\varpi\right) \tag{11}$$

Based on the likelihood function, Bayesian inference can be conducted to estimate the posterior distribution of the parameters involved in the SI, $p\left(\varpi|\mathfrak{M},\mathcal{D}\right)$, given some prior information $p\left(\varpi|\mathfrak{M}\right)$ according to Eq. (2), which quantifies both the aleatoric and epistemic uncertainties involved in SI for a more informed decision process compared to traditional SI approaches with deterministic parameters.

Based on a selected model class, all the probabilistic information for the prediction of a vector of future system responses $y^{pred}$ is contained in the posterior robust predictive PDF implied by equation image and given by the Total Probability Theorem:

$$p\left(y^{pred} \middle| \mathcal{D},\mathfrak{M}\right) = \int p\left(y^{pred} \middle| \theta,\mathcal{D},\mathfrak{M}\right) p\left(\theta \middle| \mathcal{D},\mathfrak{M}\right) d\theta \tag{12}$$

The interpretation of Eq. (12) is that it represents a weighted average of the probabilistic predictions for each model specified within the model class $\mathfrak{M}$, with each weight determined by its corresponding posterior probability $p\left(\theta \middle| \mathcal{D},\mathfrak{M}\right)$. These pioneering efforts have led scholars and practitioners in structural dynamics to embrace the advantages of Bayesian inference. Significant techniques for approximating these integrals include Laplace's asymptotic approximation and stochastic simulation methods utilizing diverse MCMC algorithms. The evolution of Bayesian learning in structural dynamics has been characterized by a continuous refinement of concepts, methodologies, and practical applications, profoundly influencing the field of structural dynamic analysis. This exploration assists in addressing uncertainties, refining models with fresh data, and enabling informed decisions grounded in probabilistic frameworks.

### 3.1.2 Hierarchical Bayesian learning

Hierarchical Bayesian learning extends the classical Bayesian framework by incorporating a multilevel probabilistic structure. This approach is particularly beneficial for modeling heterogeneous data, calibrating complex systems with multiple objectives, and introducing sparsity into statistical models [57, 58]. Hierarchical Bayesian learning introduces an additional layer of hyperparameters, which can be incorporated into either the prior or the stochastic model (also known as likelihood functions when evaluated with some given data) to produce two different models [59]. These models primarily differ in their structural dependencies among stochastic variables, as illustrated in Fig. 5, where arrows represent conditional dependencies between parameters, and nodes without incoming arrows signify the need for a hyper prior distribution.

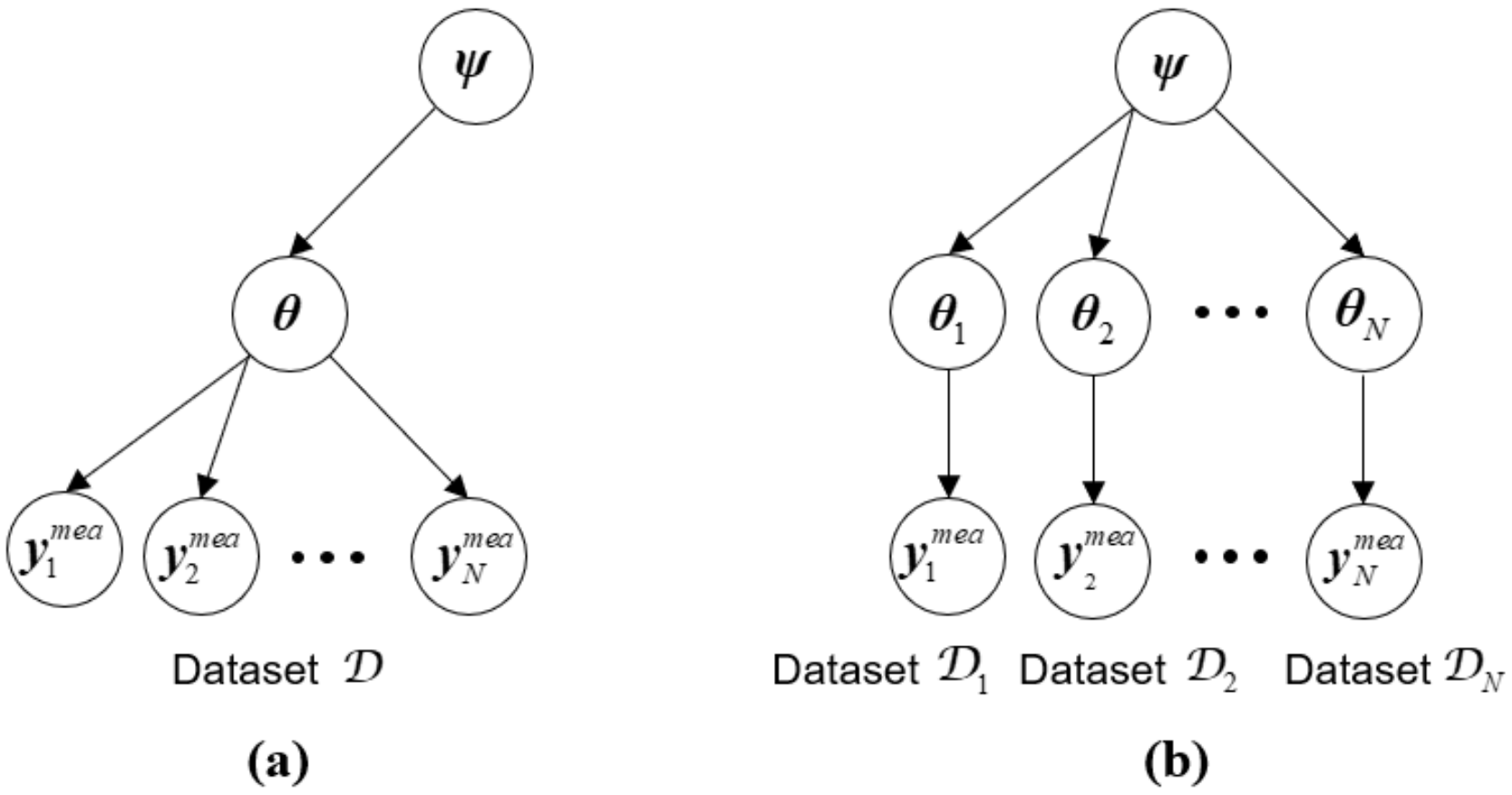

Figure 5. Graphical representation of HBM framework: (a) hierarchical prior model ; (b) hierarchical stochastic model [59].

**(1) Hierarchical prior model (HPM)**

HPM focuses on adjusting the prior distribution of model parameters by introducing hyperparameters $\psi$, thus enhancing the structured use of prior knowledge and adaptability to data. The hyperparameters $\psi$ in HPM behave as latent variables that do not change the underlying dependencies or functional forms between the model parameters $\theta$ and data $\mathcal{D}=\{y_i^{mea}\}_{i=1}^{N}$. Thus, the posterior distribution of $\theta$ can be expressed as:

$$p(\theta|\psi,\mathcal{D}) \propto p(\mathcal{D}|\theta)\cdot p(\theta|\psi) \tag{13}$$

where $p(\mathcal{D}|\theta)$ represents the likelihood function; $p(\theta|\psi)$ denotes the prior distribution of $\theta$ for a given $\psi$. In real applications, the estimation of $\psi$ is conducted by maximization of the evidence function $p(\mathcal{D}|\psi)$, known as Type-II Maximum Likelihood Approximation [37].

HPM is commonly utilized in applications that benefit from enhanced structuring of prior knowledge, such as in sparse Bayesian learning [60] or compressive sensing [61]. These applications often involve solving ill-posed problems where the addition of hierarchical priors can help impose additional constraints to guide the solution process.

**(2) Hierarchical stochastic model (HSM)**

Unlike HPM, HSM not only focuses on the prior distribution of parameters but also captures the uncertainty due to test-to-test variability, thus offering valuable insights into overall uncertainties. Let $\mathcal{D}=\{\mathcal{D}_i; i=1,2,...,\mathcal{D}_{N_t}\}$ consists of $N_t$ statistically independent measurement datasets, with the observation in the $i-th$ dataset denoted as $\{y_i^{mea}\}$. Fig. 5b illustrates the graphical representation of the HSM framework, where the dominant feature is that the model parameters vary by dataset or observation [59]. Such test-to-test variability of model parameters is caused by modelling and measurement errors, environmental and operational conditions, as well as manufacturing and assembling processes [62]. To capture this variability, HSM introduces hierarchy by assigning a prior probabilistic model for the model parameters $\theta_i$ that are governed by the hyperparameters $\psi$.

Additionally, HSM accounts for the non-stationary effect of prediction errors, which vary with input characteristics, modeling inaccuracies, and measurement noise [63]. These variations significantly impact the optimal values and associated uncertainties of model parameters. In general, a Gaussian probabilistic model is used for the $i$-th prediction errors term $\epsilon_i$ for the $i$-th dataset, characterized by zero mean and an additive variance parameter $\sigma_{\epsilon,i}^2$. The variance $\sigma_{\epsilon,i}^2$ is estimated at the same hierarchical level as the model parameters $\theta_i$ and varies across datasets. The variability of variances is captured by a Gaussian distribution defined by specific hyperparameters $\varphi$.

The hierarchical setting within HSM thus consists of two classes of parameters: those specific parameters $\theta_i$ and $\sigma_{\epsilon,i}^2$ for each dataset and the underlying hyperparameters $\psi$ and $\varphi$. Using the independent assumption among datasets or observations, the posterior distribution of all updating parameters can be expressed as [64]:

$$p(\{\theta_i\}_{i=1}^{N_t},\{\sigma_{\epsilon,i}^2\}_{i=1}^{N_t},\psi,\varphi,|\mathcal{D}) \propto \prod_{i=1}^{N_t} \underbrace{p(\mathcal{D}_i\,|\,\theta_i,\sigma_{\epsilon,i}^2)}_{\text{identification uncertainty}} \cdot \underbrace{p(\theta_i\,|\,\psi)\cdot p(\sigma_{\epsilon,i}^2\,|\,\varphi)}_{\text{uncertainy due to test-to-test variablity}} \tag{14}$$

This equation indicates that the uncertainties characterized by these hyperparameters are incorporated into the posterior distribution of the model parameters, offering profound insights into the overall uncertainties comprising both test-to-test variability and identification uncertainty from individual datasets. Despite HSM's advantages in modeling uncertainties, sampling from the posterior distribution as shown in Eq. (14) is challenging due to the extensive number of parameters involved.

### 3.1.3 Sparse Bayesian learning

As an effective strategy for alleviating ill-conditioning and ill-posed inverse problems, sparse processing techniques have garnered significant attention in recent years. Sparse Bayesian learning (SBL) is a general framework for solving sparse representations and quantifying the associated uncertainties from a Bayesian perspective. Some specific HSMs can promote sparsity in the inferred parameters, these prior models are called sparse priors. The sparse prior in the SBL framework is equivalent to the regularization term of the optimization problem. The automatic relevance determination (ARD) model with Gaussian distribution has been demonstrated to be an effective sparsity prior, capable of pruning surplus or irrelevant parameters in the model. It is defined as follows [65]:

$$p(\theta \mid \alpha) = \prod_{i=1}^{N} N(\theta_i \mid 0, \alpha_i^{-1}) \tag{15}$$

where $\alpha_i$ represents the prior precision (inverse variance) associated with the unknown model parameter $\theta_i$, serving to adjust the strength of the regularization constraint from the prior thereon. Typically, the prior precision $\alpha_i$ is modeled as a Gamma distribution. As the $\alpha_i$ approaches infinite, it implies that the corresponding unknown model parameter $\theta_i$ contributes negligibly to the model.

Likewise, the prediction error term $\epsilon$ can be treated as a Gaussian distribution, i.e., $\epsilon \sim \mathcal{N}(0, \sigma_\epsilon^2)$ .The posterior distribution of all updating parameters can be expressed as:

$$p\left(\theta, \alpha, \sigma_\epsilon^2 \mid \mathcal{D}\right) \propto p\left(\mathcal{D} \mid \theta, \sigma_\epsilon^2\right) p\left(\theta \mid \alpha\right) \tag{16}$$

where both $\alpha$ and $\sigma_\epsilon^2$ can be estimated by iterative calculation through maximizing the evidence $p\left(\mathcal{D} \mid \alpha, \sigma_\epsilon^2\right)$, which is given by:

$$p\left(\mathcal{D} \mid \alpha, \sigma_\epsilon^2\right) = \int \left(\mathcal{D} \middle| \theta, \sigma_\epsilon^2\right) \cdot p\left(\theta \middle| \alpha\right) d\theta \tag{17}$$

Note that the maximization in Eq. (17) tends to drive some prior precisions $\alpha_i$ approach positive infinity, resulting in a sparse model parameter vector $\theta$. This aligns with the principle of Bayesian Ockham's razor, which favors simpler models by penalizing unnecessary complexity. More information about SBL methods can be found in [65, 66].

### 3.1.4 Approximate Bayesian computation

In the mainstream Bayesian learning framework, the likelihood function of model parameters expressing the probability of achieving measured data given a model is essential. However, an analytical formula for the likelihood function might be difficult for some complex models, or even impossible to establish. Approximate Bayesian Computation (ABC) is proposed to bypass the explicit evaluation of the likelihood function. Instead of using this observed data vector $y$ directly in Bayes′ theorem, ABC produces sample pairs $(\theta, x) \in \Theta \times \mathcal{D} \subset \mathcal{R}^{d+l}$ for model class $\mathfrak{M}$ from a posterior probability distribution conditional on predicted outputs $x$ that are acceptably close under some metric $\rho$ and tolerance level $\varepsilon$ in the output space $\mathcal{D}$ to the observed data vector $y$.

A standard ABC method defines the approximate likelihood function given by $p_\varepsilon(\mathcal{D}|\theta, x) = p(x \in \mathcal{B}_\varepsilon(\mathcal{D})|x)$ [67]. Here, $\mathcal{B}_\varepsilon(\mathcal{D})$ is the region of the output space $\mathcal{D}$ defined as:

$$\mathcal{B}_\varepsilon(\mathcal{D}) = \{x \in \mathcal{D} : \rho(\eta(x), \eta(y)) \le \varepsilon\} \tag{18}$$

where $\eta(\cdot)$ denotes the low-dimensional vector of summary statistics for weakly comparing the closeness between $\boldsymbol{x}$ and $\boldsymbol{y}$. Based on the Bayes' theorem, the approximate posterior is given by $p_\varepsilon(\theta, x|\mathcal{D}) \propto p(x|\theta)p(\theta)\boldsymbol{I}_{\mathcal{B}_\varepsilon(\mathcal{D})}(x)$, in which $\boldsymbol{I}_{\mathcal{B}_\varepsilon(\mathcal{D})}(x)$ serves as an indicator function for the set $\mathcal{B}_\varepsilon(\mathcal{D})$ that equals one if $\rho(\eta(x), \eta(y)) \le \varepsilon$ and zero otherwise. The resulting posterior of model parameter $p_\varepsilon(\theta|\mathcal{D})$ can be obtained by marginalizing $p_\varepsilon(\theta, x|\mathcal{D})$. More information about ABC methods can be found in [68].

## 3.2 Applications to Structural Dynamics

### 3.2.1 Operational modal analysis

Modal analysis has diverse applications in structural vibration control, structural health monitoring, and structural damage detection. It focuses on identifying natural frequencies, damping ratios, and mode shapes. Significant advancements in experimental modal analysis have been achieved by incorporating measurements from both input and output data [69].

Nevertheless, when applied to large-scale civil infrastructures, these methods present challenges due to their reliance on specialized, time-consuming, costly, and disruptive experiments [70]. Consequently, there is a growing interest in ambient modal analysis, which provides a more cost-effective and efficient alternative by solely utilizing output measurements without requiring knowledge of the input [69].

In recent years, there has been a surging interest in quantifying uncertainties linked to modal parameters [71]. Among the array of methods available, Bayesian approaches offer a robust means to ascertain modal properties and their related uncertainties using both measured data and modeling assumptions [14]. In contrast to traditional modal identification techniques, Bayesian approaches establish probabilistic models for modal response within the Bayesian framework, facilitating the determination of PDFs for modal parameters. Furthermore, within the Bayesian framework, it becomes feasible to pinpoint the Power Spectral Density (PSD) of modal excitation and prediction errors. The evolution of Bayesian Operational Modal Analysis (BOMA) can be delineated into two generations. The initial iteration of BOMA was spearheaded by Yuen and Katafygiotis, who harnessed Bayesian theorem in system identification to quantify the uncertainties associated with modal parameters [72]. Au [73] achieved a significant milestone in overcoming the computational hurdles encountered by the first-generation BOMA by devising a rapid BOMA scheme.

#### ***3.2.1.1 The first generation of BOMA***

In the field of Operational Modal Analysis (OMA), several Bayesian methods have been introduced, including the Bayesian Time Domain Approach (BTDA) [74, 75], the Bayesian Spectral Density Approach (BSDA) [76], and the Bayesian Fast Fourier Transform Approach (BFFTA) [72]. Concerning ambient modal analysis, BSDA and BFFTA, which function in the frequency domain, exhibit greater potential compared to BTDA. These methodologies offer the benefit of leveraging data within specific resonance bandwidths, effectively filtering out information from entire frequency bands [73].

To integrate BOMA methodologies in the frequency domain, it is crucial to examine the statistical characteristics of Fast Fourier Transform (FFT) coefficients and PSD matrices. Yuen et al. [77] conducted a comprehensive study on the statistical properties of FFT and PSD. In the context of a linear system with $n_d$ degrees-of-freedom (DOFs), the assumption is made that discrete acceleration responses are accessible for the measured DOFs, with a sampling time interval represented as $\Delta t$. Let $y(n\Delta t) \in \Re^{n_o}, n = 1, 2, \ldots, N$ denote the discrete-time stochastic vector process corresponding to a specific set of sampled data. The FFT of $y(n)$ is defined as:

$$\boldsymbol{Y}_k = \boldsymbol{Y}_k^{\Re} + i\boldsymbol{Y}_k^{\Im} = \sqrt{\frac{\Delta t}{2\pi N}} \sum_{n=0}^{N-1} y(n) e^{(-i2\pi f_k n\Delta t)} \tag{19}$$

where $\boldsymbol{Y}_k^{\Re}$ and $\boldsymbol{Y}_k^{\Im}$ denote the real and imaginary part of $\boldsymbol{Y}_k$, respectively; $i^2 = -1$, $f_k = k\Delta f$, $k = 1, 2, \ldots, \mathrm{Int}(N/2)$, and $\Delta f = 1/(N\Delta t)$; “$k$” shown in the bracket or in the subscript denotes the frequency point $f_k$.

The research by Yuen and Katafygiotis [77] has demonstrated that as $N \to \infty$, the random vector $\boldsymbol{Y}_k$ follows a Gaussian distribution with a zero mean and covariance matrix determined by the expected value of the PSD matrix of $y(n)$. Assuming the existence of $n_s$ sets of independent and identically distributed time histories for the measured DOFs, where these sets of discrete FFT coefficients follow independent and identically complex Gaussian distributions, it can be shown that the sum of the PSD matrix follows the central complex Wishart distribution of dimension $n_o$ with $n_s$ DOFs [77].

As the dynamic measurement can be modelled as the sum of theoretical model response and the prediction error, the covariance matrix $\boldsymbol{\Sigma}_k$ could be correspondingly expressed as the sum of two terms $\boldsymbol{\Sigma}_k = \mathbb{E}\left[\boldsymbol{S}_x(k)\right] + \mathbb{E}\left[\boldsymbol{S}_\mu(k)\right]$. For high sampling rate and long duration of data, $\boldsymbol{\Sigma}_k$ over the specified frequency band can be replaced by a structured asymptotic form [72] $\mathbb{E}\left[\mathbf{S}_y(k)\right] = \boldsymbol{\Phi}\boldsymbol{\Lambda}_k\boldsymbol{\Phi}^T + \boldsymbol{S}_e\boldsymbol{I}_{n_o}$ where $\boldsymbol{S}_e$ is the spectral density of the prediction error; $I_{n_o}$ denotes the $n_o \times n_o$ identity matrix; $\boldsymbol{\Phi} = \left[\boldsymbol{\Phi}_1, \boldsymbol{\Phi}_2, \cdots \boldsymbol{\Phi}_{n_m}\right] \in \mathbb{R}^{n_o \times n_m}$ is the modal matrix composed of $n_m$ modes with $\boldsymbol{\Phi}_m$ denoting the $m$-th mode shape, comprised only of the mode shape components corresponding to the $n_o$ measured DOFs; $\boldsymbol{\Lambda}_k \in \mathbb{C}^{n_m \times n_m}$ is the theoretical spectral density matrix of the modal response given by $\boldsymbol{\Lambda}_k = \mathrm{diag}(h_k)\boldsymbol{S}_f diag(h_k^*)$, where $\boldsymbol{S}_f \in \mathbb{C}^{n_m \times n_m}$ is the spectral density matrix of the modal forces under the assumption of white noise excitation; $\mathrm{diag}(h_k) \in \mathbb{C}^{n_m \times n_m}$ denotes a diagonal matrix and its $m$-th diagonal entry can be expressed as $h_{mk} = [(\beta_{mk}^2 - 1) + \boldsymbol{i}(2\beta_{mk}\varsigma_m)]^{-1}$, where $\beta_{mk} = f_m / f_k$; here $f_m$ and $\varsigma_m$ denote the $m$-th modal frequency (in Hz) and the $m$-th modal damping ratio. The $(m, m')$-entry of $\boldsymbol{\Lambda}_k$ is given by $\boldsymbol{\Lambda}_k(m, m') = \boldsymbol{S}_{f,mm'} h_{mk} h_{m'k}^*$.

It is assumed that the FFT and spectral density set formed over the frequency band $\wp \in [k_1\Delta f, k_2\Delta f]$ is employed for ambient modal analysis. When the number of discrete data points $N \to \infty$, it has been proved that $\boldsymbol{Y}_k$ and $\boldsymbol{Y}_{k'}$ are uncorrelated as $k \neq k'$ [77]. Based on Bayes' theorem, the posterior PDF of $\theta$ is proportional to the likelihood function $p\left(\theta \,|\, \boldsymbol{Y}^{k_1,k_2}\right)$ and $p\left(\boldsymbol{S}_{k_1,k_2}^{sum} \,|\, \theta\right)$. Based on the Laplace approximation, the updated PDF of the parameter $\hat{\varpi}$ can be well-approximated by a Gaussian PDF $\mathcal{N}\left(\hat{\theta}, \boldsymbol{H}^{-1}\left(\hat{\theta}\right)\right)$. The most probable values (MPVs) $\hat{\theta}$ can be obtained by minimizing the Negative Log-Likelihood Function (NLLF) while the covariance matrix $\boldsymbol{C}\left(\hat{\theta}\right)$ can be achieved by taking the inverse of the Hessian matrix of $L(\theta)$ at $\hat{\theta}$.

The BFFT and BSDA methods present an innovative approach to managing various uncertainties, establishing a robust mathematical framework for their quantification, which proves invaluable for subsequent risk assessments. The primary aim of modal identification shifts towards determining the posterior PDF of the modal parameters based on provided data and modeling assumptions. The Bayesian probabilistic framework is explored for modal identification and identifiability based on field measurements from various structural health monitoring systems, including the Canton Tower [78, 79], TinKau Bridge [80], a 22-storey reinforced concrete building [81], among others. This methodology has found extensive application in modal identification and the investigation of time-varying trends in super high-rise buildings during severe typhoon conditions, such as the China Resources Tower and the Ping An Finance Center in Shenzhen, as well as the International Finance Centre and the International Commerce Centre in Hong Kong [82-87]. Furthermore, the BSDA method has been utilized to identify flutter derivatives of bridge sections operating in turbulent flow [88]. Findings suggest that the Bayesian approach offers a potent perspective by treating modal identification as an inference problem, utilizing probability as a metric for the relative plausibility of outcomes given a model of the system and the collected data.

#### *3.2.1.2 The second generation of BOMA*

One significant advantage of Bayesian formulations is their rigorous quantification of parameter uncertainty amidst measured data and modeling assumptions. However, despite these benefits, challenges in both computation and practicality hinder their application. The computational complexity of methods like BFFT and BSDA presents a notable obstacle, as they involve resolving high-dimensional optimization problems. The computational burden escalates significantly with more measured DOFs and modes, making direct implementation of conventional BFFT and BSDA impractical for a large number of DOFs. In BFFT, computing the posterior covariance function involves resource-intensive tasks like inverting the Hessian matrix, which becomes more challenging with increasing measured DOFs and modes. Both BFFT and BSDA require frequent computations involving the determinant and inversion of the covariance matrix for minimizing the NLLF. In high signal-to-noise ratio scenarios, the covariance matrix may become nearly rank deficient. Overcoming these computational and numerical challenges is vital to enhance the practical use and reliability of these methods.

To bridge the gap, Au [73] made substantial advancements in overcoming the computational hurdles associated with the conventional BFFT approach. This progress facilitated swift computation of the most probable values and the posterior covariance matrix using efficient algorithms. As a result, these outcomes can now be promptly obtained within a matter of seconds, including in on-site scenarios. The comprehensive framework of this newly devised approach, covering theoretical, computational, and practical dimensions, was detailed in [89].

In this section, we delve into the second generation of BOMA, categorizing it according to separated modes, closely spaced modes, multiple setups, and development of uncertainty laws.

***(1) Separated modes***

In modal parameter identification, a frequency band centered on the resonant frequency of the specific mode is typically chosen. The fast BFFT method relies on key assumptions: stationary response measurements, a substantial number of data points for precise estimation, band-limited white noise as natural excitation with constant PSD within the frequency band, uncorrelated prediction errors between channels with zero mean and equal variances, and negligible contributions from other modes in the frequency band to the identification model.

The modal parameters to be identified $\varpi = \left\{ f_s, \varsigma_s, \boldsymbol{S}_f, s_\mu, \boldsymbol{\Phi}_s \right\}$ encompass modal frequency, damping ratio, excitation spectral density, prediction error PSD, and normalized mode shape $\|\boldsymbol{\Phi}_s\| = 1$. Analytically deriving the covariance matrix's determinant and inverse is viable. Utilizing this information for the likelihood function with a uniform prior distribution ensures efficient resolution of the problem. Minimizing the NLLF enables determination of the modal parameter values and uncertainties. Au [73] skillfully approximated initial parameters using an analytical approach. Mode shapes can be inferred through singular value decomposition based on the spectral parameter MPVs. The NLLF's Hessian can be derived analytically for the free parameters and transformed to obtain the posterior covariance matrix of the modal parameters, considering the mode shape normalization constraints. Au discovered zero eigenvalues in the Hessian matrix aligned with mode shape directions. Eliminating these singular terms is crucial to remove irrelevant contributions. The modal parameters' covariance matrix can be accurately computed using its eigen-basis representation, excluding the eigenvalue with the smallest magnitude.

Another method for computing the Hessian of a function under constraints was introduced through the utilization of Lagrange multiplier concepts by Au and Xie [90], resulting in simpler expressions compared to direct differentiation approaches. Expanding on the concept of Modal Assurance Criteria (MAC) within a probabilistic framework, the posterior uncertainty of mode shapes can be quantified by measuring the expected value of the MAC between the most probable mode shape and the uncertain mode shape distributed according to the posterior PDF. Au and Zhang [91] derived an asymptotic approximation for the Expected MAC.

The pioneering concept of the fast BFFT method has inspired numerous studies that have both followed and expanded upon its foundations. Building on this motivation, a two-stage fast Bayesian spectral density approach was introduced for ambient modal analysis to tackle the challenges posed by conventional BSDA [92, 93]. In the first stage of this proposed method, spectrum variables independent of spatial information such as natural frequencies, damping ratios, modal excitation magnitudes, and prediction errors were separated from the full set of modal parameters. These variables were identified using a 'fast Bayesian spectral trace approach' (FBSTA), leveraging the sum of auto-spectral densities from all measured DOFs. Subsequently, in the second stage, mode shapes and their uncertainties are promptly extracted through a 'fast Bayesian spectral density approach' (FBSDA) utilizing the spectral density matrix. Analysis in this stage illustrates that FBSDA can be interpreted as a linear amalgamation of the 'fast Bayesian FFT approach', encompassing multiple sets of measurements. Moreover, a Bayesian spectral decomposition method has been proposed [94], which identified individual modes by leveraging their unique modal characteristics. This method decomposes the response spectrum matrix into eigenvalues (including frequency and damping) and eigenvectors (representing mode shapes) for each mode. Statistical properties of these eigenvalues and eigenvectors were then utilized to construct likelihood functions for Bayesian parameter identification, establishing posterior probability distribution functions for modal parameters by incorporating prior knowledge. Additionally, a fast computation scheme has been devised to determine the posterior uncertainties of the modal frequency, structural damping, and total damping of the Vortex-Induced Vibration (VIV) system of a bridge [95]. Various studies have further explored Bayesian frameworks for modal identification using forced vibration data [96], free vibration response of structures [97, 98], and ambient effects [99], each proposing efficient algorithms for modal parameter identification. Furthermore, Zhu et al. [100] introduced a Bayesian frequency domain method for identifying the modal properties of buried modes, offering an efficient algorithm for modal identification in challenging scenarios. Building on this progress, Zhu and Au [101-103] developed a Bayesian frequency domain method for modal identification using asynchronous 'output-only' ambient data, providing a robust approach for identifying global mode shapes while considering data quality and asynchronous nature.

The fast BFFT method has been proven to be effective in various applications. Field applications in different engineering structures such as a coupled floor slab system [104], a primary-secondary structure [105], a super-tall building under strong wind [106], Shanghai Tower [107], a boat-shaped building [108], a 250-m super-tall building situated in Shanghai [109], a high-rise multi-function building with dampers [110], Jiangyin Yangtze River bridge [111], offshore rock lighthouses [112], Xiushan suspension Bridge [113] and free vibration response of stay cables [114] have demonstrated the efficiency of these approaches successfully. The framework has also been successfully employed to determine the optimal values of modal parameters efficiently based on the structural response under earthquake excitations [115]. More importantly, the relationship between Bayesian method and frequentist approach in statistical system identification were also investigated in detail [116].

***(2) Closely-spaced modes***

Considering cases involving multiple (possibly closely-spaced) modes, numerical optimization becomes increasingly complex due to the growth in dimensionality with the number of measured DOFs and the need to repeatedly determine the inverse of ill-conditioned matrices. Consequently, the BFFTA, previously formulated for separate modes, exhibits limitations in practical applications. Au [117, 118] proposed a strategy that represents mode shapes $\boldsymbol{\Phi} \in \mathbb{R}^{n_o \times m}$ through an orthonormal basis $\boldsymbol{\Phi} = \boldsymbol{B\alpha}$, aiming to reduce optimization complexity. This approach involves defining a matrix, denoted as $\boldsymbol{B} \in \mathbb{R}^{n_o \times m'}$, which contains an orthonormal basis in the mode shape subspace with a dimension equal to the rank of $\boldsymbol{\Phi}$, while $\boldsymbol{\alpha}$ represents the coordinates of the mode shape with respect to this basis. By adopting this method, one can separate spectral and spatial effects of modal parameters and link the optimization dimensionality to the orthonormal basis dimensionality, which is typically much smaller, thereby enhancing computational efficiency. Additionally, the iterative determination of the MPV of modal parameters using Newton iteration until convergence is facilitated. The Hessian matrix for the case can also be computed analytically, enabling the derivation of the covariance matrix by inverting the Hessian of the objective function. This process provides an assessment of the uncertainty associated with modal parameters.

A recently introduced strategy for identifying multiple modes involves incorporating the Fisher Information Matrix (FIM), as proposed more recently by Zhu et al. [119]. This approach operates under the asymptotic conditions of extended data and a high signal-to-noise ratio. By capitalizing on the inherent property of the FIM being positive semi-definite, an analytical expression for the gradient of the NLLF for each Newton iteration was derived. This advancement enhances computational efficiency and ensures robust convergence. The entry in the FIM corresponding to two modal parameters was obtained through analytical derivation. Ultimately, the posterior covariance matrix can be acquired by taking the inverse of the FIM.

The Expectation-Maximization (EM) approach, as explored in the study by Li and Au [120], aims to enhance computational efficiency and numerical stability in scenarios involving multiple (closely-spaced) modes. In the BFFT framework, the measurement serves as the observed variable, while the modal response can be viewed as the latent variable. The EM algorithm iteratively computes the MPV by alternating between the Expectation (E) step and the Maximization (M) step, starting from an initial value. During the E step, the algorithm calculates the expected complete-data log-likelihood, while the M step maximizes the complete-data log-likelihood to derive new parameters. These steps are iterated until convergence is reached. Extending its application, the EM algorithm has been adapted to determine the MPV of modal parameters by incorporating data from multiple setups and considering multiple (potentially closely spaced) modes, as discussed by Zhu et al. [121]. Building on this work, Zhu and Au [122] delved into assessing the posterior uncertainty of modal parameters, particularly in relation to their posterior covariance matrix across multiple modes and setups. For the posterior covariance matrix computation, Zhu and Li [123] further explored the capability of the EM algorithm by leveraging the Louis' identity, yielding a more efficient implementation than the conventional analytical differentiation approach.

***(3) Case of multiple setups***

In full-scale operational modal tests covering a number of locations, the DOFs of interest are usually divided into several groups which are measured separately with common 'reference' DOFs present across different setups so that a set of local mode shapes are usually obtained. The difficulty of how to assemble these local mode shapes to form overall mode shapes is an important problem. Let $\varphi_r$ be the $r$-th global mode shape covering all measured DOFs which are required to be identified, while $\varphi_{r,i}$ be the components of $\varphi_r$ confined to the measured DOFs in the $i$-th setup. The local mode shape $\varphi_{r,i}$ can be mathematically related to the global mode shape $\varphi_r$ as $\varphi_{r,i} = \boldsymbol{L}_i \varphi_r$ where $\boldsymbol{L}_i \in \mathbb{R}^{n_i \times n_l}$ is a selection matrix.

An innovative technique known as the 'global least-square' method, introduced by Au [124], offers a novel approach that does not depend on selecting a reference setup. This method aims to determine the optimal global mode shape by minimizing a comprehensive measure-of-fit function. This function considers the differences between the theoretical mode shapes and those identified individually in all setups, appropriately normalized to the same scale. The methodology was extended to enable the assembly of global mode shapes for an eight-story concrete building through the analysis of ambient vibration data collected asynchronously, involving the use of different sensors with potentially varying data sampling clocks, as outlined by Zhu et al. [125]. Notably, this 'global least-square' method relies solely on the information from the most probable mode shapes identified across different setups. Therefore, there is room for refinement by incorporating the uncertainties associated with the posterior mode shapes. By assuming that accurately considering uncertainties minimizes the impact of poorly identified mode shapes in a specific setup on the final assembled mode shape due to their relative unreliability, Yan et al. [93] introduced a theory that involves assembling local mode shapes within a Bayesian statistical framework, enabling automatic consideration of data quality from various clusters. The optimal global mode shape can be efficiently obtained using a rapid iterative approach, with the associated uncertainties analytically derived.

Recognizing the potential variations in modal properties across different setups, such as the PSD of modal force and prediction errors, Au and Zhang [126] developed an advanced Bayesian method to assemble mode shapes, accommodating variations in modal properties across setups, especially for well-separated modes. This method, enhanced for computational efficiency, reconfigures the NLLF using eigenspace representation, allowing for a partial analytical solution of the MPVs and enabling the construction of an iterative scheme for obtaining the complete set of solutions. Building on this, Zhang et al. [127] explored posterior uncertainty of modal parameters by analyzing their covariance matrix, equivalent to the inverse of the Hessian of the NLLF at the most probable values, with derived analytical expressions focusing on addressing computational challenges from the norm constraint of the global mode shape. It is important to note that the Expectation-Maximization algorithm has also been utilized to address the BOMA framework for scenarios involving multiple modes identified using ambient data from various setups, as discussed by Zhu et al. [121] and Zhu and Li [123].

*(4) Uncertainty law and management*

A fundamental uncertainty law made for ambient modal analysis using fast BFFT [128, 129] was proposed recently, which opens a new path to investigate and manage the uncertainty in ambient vibration testing. The prerequisites for deriving approximations of the inverse of the Hessian matrix primarily involve: assuming a small damping ratio $\varsigma_s$ for the structure, considering a small 'noise-to-signal ratio', and ensuring a long data duration, indicating that the 'normalized data length' $N_c$ should meet certain criteria $N_c \gg 1$. The frequency band selected for analysis is $2\kappa\varsigma_s f_s$ (i.e., $f_s(1 \pm \kappa\varsigma_s)$), with $\kappa$ defined as the 'bandwidth factor'. As a result, the number of FFT ordinates contained in the selected frequency band is equal to $N_f = \mathrm{Int}(2\kappa\varsigma_s N_c)$, where $\mathrm{Int}(\cdot)$ denote round number to make sure that $N_f$ be an integral. Approximate closed-form formulas are derived analytically to express the posterior coefficient of variation (c.o.v.) explicitly in terms of different data factors, such as data duration, spectral bandwidth factor as well as the number of data segments:

$$\delta_f^2 \sim \frac{\zeta}{2\pi N_c \boldsymbol{B}_f(\kappa)};\delta_\zeta^2 \sim \frac{1}{2\pi\zeta N_c \boldsymbol{B}_f(\kappa)};\delta_S^2 \sim \frac{1}{N_f \boldsymbol{B}_S(\kappa)};\delta_{\sigma^2}^2 \sim \frac{1}{(n-1)N_f};\delta_\Phi^2 \sim \frac{(n-1)\nu\zeta}{\left[N_c \boldsymbol{B}_\Phi(\kappa)\right]} \tag{20}$$

where $\boldsymbol{B}_f(\kappa)$, $\boldsymbol{B}_\zeta(\kappa)$, $\boldsymbol{B}_S(\kappa)$ and $\boldsymbol{B}_\Phi(\kappa)$ are the factors that depend of bandwidth factor $\kappa$; $N_f = 2\zeta\kappa N_c$ and $N_c = T_d f$, where $T_d$ is the sampling duration; $\nu = S/\sigma^2$ is called the 'noise to environment' ratio. They establish the foundation for strategic planning and standardization of ambient vibration tests, allowing for the quantitative assessment of factors such as channel noise, sensor quantity, and placement, as detailed by Au et al. [130]. Building upon this ingenious idea, Xie et al. [131] developed an uncertainty law for OMA using data from multiple setups, which enables the creation of a comprehensive 'global' mode shape covering multiple locations while employing a limited number of sensors in each setup. Meanwhile, they conducted an analytical exploration of the eigenvalue characteristics of the mode shape covariance matrix, revealing how the primary uncertainty of the global mode shape depends on the arrangement of roving and reference sensors.

Building upon the approximation analysis framework initially introduced by Au [128], Yan and Katafygiotis [132] delved into investigating the analytical propagation within the two-stage BSDA for ambient analysis. The derived squared posterior c.o.v is explicitly expressed in relation to both modal parameters and data parameters. Expanding their research scope, Yan and colleagues [133] extended their inquiry to explore uncertainty propagation in cases involving mode shape assembly. The approximate covariance matrix of the global mode shape is formulated as a function of various factors, including the damping ratio, the 'bandwidth factor', the 'noise to environment' ratio, the 'normalized data length', and a selection matrix representing the sensor configuration. Inspired by Au's work, Ng and colleagues [134] developed uncertainty laws for well-separated modal parameters with a single broadband input, deriving expressions for coefficient of variation using the Fisher Information Matrix. They explored governing factors like signal-to-noise ratio, DOFs, shaker placement, and data duration, confirming their theories through analyses of synthetic and field test data.

More recently, Ma et al. [135] delved deeper into exploring the impact of noise models on OMA conducted using a Bayesian approach within the frequency domain. Apart from analyzing modal identification outcomes, they also evaluated noise models through a Bayesian evidence perspective. They devised algorithms to compute Bayesian statistics efficiently, addressing general noise models not previously studied. They demonstrated a significant advancement in OMA theory by transforming problems with general noise models into those with i.i.d. noise models. This transformation enabled the derivation of formulas to quantify identification uncertainty and introduced a new definition for the modal signal-to-noise ratio, enhancing understanding of modal parameters and noise characteristics in diverse OMA scenarios.

In modal analysis, addressing uncertainty in closely spaced modes poses a significant challenge due to the intricate relationships among mode shapes and parameters. Recent research by Au et al. [136] made a breakthrough by simplifying expressions for identification uncertainty even in scenarios with wide frequency bands. This work led to the development of an 'uncertainty law' tailored for close modes, offering analytical solutions for residual uncertainty in modal parameter extraction. The study also provided asymptotic expressions for posterior coefficients of variation of natural frequencies, damping ratios, and mode shapes for two closely spaced modes. By leveraging extended ambient data, high signal-to-noise ratios, and broad resonance bands, the research enhanced understanding and management of identification uncertainty in field test planning. Through theoretical exploration, verification exercises, and insights into influential factors in modal analysis, Au et al. [137] clarified the notion of mode proximity and its implications. The credibility of their results was bolstered through analyses involving synthetic, laboratory, and field data, strengthening the practicality of the developed methodologies.

### 3.2.2 Structural model updating

The core idea of structural model updating is to minimize the discrepancies in structural behaviors between a finite element (FE) model and its real-world counterpart by modifying the model parameters, which is essentially an optimization problem [138]. Over recent decades, Bayesian approaches have been extensively applied in model updating due to their capability to quantify the uncertainties associated with the model parameters. Bayesian model updating combines prior knowledge and measured data to infer the posterior PDF of these parameters. It effectively addresses the inverse uncertainty quantification problem, circumventing common issues such as gradient computation, ill-conditioning, and non-uniqueness in optimization methods [139]. The main innovations in Bayesian model updating can be divided into four categories: mainstream Bayesian updating, hierarchical Bayesian updating, sparse Bayesian learning, and approximate Bayesian updating. Consequently, these advancements are succinctly reviewed in this subsection.

#### 3.2.2.1 Mainstream Bayesian updating

***(1) Formulation of the likelihood function***

The likelihood function plays a pivotal role in the Bayesian model updating process by quantifying how well the FE model explains the measured data under the prediction error probability model, serving as an essential bridge between the prior and posterior PDFs. In real applications, the construction of likelihood functions must consider the characteristics of the actual structure and the available observation data. The commonly employed data include directly measured structure response (strains, displacements, accelerations, and input excitations) along with derived structural features (modal characteristics and frequency response functions) [11] for the case of linear models of systems. Additionally, sensor configurations can be included to maximize the expected information gain from the measured responses [140].

Among them, modal characteristics are most utilized in Bayesian model updating due to the relative ease in constructing the likelihood function and the more straightforward description of the structure's behavior compared to other available data. To obtain both global and local information about the structure, modal frequencies and mode shapes are generally used in combination with the likelihood function [11, 141]. Beck et al. [142] and Yuen et al. [143] subsequently introduced the concept of system mode shapes and system frequencies as instrumental variables to avoid the need for mode-matching. Despite these advances, these modal-based approaches require the careful assignment of relative weighting factors for frequencies and mode shapes, which is typically determined based on engineering experience. To this end, Christodoulou and Papadimitriou [144] employed a Bayesian formulation to optimally estimate the weights of the modal groups to be inversely proportional to the optimal values of the residuals corresponding to these modal groups, rationally accounting for both measurement and model errors in the weight estimation. Au and Zhang [145, 146] presented a two-stage approach for Bayesian model updating, which first identifies modal parameters and their uncertainties and then transfers the identification results to update the model parameters. The proposed approach weights the modal parameters according to their identification uncertainty, with more uncertain modal parameters assigning smaller weights. Nevertheless, this approach overlooked modeling errors between the two stages, that is, it was assumed that a perfect matching between the identified modal parameters in stage I and those modal parameters predicted by the structural model in stage II. Zhang et al. [147] subsequently employed a Gaussian discrepancy model to account for the mismatch between the two stages. Building on these developments, Zhu et al. [148] further improved computational efficiency in stage II by integrating the FIM and advanced eigenvalue sensitivity. They employed the Fisher scoring method to determine the MPV by substituting the Hessian matrix with the FIM, thereby eliminating repeated computations of second-order derivatives. In addition to modal data, Frequency Response Functions (FRFs) are suitable for model updating as well, which are even preferred over modal data in some applications as they bypass the errors typically associated with modal analysis [149-151].

Another class of Bayesian model updating methods directly employs measured structure responses to construct the likelihood function, avoiding the requirement of modal identification. Typically, the available data includes time histories of structural responses, such as dynamic strains, displacements, and accelerations, and sometimes combined with time histories of input excitations. In recent years, numerous methods have been developed for Bayesian model updating utilizing input-output [27, 152-155] or output-only [156] measured response. However, the use of high sampling frequencies or spatially dense datasets in Bayesian model updating introduces significant correlations in prediction errors. This deviates from the widely adopted assumption that prediction errors are zero-mean Gaussian uncorrelated, potentially leading to inaccurate Bayesian inference results [157-159]. To this end, Simeon et al. [157] investigated the impacts of temporal and spatial correlation in prediction errors on the Bayesian model updating process, illustrating the effectiveness of Bayesian model class selection in accounting for these correlations. Motivated by this advance, Koune et al. [159] employed large datasets to update a twin-girder steel plate bridge, considering spatial and temporal correlations in the model uncertainty.

***(2) Solution strategies***

With the prior knowledge and likelihood function in hand, the posterior PDF of model parameters can be derived via Bayes' theorem. The computation of the resulting posterior PDF typically involves intractable high-dimensional integrals. To overcome this, Beck and Katafygiotis [27] adopted an asymptotic approximation strategy to approximate the posterior distribution analytically. Nevertheless, this approach relies on the assumption that the posterior PDFs are unimodal and Gaussian, which may not be reliable for structural parameters with non-Gaussian distributions. Additionally, limited data and complex model classes can render the updating problem unidentifiable.

MCMC algorithms present a preferable alternative for inferring the posterior PDFs in scenarios involving non-Gaussian distribution, also covering multimodal or unidentifiable issues. The basic MCMC algorithm is the MH algorithm [41, 42]. Au and Beck [44, 160] demonstrated the utility of the MH algorithm in reliability analysis and proposed an adaptive MH algorithm by integrating simulated annealing concepts to handle highly peaked, flat, or multimodal PDFs. Subsequent research has focused on improving the stopping criteria [161], increasing the convergence speed of the adaptive MH algorithm such as Delayed rejection adaptive Metropolis (DRAM) [162], and addressing convergence issues and numerical instabilities arising in cases of high-dimensionality [163].

TMCMC, as an advanced variant of MCMC algorithms proposed by Ching and Chen [45], uses resampling to bypass kernel density estimation in adaptive MH algorithm. A series of studies have targeted reducing bias within this framework. For instance, Betz et al. [164] presented three modifications for TMCMC to mitigate bias in evidence estimation and enhance the convergence of posterior estimates. Wu et al. [165] introduced Bayesian annealed sequential importance sampling (BASIS) to eliminate the bias in the TMCMC resulting from the uneven chain lengths in each subsample, although it diminishes sampling efficiency in peaked PDFs. In addition to bias mitigation, substantial research has explored various alternative MCMC kernels to enhance TMCMC’s efficiency and applicability. Angelikopoulos et al. [166] developed X-TMCMC, which integrates a Metropolis-Adjusted-Langevin (MAL) sampler and an adaptive kriging surrogate model, significantly reducing computational costs by an order of magnitude for unimodal posterior PDFs. However, X-TMCMC's efficiency decreases in high dimensions due to a requirement for gradient estimation. Recently, Lye et al. [167] advanced the field by proposing Transitional Ensemble MCMC (TEMCMC), which employs an Affine-invariant Ensemble sampler (AIES) to sample from badly-scaled and highly-anisotropic distributions as well as an adaptive step-size tuning algorithm to maintain the acceptance rate of the sampler within the optimal bounds of 0.15 to 0.5. Later, Sengupta and Chakraborty [168] presented a modified AIES to estimate the multi-dimensional scaling (MDS) factor, which accounts for the diverse characteristics of modal data. The TMCMC sampling adaptively adopts the MDS-based tuning algorithm's plausibility value to improve its transition levels. Notably, the proposed approach circumvents the need for optimal bounds to moderate the acceptance rate as previously required.

Another class of MCMC algorithm, the HMC, was initially applied in model updating by Cheung and Beck [169, 170] to address higher dimensional challenges. HMC utilizes a molecular dynamics trajectory to circumvent the random-walk behavior of the sample, allowing the Markov chain to search more effectively for the target distribution. Its benefits are particularly notable in scenarios involving highly correlated parameters and high-dimensional problems. In real applications, HMC is mostly used for large-size structures, such as full-scale bridges [171, 172] or aircraft components [172]. Jang and Smyth [171] utilized sensitivity-based cluster analysis along with the HMC method to update a suspension bridge model. Baisthakur et al. [173] proposed an adaptive HMC to update a full-scale truss bridge. Variants like Shadow HMC (SHMC) and Separable Shadow HMC (S2HMC) have further enhanced sampling efficiency. For instance, Boulkaibet et al. [172] developed the S2HMC for updating asymmetric H-shaped beam structures and a simplified aircraft by using the measured natural frequencies. Besides, Wang et al. [174] expanded HMC's utility to Subset Simulation for reliability analysis, termed HMC-SS. This approach was later extended as the Riemannian Manifold HMC-SS by Chen et al. [175]to address the limitations of traditional Monte Carlo methods in solving reliability problems defined in highly-curved non-Gaussian spaces.

Instead of single-chain sampling method, recent advances in multi-chain algorithms, such as DiffeRential Evolution Adaptive Metropolis (DREAM) [176] and parallel-interactive methods [177], have shown promise in handling complex posterior PDFs. Although DREAM requires minimal parameter tuning, its efficiency relies on the number of user-defined parallel chains based on the actual engineering problem, which is potentially limited to large-scale structural applications [178]. To further enhance sampling efficiency and the robustness of uncertainty estimates, another recent direction involves integrating Bayesian quadrature for numerical integration. Song et al. [179] developed an adaptive Bayesian quadrature combined with a Gaussian process surrogate model for sampling-based Bayesian model updating, which avoids the need to compute the log-likelihood and to specify a postulated posterior PDF, thereby enhancing computational efficiency.

Lately, VI has been recognized as a more computationally tractable method for Bayesian inference, approximating a Bayesian posterior distribution with a simpler trial distribution by solving an optimization problem. Ni et al. [180] proposed a VI-based probabilistic model updating framework, where the likelihood is represented by adaptive Gaussian process regression and the posterior PDF is approximated by a Gaussian mixture model. They applied this method to assess the post-earthquake reliability of a two-story, two-bay reinforced concrete structure [181] and further enhanced it by integrating a substructure technique to address high-dimensional challenges in large-scale structures [182]. By combining Bayesian optimization and Bayesian quadrature, Hong et al. [183] developed a novel Bayesian active learning method and integrated it into VI-based Bayesian model updating, which achieves strong global convergence with significantly fewer simulator calls. In addition, scholars also have attempted to explore more possibilities for VI applications. Acerbi [184, 185] developed Variational Bayesian Monte Carlo (VBMC), a method that combines VI with Gaussian-process-based, active-sampling Bayesian quadrature. The proposed method produces both an approximation of the posterior PDF and an approximate lower bound of the model evidence, facilitating model selection. Nevertheless, VBMC struggles to capture multi-modal posteriors as it cannot escape from previously discovered regions of high posterior density. To this end, Igea and Cicirelio [186] improved VBMC by introducing cyclical annealing, improving the algorithm's phase of exploration and finding high-probability areas in the multi-modal posteriors throughout the different cycles. More recently, A novel deep network which comprises an affine-embedded reparameterization subnetwork and a complex system metamodeling subnetwork is proposed which leverage maximum mean discrepancy (MMD) as the similarity metric of variational distribution and actual distribution [187]. Comparative studies have shown that cyclical VBMC outperforms the standard VBMC, monotonic VBMC, and TEMCMC in handling multi-modal posteriors and limited physics-based model runs.

***(3) Model updating of nonlinear systems***

Most studies previously discussed focus on linear structural models; however, numerous nonlinear dynamic behaviors observed in various engineering applications, such as structural hysteresis during earthquakes, dry-friction effects in mechanical systems, and airplane wing flutter, necessitate a different analytical approach. These nonlinearities stem from various sources, including material and geometric nonlinearities, nonlinearities in force and displacement boundary conditions, friction, and fracture [188, 189]. Due to the absence of a generalized mathematical representation for the input-output mapping in nonlinear systems, identifying nonlinear systems poses a considerably more challenging task than that of their linear counterparts [188].

Well-presented contributions to Bayesian model updating of nonlinear systems were made by Beck and Katafygiotis [27], Yuen et al. [190], Conte and Ebrahimian [189, 191-193], who pioneered the use of batch Bayesian estimation in this context. Subsequent advancements include the development of Bayesian updating and model selection methods by Muto and Beck [194], who employed the TMCMC algorithm to estimate hysteretic structural model parameters. Worden and Hensman [195] highlighted the benefits of the Bayesian approach to address nonlinear system identification, applying it to the Bouc-Wen hysteretic system. Green [196] contributed with a Data Annealing-based MCMC algorithm for Bayesian model updating of a nonlinear dynamical system. In a focused study on seismic structural health monitoring, Ceravolo et al. [197] employed a Bayesian uncertainty quantification framework to identify hysteretic parameters of masonry structures with consideration of the model discrepancy.

Ebrahimian et al. presented a Bayesian framework for parameter estimation in structures with material nonlinearities [191], later extending it to joint estimation of parameters and seismic inputs[192]. Ramancha et al. [189] investigated identifiability issues in Bayesian model updating of nonlinear systems. Addressing localized nonlinearities, Giagopoulos et al. [198] employed a Bayesian approach to identify nonlinear components of a small-scale vehicle model. Motivated by the development of nonlinear normal modes (NNMs), Song et al. [199, 200] proposed a two-phase Bayesian model updating methodology for structures with localized stiffness nonlinearities and demonstrated its efficiency on a cantilever beam and a wing engine with nonlinear connections. More recently, Teloli et al. [201] demonstrated the effectiveness of the use of both high-order FRFs and the Bouc-Wen model in addressing structures assembled by bolted joints.

***(4) Model updating with Bayesian Kalman filter***

Kalman filter (KF) is the representative recursive Bayesian inference technique for state estimation, structure parameter estimation, and input excitation back-calculation. Unlike traditional Bayesian model updating, the KF enables near real-time structural model updating by recursively estimating the posterior probability density function from limited measured data. Prominent extensions of the KF, such as the Extended Kalman Filter (EKF) [202, 203], the Unscented Kalman Filter (UKF) [204-207], and the Particle Filter (PF) [204], have attracted considerable attention.

Significant contributions to KF-based model updating have been reported by several research groups: Lombaert et al. [208-210], Chatzi et al. [204, 211, 212], Naets et al. [202, 213], Yuen et al. [214, 215], Conte et al. [205, 216], Moaveni et al. [217, 218], Ebrahimian et al.

[219-222].

For simultaneous state-parameter estimation, this problem was formulated to estimate an augmented state vector that includes both unknown structural states and model parameters. This formulation enables the merging of state updates and parameter updates into a single recursive filtering framework, applicable to both linear and nonlinear filters. Yuen et al. [214, 215] proposed a series of EKF algorithms for online (or real-time) parameter estimation for large-scale structural systems based on the substructural approach. However, EKF is not effective for highly nonlinear structural systems. Using non-collocated heterogeneous data, Chatzi and Smyth [204] compared the performance and efficacy of the UKF and PF applied to parameter estimation of the nonlinear constitutive model. To mitigate computational demands of PF in the high-dimensional state-space model, Eftekhar Azam and Mariani [223] proposed a coupled extended Kalman particle filter (EK-PK) for online joint state-parameter estimation of a nonlinear shear building model suffering strength degradation under external actions.

Nevertheless, structural model updating remains particularly challenging in the absence of input force information, which motivates the development of joint input–state–parameter estimation methods. Naets et al. [213] combined an augmented discrete EKF, which incorporates unknown forces and parameters into the state vector, with a parameter model reduction technique to enable online joint input–state–parameter estimation. Meas et al. [224] further introduced a nonlinear smoothing algorithm for joint input-state-parameter estimation in linear systems, significantly reducing the estimation uncertainty introduced by measurement noise when data originates from sensors that are not collocated with the estimated inputs. On the other hand, Lei and his co-workers [225, 226] developed analytical recursive solutions of the EKF with unknown input for online linear/nonlinear systems identification and the UKF with unknown input for nonlinear systems identification. Unlike the direct extension of UKF in [226], Dertimanis et al. [227] combined the dual Kalman filter (DKF) and UKF to address the joint input-state-parameter estimation of linear systems, first estimating unknown inputs via KF, then updating the augmented state vector via UKF.

Despite these advances, estimating input forces and displacements from noisy acceleration data remains susceptible to low-frequency drifts. To address this issue, scholars employed several strategies: (1) L-curve regularization scheme for calibrating the covariance matrix of the input forces [211], (2) dummy displacement measurement schemes imposing prior information on unmeasured quantities [212, 213], (3) data-fusion strategies combining displacement/strain and acceleration measurements [202, 225], and (4) parameterization of unknown inputs via through time-varying autoregressive models [228], Gaussian process models [229], and modulated colored noise [230].

Despite their extensive application, a persistent challenge in implementing Bayesian Kalman filters is determining the covariance matrices of the process and measurement noise, which are often assumed to be constant and determined through trial and error [231]. Such an approach can lead to suboptimal state estimation, as the actual noise covariance matrices are unknown and generally time-varying. Recent advancements have focused on developing methods to more accurately estimate these matrices, categorized into covariance matching [232, 233], correlation [234, 235], and Bayesian techniques [231, 236-239]. By integrating Bayesian inference with the KF, Yuen et al. [236] estimated the covariance matrices of the process and measurement noise through optimizing their posterior PDF with a half-or-double algorithm;

however, the high computational cost limited this approach to offline applications. Yuen and his colleagues [231, 237, 238] then extended these Bayesian algorithms for online estimation of the noise covariance matrices in the KF, EKF, and UKF. The latter two methods can be applied to non-stationary scenarios as well. On the other hand, Huang et al. [240] assigned conjugate prior distributions to the noise covariance matrices and derived explicit update formulations for the noise parameters. More recently, Teymouri et al. [239] proposed a Bayesian EM methodology for input-state-parameter-noise identification. The proposed method employed the augmented EKF and fixed-point smoother to perform online estimation, subsequently updating the noise covariance matrices via explicit formulation at the end of Bayesian EM iterations.

#### 3.2.2.2 Hierarchical Bayesian updating

The mainstream Bayesian model updating framework overlooks the parameters' uncertainties due to variabilities arising from modeling and measurement errors, environmental and operational conditions, as well as manufacturing and assembling processes [241]. To address this issue, the hierarchical Bayesian modeling framework was developed, offering extra modeling flexibility to account for the variability of inferred model parameters [242]. Behmanesh and Moaveni [243] presented a hierarchical Bayesian framework for a footbridge to account for ambient temperature and excitation amplitude. Song et al. [62, 244] improved this methodology for a 10-story shear building mode, highlighting the impact of significant model error. More recently, they extended this application to a long-span arch bridge, taking into account the influence of temperature and traffic loads on structural properties using weigh-in-motion (WIM) data [245]. Moreover, Nagel and Sudret [246, 247] established a hierarchical

Bayesian updating framework to specialize in noise-free vibration measurements. Applications of hierarchical Bayesian inference have also been explored in calibrating hysteretic reduced order structural models for earthquake engineering based on time history response data [248], in calibrating the uniaxial steel constitutive model with multiple experiments to account for both epistemic uncertainty and aleatoric uncertainty arising from specimen-to-specimen variabilities [249], and in inferring the material properties of multiscale material systems [250]. However, the abovementioned studies predominantly leverage sampling-based techniques within the model updating framework.

To enhance computational efficiency, asymptotic approximation methods have been introduced within the framework for linear model updating based on the time domain [63, 251, 252] and frequency domain [241], as well as nonlinear model updating using the time domain [253]. Jia et al. [254] expanded upon the framework detailed in [241] to handle uncertainty variability stemming from the assembly process in hierarchical structures, utilizing multiple datasets from various hierarchical levels. Addressing the linear increase in hyperparameters with model parameters, Jia et al. [64] proposed a VI scheme within the framework to obtain analytically tractable solutions for the posterior distributions of the hyperparameters and the predictive distribution of model parameters, thereby improving computational efficiency. Also based on the VI scheme, insightful expressions they proposed for the hyper mean and the hyper covariance of a Gaussian PDF, quantifying the uncertainty in the model parameters, as the sum of epistemic/identification uncertainty and the aleatoric uncertainty. More recently, Teymouri et al. [255] improved the weighting scheme of modal residual in [256] and made an application to FINO3 offshore platform. Another noteworthy study by Sedehi et al. [252] introduced a new kernel covariance to promote physical-informed GP models using time-domain vibration data, employing the hierarchical Bayesian modeling to account for the temporal variability.

Moreover, Ping et al. [257] have extended these developments to model Gaussian processes or fields quantified by the improved orthogonal series expansion (iOSE) method, a novel Gaussian process simulation technique replacing the K-L expansion. This advancement not only addresses identification bias and enhances computational efficiency in previously developed methods but also facilitates the identification of non-Gaussian processes. Furthering this innovation, Ping et al. [258] integrated the iOSE and Polynomial Chaos Expansion (PCE) methods within the hierarchical Bayesian modeling framework to address both stationary and non-stationary non-Gaussian processes. More recently, an integrated hierarchical Bayesian framework was developed to concurrently address the dual challenges of quantifying non-stationary uncertainties in predictive errors through adaptive Gaussian process regression and identifying latent physical parameters via systematic model-data fusion [259].

#### 3.2.2.3 Sparse Bayesian updating

SBL, founded on Bayesian and sparse representations, is recognized as a promising framework that incorporates a sparse prior to constrain the range of plausible solutions in Bayesian inference. A key challenge in the SBL framework is that the integral in the evidence in Eq. (17) is analytically intractable since the model output is a nonlinear function of the model parameters. To address this, Hou et al. [260] implemented an iterative EM technique to address nonlinear eigenvalue problems, relying on the assumption that the posterior PDF followed a standard Gaussian distribution for sampling in expectation calculations. Wang et al. [261] applied the Laplace approximation by assuming the nonlinear posterior PDF as Gaussian distribution so that the solution of unknown parameters and hyperparameters can be derived in an analytical form. Wang et al. [262] further applied variational inference and DRAM algorithm to the SBL framework, which not only demonstrated more computationally efficient than the EM technique [260] in comparative studies but also proved adaptability to both standard and nonstandard probability distributions.

Given that the DOFs measured in a structure are typically fewer than those in the actual structural model, the hyperparameter (prior precision) in SBL can exhibit multiple local maxima, leading to non-robust Bayesian inference [263]. To address this, Huang and Beck [66] developed a specific hierarchical SBL framework that introduces system modal parameters as extra variables and imposes sparsity on the change of model parameters. The proposed framework utilized an iterative scheme with a series of coupled linear regression problems to effectively circumvent the direct treatment of nonlinear inverse problems, resulting in decreased identification errors compared to those reported by Yuen et al. [143]. Building upon this hierarchical framework, Huang and Beck [263] employed the full Gibbs sampling to fully characterize the posterior uncertainty of all unknown parameters and hyperparameters, particularly beneficial when the model class is not globally identifiable as the Laplace approximation in the earlier hierarchical SBL methods [60, 66, 264] are not accurate. Expanding further, Li et al. [265] introduced the model reduction technique to alleviate the ill-conditioning phenomenon encountered during the estimation of high-dimensional system mode shapes from far fewer measured DOFs, which improves the applicability of the hierarchical SBL method.

Moreover, several multi-task SBL methods have been proposed to exploit redundancy data between multiple setups. Huang et al. [266] used a multi-task SBL to adaptively borrow the respective strengths of two fractal dimension-based damage indices to acquire a unifying damage identification index. Later, Huang et al.[267] presented a multi-task SBL method by assigning a shared hyper-prior and prediction error precision parameter, which characterizes the common sparseness profile across multiple tasks. This approach was applied to identify structural stiffness losses by exploiting commonality among stiffness reduction models in the temporal domain. To improve the reliability of the identification result, the algorithm marginalized the common prediction error precision parameter instead of merely finding its MAP value. In a novel study by Xue et al. [268], phase information from frequency domain signals was utilized to construct a sparse inverse problem in complex space, culminating in the development of a Multi-task Complex hierarchical SBL model.

Besides these methodological advancements, there is growing interest in incorporating prior physical information and employing other prior distributions within the SBL framework. Huang et al. [240] introduced additional prior physical information on structure to improve identification accuracy. They integrated dual Kalman filters with SBL to simultaneously track time-varying, spatially-sparse stiffness parameters change and inputs. Xie et al. [269] developed a Laplace prior-based SBL approach for modal-based damage identification and uncertainty quantification, where the Laplace-prior can be obtained by further enforcing the Gamma prior on the ARD Gaussian hyperparameters. Due to the Laplace prior involving much fewer hyperparameters, the proposed SBL-Laplace is more efficient and less dataset-demanding than SBL-ARD.

#### 3.2.2.3 Approximate Bayesian updating

ABC offers an alternative to solve the inverse problem by directly evaluating the agreement between simulations and measurements to infer unknown posterior distributions without explicit likelihood functions. ABC's effectiveness depends on the key hyperparameters such as the summary statistic, distance metric, and especially the tolerance level, which trades off posterior approximation accuracy and computational cost.

Over the past decades, numerous strategies have emerged to address this trade-off through the integration of ABC principles with efficient sampling methodologies, including ABC-MCMC [270], ABC-Population Monte Carlo (ABC-PMC) [271], ABC-sequential Monte Carlo (ABC-SMC) [272], and ABC-Subset Simulation (ABC-SubSim) [67] and ABC-Nested Sampling (ABC-NS) [273]. In structural dynamics, Chiachio et al. [67] pioneered the integration of subset simulation techniques within the ABC framework to update model parameters, demonstrating superior efficiency and accuracy compared to other ABC-SMC algorithms. Nevertheless, manual calibration and preliminary trials were required to optimally scale the standard deviation of the PDFs for the evolved values in each Markov chain, thereby increasing certain computational demands. To this end, Vakilzadeh et al. [274] developed a self-regulating ABC-SubSim using the hierarchical state-space model to dynamically adjust the proposal standard deviation within the Modified MH algorithm per simulation level, aiming to enforce the mean acceptance probability of candidate samples close to a desired target value, thereby enabling automatic termination and selection of the tolerance parameters. Notably, this method also introduced three new hyper-parameters into ABC-SubSim, requiring further intervention by the modeler. Barros et al. [275] later proposed an adaptive version of ABC-SubSim for nonlinear structural model updating, which indirectly scales the proposal standard deviation by tuning the conditional probability hyper-parameter. However, this adaptation method demonstrated limited efficiency in avoiding high sampling rejection within small-size subsets. Chiachio et al. [276] further improved the original ABC-SubSim based on the sample's values distribution to automatic scaling for the proposal standard deviation, thereby avoiding manual adjustments. Despite these advancements, obtaining a refined optimal value via ABC-SubSim could be challenging due to the stochastic nature of subset simulation. Feng et al. [277] proposed a hybrid optimization scheme that combines the subset simulation with a faster local optimization technique to obtain a more refined optimal value of the model parameter vector. During the same period, Abdessalem et al. [272] demonstrated that ABC-SMC dealt well with the model selection and parameter estimation for nonlinear dynamical systems in a straightforward way. They later developed a more efficient algorithm ABC-NS, which combines the ABC principles with the technique of ellipsoidal nested sampling for accurate calculation of model evidence [273]. This innovation has prompted a series of follow-up studies about model selection [278].

Recent studies have also explored various stochastic distance metrics within the ABC framework, including Euclidean, Bhattacharyya, Bray-Curtis distances, and 1-Wasserstein distance [279, 280]. Among them, the Euclidean distance is the simplest and most used to quantify the dissimilarity between the distributions of simulated model outputs and measured output [281]. Another stochastic distance metric that has garnered significant attention is the Bhattacharyya distance, recognized for its ability to capture higher-level statistical information. Bi et al. [282] formulated a two-step ABC updating framework that uses the Euclidean distance function to ensure an overlap between the distribution of the simulated output and the measured output and further updated the model with the Bhattacharyya distance function. Notably, the application of Bhattacharyya distance in high-dimensional dynamic responses is limited as the curse of dimensionality. To overcome this, Kitahara et al. [283] combined Bayesian updating with structural reliability and adaptive Kriging modeling to implement a two-step ABC updating for a seismic-isolated bridge. Additionally, the Bray–Curtis distance function, recently introduced into the ABC framework [284], offers a unique advantage. Unlike other metrics, the Bray-Curtis distance is normalized within an interval of [0,1], providing a hard upper bound that ensures its efficacy in quantifying statistical differences across varied applications [280].

### 3.2.3 Model-driven structural damage diagnosis

The performance of engineering structures inevitably deteriorates during the operational stage due to severe loading events or progressive damage from long-term environmental effects. Structural damage identification is typically classified into four levels: detection, localization, quantification, and prognosis of damage [285]. Over the past few decades, various Bayesian damage identification methods have been developed, relying on structural model updating methodologies to find plausible extents of structural damage and their associated probabilities based on measured data. These methods can be classified into two categories: vibration-based and ultrasonic-guided wave methods. Consequently, these advancements are succinctly reviewed in this subsection.

#### 3.2.3.1 Damage diagnosis based on vibration measurement

Vibration-based data are considered highly suitable for damage identification. Such data offer insights into both the global and local behaviors of the structure and can be measured during the operational state [11]. The basic idea of vibration-based damage identification lies in that structural damage may induce changes in dynamic properties.

An earlier contribution involved a probabilistic damage identification method that calculates the probability of damage using identified modal parameter data, thereby realizing damage identification of a 10 DOF shear structure [141]. This method assesses the probability that recent stiffness values are less than a specified fraction of those based on past data, rather than identifying absolute changes in structural stiffness. The approach was refined by incorporating system mode shapes to address issues with noisy and incomplete data [142], though this addition significantly increased the problem's dimensionality in large-scale complex structural models. Motivated by this, Yin et al. [286] incorporated the dynamic model reduction technique into the probabilistic damage identification framework to identify the location and severity of bolt connection damage in a 2D frame structure. To address the challenges of insufficient modal information and insensitive to local damage, Hou et al. [287] offered an alternative approach by proposing a damage identification method that utilized additional virtual masses to enrich modal information.

On the other hand, most studies available require the response to be stationary, rendering difficulties when dealing with non-stationary excitation. Based on the use of a transmissibility matrix and statistical properties of fast Fourier transform coefficients, Yuen et al. [152] proposed a Bayesian frequency-domain substructure identification approach, allowing for the identification of model parameters for some critical substructure subjected to non-stationary excitation. Yan et al. [288] further enhanced this approach's computational efficiency by incorporating the transmissibility matrix and statistical properties of the trace of the Wishart matrix. In addition, Yuen and Huang [215] proposed an improvement to the previous work [152] by modelling boundary forces as filtered white noise, thereby imposing additional constraints and significantly enhancing inverse problem identifiability.

Additionally, the applications of SBL, and ABC in structural damage are also noteworthy. Huang and his coworkers [60, 66, 263, 264] conducted a series of damage identification for the IASC-ASCE Phase II problem in the context of specific hierarchical SBL. The sparseness of the inferred structural stiffness loss is promoted automatically, allowing more robust damage localization with higher resolution. Xia and his colleagues [260-262, 289]integrated techniques such as EM, Laplace approximation and VI into the SBL framework, successfully identifying the damage in a small-scale cantilever beam and frame structure. Ni et al. [290] developed an SBL framework for online damage identification of railway wheel conditions using Fiber Bragg Grating (FBG)-based track-side strain monitoring. Filippitzis et al. [291]employed the SBL framework to identify the location and severity of weld fracture damage in a 15-story steel frame building. Wang et al. [292]improved the SBL framework for fully probabilistic modeling of stationary ambient vibration data and non-stationary decay vibration data, enabling accurate probabilistic predictions for a real bridge across both time and frequency domains. Moreover, Fang et al. [270, 271] combined ABC with MH sampling and the stochastic response surface (SRS) method for rapid probabilistic damage identification in reinforced concrete beams, and later introduced a grey Bayesian inference strategy with ABC-PMC and SRS for multi-damage identification.

#### 3.2.3.2 Damage diagnosis based on ultrasonic guided wave

Ultrasonic Guided Wave (UGW) techniques are increasingly favored in damage identification due to their ability to detect local damage with high sensitivity compared to traditional vibration-based methods. Moreover, UGWs can propagate over long distances and through complex geometries, making them ideal for inspecting structures like plates, beams, pipes, joint connections, wind turbine blades, and aircraft wings. Typically, UGW employs several acoustic features for damage diagnosis, such as reflection/transmission coefficients, phase changes, time of flight (ToF) of the wave packet, and mode conversion [293].

The significant modeling uncertainties in UGW-based propagation distance recognition and damage localization have prompted extensive research from a Bayesian probabilistic perspective. Yan et al. [294] utilized ToF data from scattered Lamb waves for Bayesian inference to identify damage location and wave velocity in plate-like structures. Ng and his coworkers [295] made a significant advancement in the identification of multiple cracks in isotropic beam-like structures. They employed Bayesian model class selection to determine the number of cracks, followed by the identification of crack parameters and associated uncertainties within a Bayesian statistical framework. Chiachío et al. [296] proposed a multilevel Bayesian framework to address these sources of uncertainty in UGW-based damage identification. Further contributions include those by Yang et al. [297], who developed a Bayesian method to quantify crack sizes using in-situ tests of Lamb waves. Their findings demonstrated that phase change and normalized amplitude are effective indicators for crack size quantification in lap-joint components. Abdessalem et al. [298] employed a Bayesian approach to quantify uncertainty in parameter estimates of an ultrasonic inspection system from limited signal measurements and prior information to enhance confidence in the probability of the detection curve. Yan et al. [299, 300] developed a Bayesian inference framework for model updating and damage identification in composite beam structures based on an analytical probabilistic model of the scattering coefficient, in which a combination scheme of Wave and Finite Element (WFE) and surrogate models was utilized to quantify damage interaction. This idea was further extended to the crack identification of railway [301]. Cantero-Chinchilla et al. [302] introduced a Bayesian framework to identify and localize damage in composite beam structures based on a transient wave propagation model. Their work distinguished itself by directly utilizing complete time-domain UGW signals, which minimized additional uncertainties typically introduced by baseline comparisons or further transformations. More recently, Wu et al. [303] proposed a fast Lamb wave-based physics-informed Bayesian damage identification framework in plate-like structures and employed a semi-analytical forward model to perform rapid computations of wave-damage interaction without introducing more epistemic uncertainty.

In addition to these developments, the integration of SBL and ABC into structural damage identification has shown significant promise. For instance, Wu et al. [304] developed a signal-processing method that employs a robust SBL to process noisy UGWs for flaw detection, where the dictionary parameters are chosen with the energy distribution of the signal. Zhang et al. [305] employed a multiple SBL scheme to high-efficiency decompose the residual signals into the sparse matrix of location-based components, followed by damage imaging in composite laminates using the delay-and-sum method with sparse coefficients in the time domain. Zhao et al. [306] proposed a two-stage SBL approach for propagation distance recognition and damage localization in plate-like structures, which significantly reduced the uncertainties about wave packet occurrences and overlapping waveforms, thereby enhancing the reliability of damage localization. Subsequent studies have demonstrated the robustness of the SBL method in solving inverse problems for multi-damage localization with data from a few sensors [307]. Similarly addressing the challenge of minimal sensor actuator setups, Fakih et al. [308] combined the ABC-SubSim method with ANN-based surrogate models for damage identification in welded structures. Additionally, Zeng et al. [293]presented an ABC-SubSim method to characterize damage in pipe-like structures using torsional-guided waves, where a time-domain Spectral Finite Element (SFE) method coupled with a cracked FE model was employed to enhance computational efficiency.

### 3.2.4 Model class selection

In structural dynamics, a class of candidate model needs to be selected for structural analysis and model updating. It is obvious that a more complicated model class often fits the data better than one which has fewer adjustable uncertain parameters, which is likely to lead to over-fitting and thereby causing poor results in future prediction [14]. Therefore, it is necessary to penalize a complicated model in model class selection. Among various methods, the Bayesian approach has been developed by showing that the evidence for each model class provided by the data automatically enforces a quantitative expression of a principle of model parsimony or of Ockham's razor [309].

Use $\mathcal{D}$ to denote the input-output measurements of a physical system. The goal is to use $\mathcal{D}$ to select the most suitable class of models representing the system among $N_M$ prescribed model classes $\mathfrak{M}_1, \mathfrak{M}_2, \cdots, \mathfrak{M}_{N_M}$. Based on the Bayes' theorem, the plausibility of a model class conditional on the data $\mathcal{D}$ can be obtained by [34]:

$$p(\mathfrak{M}_i|\mathcal{D}) = \frac{p(\mathcal{D}|\mathfrak{M}_i)\cdot p(\mathfrak{M}_i)}{p(\mathcal{D})}, \quad i = 1, 2, \ldots, N_M \tag{21}$$

where $p(\mathcal{D}) = \sum_{i=1}^{N_M} p(\mathcal{D}|\mathfrak{M}_i)\cdot p(\mathfrak{M}_i)$ is calculated by the law of total probability and $p(\mathfrak{M}_i)$ expresses the user's judgement on the initial plausibility of the model classes with $\sum_{i=1}^{N_M} p(\mathfrak{M}_i) = 1$. $p(\mathcal{D}|\mathfrak{M}_i)$ denotes the evidence of the model class $\mathfrak{M}_i$ provided by the data $\mathcal{D}$. The most plausible model class is the one which maximizes $p(\mathcal{D}|\mathfrak{M}_i)\cdot p(\mathfrak{M}_i)$.

The evidence for $p(\mathfrak{M}_i)$ provided by the data $\mathcal{D}$ can be calculated by:

$$p(\mathcal{D}|\mathfrak{M}_i) = \int_{\boldsymbol{\Theta}_i} p(\mathcal{D}|\theta_i, \mathfrak{M}_i)\, p(\theta_i|\mathfrak{M}_i)\, d\theta_i, \quad i = 1, 2, \ldots, N_M \tag{22}$$

where $\theta_i$ is the parameter vector in the parameter space $\boldsymbol{\Theta}_i \subset \mathbb{R}^{N_i}$ and it uniquely determines each model in $\mathfrak{M}_i$. The parameter vector $\theta_i$ and the parameter space $\boldsymbol{\Theta}_i$ depend on the model class $\mathfrak{M}_i$. The prior PDF $p(\theta_i|\mathfrak{M}_i)$ is given by the user based on engineering judgement and $p(\mathcal{D}|\theta_i, \mathfrak{M}_i)$ is the likelihood function that represents the contribution of the data.

In globally identifiable cases [27], the posterior PDF, $p(\theta_i|\mathcal{D}, \mathfrak{M}_i)$, of the parameters given a large amount of data may be approximated accurately as Gaussian, hence the evidence, $p(\mathcal{D}|\mathfrak{M}_i)$, can be approximated by asymptotic approximation [30]:

$$p(\mathcal{D}|\mathfrak{M}_i) \approx p(\mathcal{D}|\theta_i^*, \mathfrak{M}_i)\, p(\theta_i^*|\mathfrak{M}_i)(2\pi)^{\frac{N_i}{2}} \left|\boldsymbol{H}_i(\theta_i^*)\right|^{-\frac{1}{2}}, \quad i = 1, 2, \ldots, N_M \tag{23}$$

where $N_i$ means the number of parameters of the model class $\mathfrak{M}_i$. The optimal parameter vector $\theta_i^*$ is the most probable value that maximizes $p(\theta_i|\mathcal{D},\mathfrak{M}_i)$ in domain $\boldsymbol{\Theta}_i$ , and $\boldsymbol{H}_i(\theta_i^*)$ is the Hessian matrix of the objective function $-\ln\left[p(\mathcal{D}|\theta_i,\mathfrak{M}_i)p(\theta_i|\mathfrak{M}_i)\right]$ with respect to $\theta_i$ evaluated at $\theta_i^*$.

In above equation, the likelihood factor $p\left(\mathcal{D}\middle|\theta_i^*,\mathfrak{M}_i\right)$ represents the goodness of data fit, and will have the largest value for the model class $\mathfrak{M}_i$ that gives the smallest least-squares fit to the data. The other factors $p(\theta_i^*|\mathfrak{M}_i)(2\pi)^{\frac{N_i}{2}}\left|\boldsymbol{H}_i(\theta_i^*)\right|^{-\frac{1}{2}}$ is called the Ockham factor which denotes a penalty against parameterization. The logarithm of the Ockham factor can be simplified as:

$$\beta_i = -\frac{1}{2}N_i \ln N + R_i \tag{24}$$

where $N$ is the number of data points; and $R_i$ depends mainly on the prior PDF and is $\mathcal{O}(1)$ for large $N$.

Then, the logarithm of the evidence can be written as bellow:

$$\ln p(\mathcal{D}|\mathfrak{M}_i) = \ln p\left(\mathcal{D}\middle|\theta_i^*,\mathfrak{M}_i\right) - \frac{1}{2}N_i \ln N + R_i \tag{25}$$

In other words, the model evidence prefers a model class giving rise to better fit to the data, while penalizing a model class as the number of parameters increases.

For general cases where the posterior PDF cannot be approximated by Gaussian distribution, the asymptotic expansion is not applicable. In general, the log evidence is equal to the difference of two integrals [310]:

$$\ln p(\mathcal{D}|\mathfrak{M}_i) = \int_{\boldsymbol{\Theta}_i}\left[\ln p(\mathcal{D}|\theta_i,\mathfrak{M}_i)\right]p(\theta_i|\mathcal{D},\mathfrak{M}_i)d\theta_i - \int_{\boldsymbol{\Theta}_i}\left[\ln\frac{p(\theta_i|\mathcal{D},\mathfrak{M}_i)}{p(\theta_i|\mathfrak{M}_i)}\right]p(\theta_i|\mathcal{D},\mathfrak{M}_i)d\theta_i \tag{26}$$

The first integral measures the average log goodness-of-fit of the model class $\mathfrak{M}_i$ and the second denotes the relative entropy between the prior and posterior PDFs. Hence, the log evidence of a model class equals the average log goodness-of-fit, penalized by the information gained on the model parameters.

For locally identifiable case or unidentifiable case [27], the asymptotic expansion is not applicable and the TMCMC algorithm [45] which is developed under the concept of an adaptive MCMC simulation procedure [44] can be used for evidence computation. Cheung and Beck [311] proposed a Bayesian model class selection and averaging method, in which the model evidence is computed using posterior samples obtained from MCMC sampling. Worden et al. [195] applied the Bayesian approach to the model selection for a class of hysteretic systems. DiazDelaO et al. [312] proposed a revised formulation of BUS, enabling subset simulation without knowing the multiplier required by the BUS; and the method was applied to model selection. Abdessalem et al. [313] explored the utilization of ABC algorithm for model selection and parameter estimation in structural dynamics, where different metrics and summary statistics representative of the data were investigated. In the work by Yuen, Kuok, et al. [314], a novel Bayesian framework is proposed for real-time system identification using calibratable model classes. Rather than relying on a predetermined model class that adequately represents the system, this approach begins with simpler model classes and incorporates a self-healing mechanism during the identification process. The framework chooses from various candidate model classes and adaptively reconfigures them to address deficiencies in real-time. Additionally, it enables tracking of the time-varying properties of the system based on these model classes.

### 3.2.5 Reliability updating

Probabilistic structural analysis tools are utilized for computing the structural reliability during the design stage in conventional structural reliability assessment [315, 316]. For this, the uncertainties in the structural model parameters and the spatio-temporal variability of external loads are quantified using probability distributions, stochastic processes and random fields. Monitoring data collected during system operation from installed sensor networks at the healthy condition are used to provide useful information for updating the prior uncertainties in the model parameters assigned at the design phase, and thus updating structural reliability based on data. Moreover, during long-term operation, the condition, and consequently the reliability, of the structure may deteriorate from fatigue or corrosion, or from damage. Monitoring data also offers valuable information for continually tracing structural changes and updating uncertainties in the model parameters due to deterioration or damage. These updated uncertainties along with the expected load uncertainties can be used to update structural reliability.

To compute the structural reliability, the uncertain parameters are separated into two sets. The first set $\theta \in \mathbf{\Theta}$ consists of the parameters inferred from the data, while the second set $\varphi \in \mathbf{\Phi}$ consists of the non-updatable parameters, where $\mathbf{\Theta}$ and $\mathbf{\Phi}$ are regions in multi-dimensional parameter spaces that $\theta$ and $\varphi$ may assume values. Non-updatable parameters may include the large number of Gaussian and/or non-Gaussian uncertain variables arising from stochastic processes and random fields introduced to model the spatio-temporal variability of applied loads. Let the failure domain $\mathcal{F}$ be a subregion in the augmented parameter space $(\mathbf{\Theta}, \mathbf{\Phi})$ where the structure fails when $(\theta, \varphi) \in \mathcal{F}$. The probability of failure, the complement of structural reliability, taking into account structural model and external load uncertainties, is given by the multi-dimensional integral as:

$$p(F|\mathfrak{M}) = \int_{\boldsymbol{\Theta},\boldsymbol{\Phi}} I_{\mathcal{F}}(\theta,\varphi|\mathfrak{M})\, p(\theta|\mathfrak{M})\, p(\varphi|\mathfrak{M})\, d\theta d\varphi \tag{27}$$

where $I_{\mathcal{F}}(\theta,\varphi|\mathfrak{M})$ is the non-smooth indicator function which is one if the parameter values $(\theta,\varphi)\in\mathcal{F}$ and zero otherwise, and $p(\theta|\mathfrak{M})$ and $p(\varphi|\mathfrak{M})$ are the PDF introduced based on engineering experience and judgment to quantify the prior uncertainties in the model and load parameters in the absence of monitoring data. Using the two distinctive sets of parameters, the integral can be alternatively reformulated as a multi-dimensional integral involving the product of the conditional failure probability $p(F|\theta,\mathfrak{M})$ given a value of the updatable model parameter set $\theta$, weighted by the prior probability distribution $p(\theta|\mathfrak{M})$ of the updatable model parameter set $\theta$, as follows [30]:

$$p(F|\mathfrak{M}) = \int_{\boldsymbol{\Theta}} p(F|\theta,\mathfrak{M})\, p(\theta|\mathfrak{M})\, d\theta \tag{28}$$

where $p(\mathcal{F}|\theta,\mathfrak{M})$ is now a smooth function of the model parameters given by:

$$p(F|\theta,\mathfrak{M}) = \int_{\boldsymbol{\Phi}} I_{\mathcal{F}}(\theta,\varphi|\mathfrak{M})\, p(\varphi|\mathfrak{M})\, d\varphi \tag{29}$$

The dimension of the second integral is substantially reduced to the number of updatable parameters which is often significantly smaller than the number of non-updatable parameters. The alternative formulation allows independent developments for the two integrals, the conditional probability of failure integral $p(F|\theta,\mathfrak{M})$ in Eq. (29) accounting mostly for the uncertainties in the applied loads, and the robust probability integral $p(F|\mathfrak{M})$ in Eq. (28) over the updatable parameter set accounting for structural model uncertainties. In particular, it is useful when closed form solutions of $p(F|\theta,\mathfrak{M})$ are available [30].

In the case where dynamic data $\mathcal{D}_N$ are available, the robust failure probability of the system should be updated. The updated PDF $p(\theta|\mathfrak{M},\mathcal{D}_N)$ described earlier replaces the initial PDF in the failure probability integrals Eq. (28) and Eq. (29), and the law of total probability based on Eq. (27) gives [317]:

$$p(F|\mathfrak{M},\mathcal{D}_N)=\int_{\mathbf{\Theta}} p(F|\theta,\mathfrak{M})\,p(\theta|\mathfrak{M},\mathcal{D}_N)\,d\theta \tag{30}$$

where $\mathbf{\Theta}$ is a subregion of $R^m$. Computationally efficient algorithms have been employed to compute either the reliability integral Eq. (28) or the alternative integral Eq. (29) for the cases of prior uncertainties, as well as the probability integral Eq. (27) or the probability integral Eq. (28) by replacing $p(\theta|\mathfrak{M})$ by $p(\theta|\mathfrak{M},\mathcal{D}_N)$ for the case where data are used to infer the model parameter set $\theta$.

Papadimitriou et al. [317] introduced a general framework for a robust measure of structural reliability and a methodology for updating it using dynamic test data, where a Bayesian probabilistic framework for system identification [27] is integrated with probabilistic structural analysis tools. In [317], without dynamic data, the robust failure probability is to be computed where the initial (prior) PDF $p(\theta|\mathfrak{M})$ of the model parameters is involved. It is assumed that the probability of failure of a structure with known model parameters, $p(\mathcal{F}|\theta,\mathfrak{M})$, is available or it can be evaluated using probabilistic structural analysis tools and the prior probability of failure is computed using the probability integral Eq. (28). This integral is difficult to calculate using numerical integration, and an asymptotic result that has proved effective for efficient approximate calculation of reliability integrals may be used [30].

In the case where dynamic data $\mathcal{D}_N$ are available, the updated robust failure probability integral Eq. (30) of the system is evaluated using the updated PDF $p(\theta|\mathfrak{M},\mathcal{D}_N)$ derived using

Bayesian inference. This integral is also difficult to evaluate, unless the number of model parameters $\theta$ is small. Hence, an efficient asymptotic approximation for large $N$ is used, provided that the parameter set is identifiable. In these identifiable cases and for more than one mode of the posterior PDF $p(\theta|\mathfrak{M},\mathcal{D}_N)$, the integral for the robust failure probability may be decomposed into a finite sum of integrals over $K$ disjoint subregions of the parameter space, where each subregion contains only one optimal point. In each subregion, the asymptotic result must be employed. The robust failure probability of the structure can be approximated as a weighted sum of the conditional probabilities of failure corresponding to each optimal value of the parameters. In identifiable cases, searching for optimal parameter points is a nontrivial nonconvex global optimization problem. In unidentifiable cases, an asymptotic approximation to the integral requires locating the representative points on the manifold $S$ containing the maxima of the posterior, which is complex and computationally expensive [317].

Similar methods for approximating the posterior probability of failure were introduced by other researchers. Combining first-order reliability methods with Bayesian mixture model, the work by Ni and Chen [318] estimated the bridge reliability from strain measurement data. A similar asymptotic approximation for the posterior PDF combined with first order reliability method was employed by Guan et al. [319] to provide an approximate estimate of the component reliability. Der Kiureghian [316] combined Bayesian parameter estimation framework with reliability methods to develop expressions for the reliability index that account for model uncertainties in structural reliability updating using data.

Methods based on full sampling the posterior distribution of the model parameters have also been introduced to estimate the reliability integral in Eq. (30) in the space of updatable parameters or to estimate the reliability integral Eq. (27) in the augmented parameter space including updatable and non-updatable parameters. Specifically, Beck and Au [44] proposed a computationally efficient adaptive simulation method to evaluate the reliability integral based on TMCMC [45] that is similar to simulated annealing, addressing both identifiable and non-identifiable cases. Jensen et al. [320] and Hatzidoukas et al. [321] used the subset simulation algorithm [49] to estimate the updated failure probability. To update the reliability of linear systems under stochastic Gaussian loading and output data, Ching and Beck [322] employed an efficient IS technique. Bansal and Cheung [323] combined Gibbs sampling for Bayesian model updating with subset simulation to estimate the probability of failure using data, treating the case of linear systems under stochastic Gaussian loads.

Over the last decade, new developments have emerged in the field of reliability updating. Straub et al. [47, 324] developed the BUS to calculate the posterior failure probability without using posterior samples of uncertain parameters. In terms of rare posterior failure probability with equality observation, Wang et al. [325] proposed a new line sampling-based perspective on reliability updating, where the posterior failure probability can be expressed by a more concise mathematical form and completed in the original random input space without auxiliary variable. Instead of focusing exclusively on the expected value of the posterior failure reliability, Li et al. [326] proposed an efficient framework to fully characterize the statistical properties of the posterior failure probability. Recently, using a two-stage cross entropy-based IS algorithm [327], Kanjilal et al. [328] presented an efficient method for reliability updating.

The aforementioned methods for updating reliability based on data focus mostly on the case of epistemic uncertainties. However, research is needed to handle the aleatoric uncertainties arising from environmental, manufacturing, operational, material and component variabilities as well as modeling errors. Such developments can be based on hierarchical Bayesian frameworks [62, 242] proposed to quantify aleatoric and epistemic uncertainties from data. Up to date, reliability analyses that incorporate these uncertainties are scarce [241]. Methods could exploit the special structure of posterior uncertainties obtained using asymptotic, variational inference and MC simulation techniques within HBM, to explore computational efficient algorithms for handling these uncertainties within the updating reliability analyses. Developments should also incorporate other types of non-updatable uncertainties such as spatio-temporal variability of external loads, as well as take into account the structural health monitoring information for continually updating reliability based on deterioration/damage inferred by monitoring data. The work by Jia et al. [329] is a first attempt along this direction, using sampling algorithms to estimate the updating robust failure probability taking into account, though a hierarchical Bayesian modeling framework, both the aleatoric and epistemic uncertainties.

# 4 Bayesian Inference Schemes for Data-centric Statistical Models and Applications to Structural Dynamics

Apart from constructing a physical law-based model, another major type of structural dynamic analysis approaches is directly establishing a data-centric statistical model based on measured or simulated structural responses and identifying structural parameters or conditions through statistical pattern recognition [1]. These approaches are designed for scenarios where the physical mechanisms are not fully investigated or understood. In the past few decades, the rapid advancements in sensor systems, computing resources, and algorithmic improvements have resulted in an explosive growth in both the quantity and quality of available data, thereby promoting the prevalence of these data-centric statistical model-based approaches. Meanwhile, ML has emerged as a pivotal analytical technique within these methodologies for statistical pattern recognition [1]. Throughout the years, a vast array of ML techniques, including supervised, unsupervised, and reinforcement learning approaches, have been developed and extensively applied to analyze voluminous and intricate datasets in the context structural dynamics. These techniques have made significant inroads in this area and have demonstrated success across various tasks, as extensively reviewed in numerous studies [7, 9, 330-333].

Despite these accomplishments of data-centric statistical modeling and ML in structural dynamics, it is widely acknowledged that ML-based methods, particularly deep learning (DL) methods, occasionally make unexpected, incorrect, but overconfident predictions, especially in a complex real-world environment [52, 334]. Consequently, it is crucial for ML-based methods to recognize the degree of uncertainty in their predictions to prevent overconfident results, particularly in high-stake fields such as structural dynamics, which has motivated the growing popularity and rapid developments in Bayesian inference for data-centric statistical models over time due to their advantages in handling uncertainties. In light of this, this section summarizes the current state of research on Bayesian inference for data-centric statistical models and its application in structural dynamics, with a focus on four prevalent Bayesian ML techniques including Bayesian nonparametric mixture models (BNPMMs), Gaussian processes (GPs), Bayesian dynamic linear models (BDLMs), and Bayesian neural networks (BNNs).

Particular emphasis will be placed on how these techniques estimate the posterior distribution of model parameters and formulate the posterior predictive distribution.

## 4.1 Bayesian Inference Approaches for Data-centric Statistical Models

### 4.1.1 Bayesian nonparametric mixture models

Bayesian nonparametric (BNP) models are Bayesian models defined on an infinite-dimensional parameter space, which allows them to adapt their complexity to the observed data [335, 336]. A key advantage of these models is that the model complexity is part of the posterior distribution determined during data analysis, rather than being prespecified [336]. Among the various BNP models, two prevalent approaches are Gaussian process for classification and regression, in which the correlation structure is refined with growing sample size, and Bayesian nonparametric mixture models (also referred to as Dirichlet process mixture models) for clustering and density estimation, which adapt the number of mixing components to the complexity of the data [335]. These approaches will be further discussed in this section.

Clustering methods are designed to partition observed data into distinct groups based on their inherent similarities, which can be flexibly applied in both unsupervised and semi-supervised settings [337, 338] and have been widely used in structural dynamic analysis [339-343]. However, a major limitation of most clustering approaches is the need to prespecify the number of clusters, which is often challenging in engineering applications and highlights their weakness in handling the uncertainty arising from the choice of model complexity. The Dirichlet process mixture model (DPMM) is developed to address this issue, which is a BNP extension of conventional mixture models [344]. Specifically, the DPMM employs a Dirichlet process prior that encompasses both the component number and the parameters of each component, and the posteriors can then be estimated according to the Bayes' theorem. In short, the Dirichlet process (DP) is a stochastic process in which sample paths are probability measures with probability one. Each sample drawn from a DP can be interpreted as a random distribution whose marginal distributions are Dirichlet distributions [344]. For a random distribution $G$, if for any finite measurable partition $\mathbf{A}=\{A_1, A_2,...,A_r\}$ of the domain $\boldsymbol{\Theta}$, the following equation holds:

$$\left(G(A_1),G(A_2),...,G(A_r)\right)\sim \mathrm{Dir}\left(\alpha G_0(A_1),\alpha G_0(A_2),...,\alpha G_0(A_r)\right) \quad (31)$$

Then $G$ follows a DP, denoted by $G\sim \mathrm{DP}(\alpha,G_0)$, where $G_0$ is a random distribution defined over $\boldsymbol{\Theta}$ known as the base distribution; $\alpha$ is a positive number referred to as the concentration parameter. As $G$ is a distribution itself, a set of i.i.d. samples $\{\theta_1,\theta_2,...,\theta_N\}$, referred to as atoms, can be drawn from $G$. Conditioned on these samples, the DPMM that utilizes DP as a prior can be formulated straightforwardly by treating $\boldsymbol{\theta}_n$ as the parameters of the distribution of the $n$th observation $x_n$, i.e., $x_n\sim F(x_n\,|\,\theta_n)$, while the distinct values $\{\theta_1^*,\theta_2^*,...,\theta_K^*\}$ in these samples intuitively induce a partitioning of $\{\theta_1,\theta_2,...,\theta_N\}$ with each $\theta_k^*$ representing an independent component in the mixture model.

In practice, the generative process of the DPMM is often described using the stick-breaking construction as follows [53, 345]: a unit-length stick is successively broken into infinite segments. At each step, a fraction $v_k$ is drawn from the Beta distribution $\mathrm{Beta}(1,\alpha)$, and the mixing proportion for each components is $\pi_k=v_k\prod_{i=1}^{k-1}(1-v_i)$. This process results in a mixture model with an infinite number of components. Each data point $x_n$ is then drawn from the distribution $F\left(x\,|\,z_n,\theta_{z_n}^*\right)$, where $z_n$ is the latent variable denoting cluster assignment.

Based on the stick-breaking construction, the likelihood function for a set of i.i.d. samples $\mathcal{D}=\{x_1,x_2,...,x_n\}$ can be expressed as [53]:

$$p(\mathcal{D}|\varpi)=\prod_{n=1}^{N}p(x_n|\varpi)=\prod_{n=1}^{N}\prod_{k=1}^{K}\left[F(x_n|\theta_n)\right]^{\mathbf{1}[z_n=k]},\ K\rightarrow\infty \tag{32}$$

where $\mathbf{1}[z_n=k]$ is an indicator function that equals 1 if $z_n=k$ and 0 otherwise; $\varpi=\{v,z,\theta\}$ denotes the parameters of the DPMM. The posterior distribution of $\varpi$ can be inferred accordingly using Bayes' theorem:

$$p(\varpi|\mathcal{D})=\frac{p(\varpi)p(\mathcal{D}|\varpi)}{\int p(\varpi)p(\mathcal{D}|\varpi)d\varpi}=\frac{\prod_{k=1}^{K}p(v_k)p(\theta_k)\prod_{n=1}^{N}p(z_n|v)p(x_n|z_n,\theta)}{p(\mathcal{D})} \tag{33}$$

To estimate the posterior distribution of the DPMM, various approximation methods, including collapsed Gibbs sampling [339, 346], blocked Gibbs sampling [53], truncated VI [53], stochastic VI [51], have been developed over the years, while the posterior predictive distribution can be derived by marginalizing out the model parameters of the DPMM to provide a probabilistic clustering result [339]. This Bayesian nonparametric clustering method explicitly addresses the epistemic uncertainty stemming from model selection and model parameters by incorporating both the number of components and their respective parameters into a nonparametric prior. Meanwhile, aleatoric uncertainty is captured through the soft assignment strategy [43]. Consequently, DPMMs offer a systematic framework for elucidating and quantifying predictive uncertainty in clustering, resulting in greater robustness compared to traditional mixture models. For more details about DP and DPMMs, one can refer to [53, 344]. Despite the advantages of DPMMs in handling uncertainties, their application in structural dynamics is relatively limited, which could be attributed to the complexity associated with estimating their posterior distribution. Current applications of DPMMs in structural dynamics primarily focus on structural damage diagnosis, particularly in addressing the effects of various environmental and operational conditions [339, 346, 347].

### 4.1.2 Gaussian processes

For ML-based methods in structural dynamics, raw response measurements are normally transformed into feature vector representations through a user-specified feature map to train the ML model. However, it can be difficult to appropriately capture the properties of structural responses through human-specified fixed-sized feature vectors sometimes. An alternative approach to address this issue is to use a kernel function, denoted by $\kappa(x,x')=\phi(x)^T\phi(x')$, to operate the data in a high-dimensional feature space implicitly, which is known as *kernel methods* [348]. Among various kernel methods, the GP is popular as a nonparametric Bayesian kernel method with superior uncertainty quantification capability for both classification and regression tasks [335], and it has also been widely applied in structural dynamics for various tasks [349-351].

Specifically, the GP performs Bayesian inference over the mapping functions $f$ from the inputs $x$ to target variables $y$ directly, rather than inferring parameters for these functions [348]. A function $f(x)$ distributed as a GP with the mean function $m(x)$ and covariance function $\kappa(x,x')$ can be denoted by:

$$f(x) \sim \mathrm{GP}\big(m(x), \kappa(x,x')\big) \tag{34}$$

where $m(x)=\mathbb{E}\left[f(x)\right]$ and $\kappa(x,x')=\mathbb{E}\left[\big(f(x)-m(x)\big)\big(f(x')-m(x)\big)^T\right]$ is a positive definite kernel function. For any finite set of data points $X=\{x_1,x_2,...,x_N\}$, this process defines a joint Gaussian distribution $p(f\,|\,\boldsymbol{X})=\mathcal{N}\left(f\,|\,\mu(X),\boldsymbol{\Sigma}(\boldsymbol{X})\right)$ with the mean vector and covariance matrix given by $\mu(\boldsymbol{X})=\{m(x_1),m(x_2),...,m(x_N)\}$ and $\boldsymbol{\Sigma}_{ij}(\boldsymbol{X})=\kappa(x_i,x_j)$, respectively. This formulation explicitly demonstrates the nonparametric nature of GP: each sample of a GP is itself a distribution and every finite collection of these samples follows a multivariate Gaussian distribution [348]. In practice, the GP is normally used as a nonparametric prior of the classification or regression function. Take a regression task for example, suppose the training data is denoted by $\mathcal{D}_{tr} = \{\boldsymbol{X}_{tr}, \boldsymbol{Y}_{tr}\} = \{x_i, y_i\}_{i=1}^{N}$ with the targets governed by some mapping function and some independent Gaussian noise, namely $y = f(x) + \epsilon, \epsilon \sim \mathcal{N}(0, \sigma^2)$, the aim is to predict the target $Y$ given new observations $X$. Based on the definition of GP, it has:

$$\begin{pmatrix} f(\boldsymbol{X}_{tr}) \\ f(\boldsymbol{X}) \end{pmatrix} \sim \mathcal{N}\left( \begin{pmatrix} \mu(\boldsymbol{X}_{tr}) \\ \mu(\boldsymbol{X}) \end{pmatrix}, \begin{pmatrix} \boldsymbol{\Sigma}_{tr} & \boldsymbol{\Sigma}' \\ (\boldsymbol{\Sigma}')^T & \boldsymbol{\Sigma} \end{pmatrix} \right) \tag{35}$$

where $\boldsymbol{\Sigma}_{tr} = \kappa(X_{tr}, X_{tr})$, $\boldsymbol{\Sigma}' = \kappa(X_{tr}, X)$, and $\boldsymbol{\Sigma} = \kappa(X, X)$. For computational simplicity, it is common to assume the mean is zero and use the squared exponential kernel [352]. As a result, Eq. (35) can be rewritten as $\begin{pmatrix} f(\boldsymbol{X}_{tr}) \\ f(\boldsymbol{X}) \end{pmatrix} \sim \mathcal{N}\left( 0, \begin{pmatrix} \boldsymbol{\Sigma}_{tr} & \boldsymbol{\Sigma}' \\ (\boldsymbol{\Sigma}')^T & \boldsymbol{\Sigma} \end{pmatrix} \right)$. Since the model prediction $\boldsymbol{Y}$ can be represented by the sum of $f(\boldsymbol{X})$ and the noise $\epsilon$, the joint distribution of the model predictions and the targets in the training data is also a Gaussian distribution given by $\begin{pmatrix} \boldsymbol{Y}_{tr} \\ \boldsymbol{Y} \end{pmatrix} = \begin{pmatrix} f(\boldsymbol{X}_{tr}) \\ f(\boldsymbol{X}) \end{pmatrix} + \begin{pmatrix} \epsilon \\ \epsilon \end{pmatrix} \sim \mathcal{N}\left( 0, \begin{pmatrix} \boldsymbol{\Sigma}_{tr} + \sigma^2 \boldsymbol{I} & \boldsymbol{\Sigma}' \\ (\boldsymbol{\Sigma}')^T & \boldsymbol{\Sigma} + \sigma^2 \boldsymbol{I} \end{pmatrix} \right)$. Therefore, the predictive distribution of the GP can be analytically derived by marginalizing out the function $f$ [348]:

$$p(\boldsymbol{Y} | \boldsymbol{X}, \mathcal{D}_{tr}) = \mathcal{N}\left( f(\boldsymbol{X}) \,|\, \mu^*, \boldsymbol{\Sigma}^* \right) \tag{36}$$

where $\mu^* = (\boldsymbol{\Sigma}')^T \left[ \boldsymbol{\Sigma}_{tr} + \sigma^2 \boldsymbol{I} \right]^{-1} \boldsymbol{Y}_{tr}$ and $\boldsymbol{\Sigma}^* = (\boldsymbol{\Sigma} + \sigma^2 \boldsymbol{I}) - (\boldsymbol{\Sigma}')^T \left[ \boldsymbol{\Sigma}_{tr} + \sigma^2 \boldsymbol{I} \right]^{-1} \boldsymbol{\Sigma}'$, respectively [348]. Similarly, for classification tasks, the GP can be employed in a similar manner to regression tasks by replacing the multinomial logit function with the multinomial probit function [348].

Compared to other Bayesian methods, a significant advantage of GPs is that their posterior predictive distribution can be analytically derived [353, 354], as the nonparametric nature of GPs enables the marginalization in the Bayes' formula. This characteristic considerably facilitates uncertainty quantification and inference processes with GP-based methods. Additionally, GPs offer high flexibility in modeling data with complex structures and correlations, as their nonparametric property implicitly involves a model selection process. However, GPs often incur significant computational and storage costs owing to their complex kernel matrices [43]. Moreover, the performance of GPs in terms of accuracy and robustness is sensitive to the choice of hyperparameters, such as the noise variance and the kernel parameters. This motivates researchers to treat these hyperparameters as uncertain variables and optimize them through a hierarchical Bayesian framework [348, 355]. However, this approach further increases the computational complexity of GPs. In the field of structural dynamics, GPs and their variants, particularly GPR (also referred to as kriging), have been extensively applied to tasks such as structural damage detection [356], localization [357], classification [350], and quantification [351]. These applications demonstrate the superior performance of GPs in addressing uncertainties and enhancing the robustness of structural dynamic analysis results compared to traditional methods.

### 4.1.3 Bayesian dynamic linear models

Bayesian dynamic linear models are state space models with the states described using uncertain variables following Gaussian distributions, and the state transitions are defined by linear functions. Specifically, the BDLM is comprised of two equations, namely observation equation and system equation [358]. The first one describes the observation $y_t$ at time step $t$, which can be expressed as:

$$y_t = F_t x_t + v_t, \quad x_t \sim \mathcal{N}\left(\mu_t, \mathbf{\Sigma}_t\right), \quad v_t \sim \mathcal{N}\left(0, \mathbf{V}_t\right) \tag{37}$$

where $x_t$ denotes the hidden state described by the following system equation:

$$x_t = G_t x_{t-1} + w_t, \quad w_t \sim \mathcal{N}(0, \boldsymbol{W}_t) \tag{38}$$

The variables $v_t$ and $w_t$ denote the measurement error and the modeling error, respectively, which are assumed to be internally and mutually independent and Gaussian distributed. Consequently, BDLMs address the uncertainty from the hidden states of a system by modeling them as uncertain variables and estimating the posterior distribution of them based on observed data through the Bayes' theorem, i.e., $p(x_t|y_t) \propto p(x_t)\,p(y_t|x_t)$. This strength makes them attractive and widely applied in modeling and forecasting dynamic systems and time series in various fields including structural dynamics [359-364]. However, despite their advantages in handling uncertainty and successful applications in structural dynamics, BDLMs inherently involve a linear assumption of the underlying system and the observation process, which can restrict their applicability to real-world complex structural systems. Additionally, both modeling error and measurement noise are assumed to be Gaussian distributed in BDLMs, which potentially lead to inaccurate state estimates and predictions. These limitations highlight the need for further efforts to improve the performance and robustness of BDLMs.

#### 4.1.4 Bayesian neural networks

In modern ML literature, neural network (NN)-based models, especially deep neural networks (DNNs), dominate both classification and regression tasks due to their superior feature extraction and nonlinear mapping capabilities [365], which have made them highly applicable in various areas including structural dynamics [7, 9, 332]. Despite the success of DL methods, it is widely recognized that DNNs are vulnerable to overconfident predictions in complex, real-world problems with limited training data [52, 334], which has motivated increased efforts to enhance the robustness and uncertainty quantification of DNN-based methods in recent years. Among the various approaches to address uncertainties in DNNs, BNNs have emerged as a universal and prevalent paradigm by integrating Bayesian inference into DNNs to formulate a more systematic framework to capture and quantify uncertainty. Therefore, this section will delve into BNNs, with a focus on different types of BNNs and approaches to inferring their posterior distributions.

#### 4.1.4.1 Fundamentals of Bayesian neural networks

As a Bayesian approach, BNNs treat the parameters of NNs (typically the network weights) as uncertain variables with a specific prior, as illustrated in Fig. 6, which induces a distribution over a parametric set of functions. Subsequently, the posterior distribution of network parameters can be estimated using Bayes' rule. Meanwhile, the posterior predictive distribution of BNNs can be formulated by taking expectations over the posterior of network parameters, which is empirically estimated through Monte Carlo integration in practice:

$$p\left(y_{te} \middle| x_{te}, \mathcal{D}_{tr}\right) = \int p\left(\varpi \middle| \mathcal{D}_{tr}\right) p\left(y_{te} \middle| x_{te}, \varpi\right) d\varpi \approx \frac{1}{M} \sum_{i=1}^{M} p\left(y_{te} \middle| x_{te}, \varpi^{(i)}\right) \tag{39}$$

where $\mathcal{D}_{tr} = \{x_{tr}, y_{tr}\}$ and $\mathcal{D}_{te} = \{x_{te}, y_{te}\}$ denote the training and testing sets, respectively; $\varpi$ is the parameters of the NN; $\varpi^{(i)}$ denotes the $i$-th sample drawn from the posterior distribution $p\left(\varpi \middle| \mathcal{D}_{tr}\right)$. Consequently, BNNs provide a systematic framework for uncertainty quantification in NN predictions, particularly addressing the epistemic uncertainty arising from network parameters, which, in turn, makes them more robust against overfitting and mitigates the risk of overconfident predictions [52].

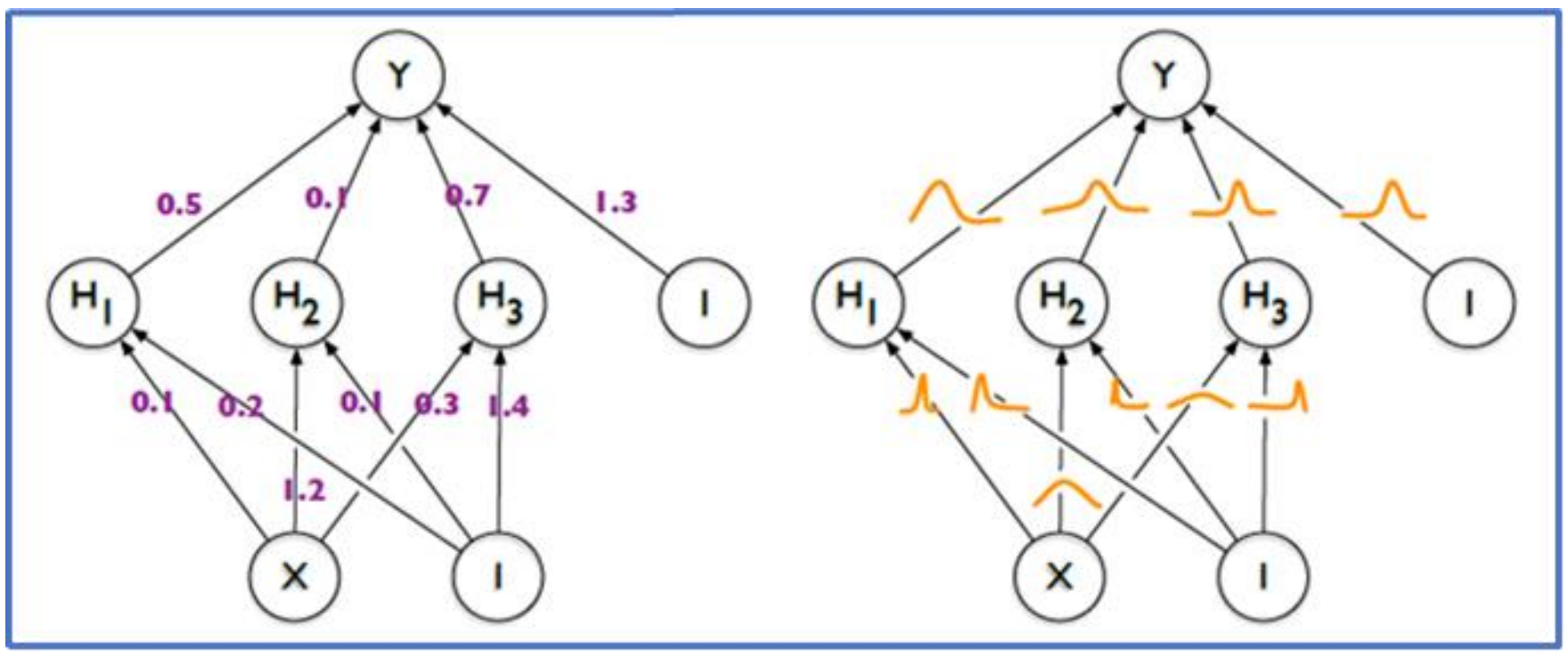

Figure 6. Comparison of BNNs and traditional DNNs. Left: a traditional NN with each weight being a fixed value. Right: a BNN with each weight assigned a probability distribution over possible values (reproduced from [54])

However, despite BNNs combining the advantages of DNNs in nonlinear mapping and Bayesian inference in uncertainty quantification, they typically suffer from computational issues even worse than other Bayesian methods, due to the substantial amount of parameters involved in modern DNNs. Therefore, considerable efforts have been devoted to developing efficient yet accurate approaches to estimate the posterior distributions of the parameters in BNNs over the years, with the most remarkable ones including Laplace approximation [366], Hamiltonian Monte Carlo sampling [367], VI with full covariance matrices [368]. While these methods represent significant initial steps towards practical BNNs, they face challenges due to either prohibitive computational costs or compromised predictive performance resulting from simplifications, which makes it challenging to adapt them to contemporary DNNs with intricate structures and numerous parameters. More recently, a trend has emerged in estimating the posterior of BNNs by integrating Monte Carlo sampling with VI. This integration aims to leverage their unique advantages to provide efficient solutions for estimating the posterior and predictive distributions of complex BNNs without sacrificing model performance, among which the most prominent approaches include Bayes by Backprop [54] and MC dropout [16].

***(1) Bayes by Backprop***

Bayes by Backprop is a principled and backpropagation-compatible algorithm for learning a probability distribution over the parameters of NNs [54], which estimates the posterior distribution through VI assisted by the reparameterization trick. Specifically, the loss function (negative ELBO, also known as the variational free energy) for a BNN with VI can be expressed as:

$$\mathcal{L}(\phi) = -\mathbb{E}_{q_\phi(\varpi)}\left[\log p\left(\mathcal{D}_{tr}\,|\varpi\right)\right] + D_{KL}\left(q_\phi(\varpi)\|p(\varpi)\right) \tag{40}$$

This equation demonstrates that the loss function comprises of a data-dependent part $-\mathbb{E}_{q_\phi(\varpi)}\left[\log p\left(\mathcal{D}_{tr}\,|\varpi\right)\right]$, known as the likelihood loss, and a prior dependent part $D_{KL}\left(q_\phi(\varpi)\|p(\varpi)\right)$, referred to as the complexity loss. Direct Minimization of Eq. (40) is computationally prohibitive, thus in Bayes by Backprop, the loss function is approximated using Monte Carlo integration:

$$\begin{aligned}\mathcal{L}(\phi) &\approx -\sum_{i=1}^{M}\log p\left(\mathcal{D}_{tr}\left|\varpi^{(i)}\right.\right) + D_{KL}\left(q_\phi\left(\varpi^{(i)}\right)\middle\|p\left(\varpi^{(i)}\right)\right)\\ &= \sum_{i=1}^{M}\log q_\phi\left(\varpi^{(i)}\right) - \log p\left(\varpi^{(i)}\right)\log p\left(\mathcal{D}_{tr}\left|\varpi^{(i)}\right.\right)\end{aligned} \tag{41}$$

where $\varpi^{(i)}$ is the $i$-th Monte Carlo sample drawn from the variational posterior $q_\phi\left(\varpi^{(i)}\right)$. Suppose the prior and variational posterior are diagonal Gaussian distributions, a sample $\varpi^{(i)}$ can be obtained using the local reparameterization trick [55] and is given by $\varpi^{(i)} = \mu + \sigma \odot \epsilon^{(i)}$ with $\epsilon^{(i)} \sim \mathcal{N}(0, \boldsymbol{I})$ and $\odot$ denoting element-wise multiplication. Based on this construction, the unbiased gradient estimates of the loss function with respect to the variational parameters $\phi = \{\mu, \sigma\}$ can be analytically derived [54], which enables the model to be trained by the usual backpropagation algorithm. The posterior predictive distribution of the BNN can be formulated by substituting the variational distribution for the true posterior in Eq. (39), which is also applicable to other VI-based BNNs:

$$\begin{aligned} p\left(y_{te} \mid x_{te}, \mathcal{D}_{tr}\right) &= \int p\left(\varpi \mid \mathcal{D}_{tr}\right) p\left(y_{te} \mid x_{te}, \varpi\right) d\varpi \\ &\approx \int q_{\phi}(\varpi) p\left(y_{te} \mid x_{te}, \varpi\right) d\varpi \\ &\approx \frac{1}{M} \sum_{i=1}^{M} p\left(y_{te} \mid x_{te}, \varpi^{(i)}\right) \Big|_{\varpi^{(i)} \sim q_{\phi}(\varpi)} \end{aligned} \tag{42}$$

Furthermore, to enhance the flexibility of Bayes by Backprop beyond the restrictions of Gaussian priors, this algorithm is further refined by employing a scale mixture prior composed of two Gaussian distributions in [54], which improves the performance of Bayes by Backprop-based BNNs. Case studies on some benchmark datasets demonstrate that Bayes by Backprop delivers performance comparable to some state-of-the-art techniques with an improved capability in interpreting and quantifying predictive uncertainty for better generalization [54].

However, despite of the efficient estimation of the Monte Carlo gradients with respect to the variational parameters, Bayes by Backprop can be computationally expensive sometimes, as the Gaussian variational posterior doubles the number of network parameters without significantly increasing the model's capacity [16], which could limit its application to large, complex DNNs. Moreover, using a diagonal Gaussian variational distribution ignores the correlations among the network parameters, which suggests a trade-off between learning capacity and computation efficiency in Bayes by Backprop. To mitigate the computational demands of BNNs, Gal and Ghahramani [369, 370] introduced a more efficient algorithm known as Monte Carlo dropout, whose computational complexity during the training phase is equivalent to that of dropout NNs. Furthermore, this method can be implemented with a Gaussian variational distribution that factorizes the distribution for each row of the network parameters to alleviate the issue of losing correlations.

***(2) Monte Carlo dropout***

The development of Monte Carlo dropout is inspired by theoretical findings that suggest the loss function of dropout NNs is equivalent to that of BNNs with VI when the prior satisfies the KL condition [16]. Specifically, dropout randomly omits each hidden unit of the network with a certain probability (known as dropout rate) for each training point [371], as shown in Fig. 7, which constrains model complexity to alleviate overfitting of NNs. Conditioned on a NN with $M$ layers, the loss function with dropout and L2 regularization can be expressed as:

$$L_{dropout} = \frac{1}{N}\sum_{i=1}^{N} L(y_i, \hat{y}_i) + \lambda \sum_{i=1}^{M}\left(\|\boldsymbol{W}_i\|^2 + \|b_i\|^2\right) \tag{43}$$

where $L(y_i, \hat{y}_i)$ denotes a certain type of loss function; $W_i$ and $b_i$ are the weight matrix and bias vector of the $i$-th layer, respectively; and $\lambda$ is the weight decay. Dropout is applied by sampling binary variables from a Bernoulli distribution with the dropout rate $p_i$ for every training point and for every network unit in each layer (apart from the last one), thus the weight matrix of the NN can be expressed as:

$$\boldsymbol{W}_i = \boldsymbol{M}_i \cdot \mathrm{diag}\left(\left[z_{ij}\right]_{j=1}^{K_i}\right),\ z_{ij} \sim \mathrm{Bernoulli}(p_i) \text{ for } i = 1, \ldots, L,\ j = 1, \ldots, K_{i-1} \tag{44}$$

where $M_i$ denotes the weight matrix without dropout of the $i$-th layer and $K_i$ is the number of neurons in the $i$-th layer. On the other hand, for a BNN with VI, the objective is minimizing the KL divergence between the true posterior over the weights $p(\varpi|\mathcal{D}_{tr})$ and a variational distribution $q_\phi(\varpi)$ with $\varpi = \{\boldsymbol{W}, b\}$, which can be expressed as:

$$L_{VI} = \mathcal{L}(\phi) = -\mathbb{E}_{q_\phi(\varpi)}\left[\log p(\mathcal{D}_{tr}|\varpi)\right] + D_{KL}\left(q_\phi(\varpi)\|p(\varpi)\right) \tag{45}$$

According to Gal and Ghahramani [370], the loss functions shown in Eq. (43) and Eq. (45) are equivalent given appropriate variational parameters and hyperparameters, which demonstrates that the standard training techniques for NNs with dropout can be directly implemented to approximate a BNN with VI, without introducing additional computational burden. The predictive distribution can be estimated with Monte Carlo integration by applying dropout during the testing phase, thus this method is referred to as Monte Carlo dropout. Compared to other methods, Monte Carlo dropout significantly alleviates the computational burden of training and inferring with BNNs, which enables this method to be integrated into large scale DNNs for uncertainty quantification and robustness improvement.

Despite the efficiency and flexibility, it is also acknowledge that Monte Carlo dropout has certain limitations [16]. Firstly, Monte Carlo dropout applies dropout in both training and testing phases, which introduces additional computational costs in the testing phase to empirically estimate the statistics of the predictive distribution. Secondly, the uncertainty estimates of Monte Carlo dropout-based BNNs are not well-calibrated [16, 372], which means that the predictive uncertainty can increase for data points of large magnitude or may vary in scale for different datasets. Thirdly, like most other approaches to training BNNs, Monte Carlo dropout employs a Gaussian or Bernoulli variational distribution for efficiency and scalability. However, this simplification inevitably restricts the capacity of BNNs, as the Gaussian or Bernoulli variational posterior cannot adequately capture the multimodal properties of the true posterior distribution of network parameters. Consequently, in practice, Monte Carlo dropout is sometimes combined with some other programming techniques, such as distributed computing, to train BNNs for a better model performance and uncertainty estimate [16].

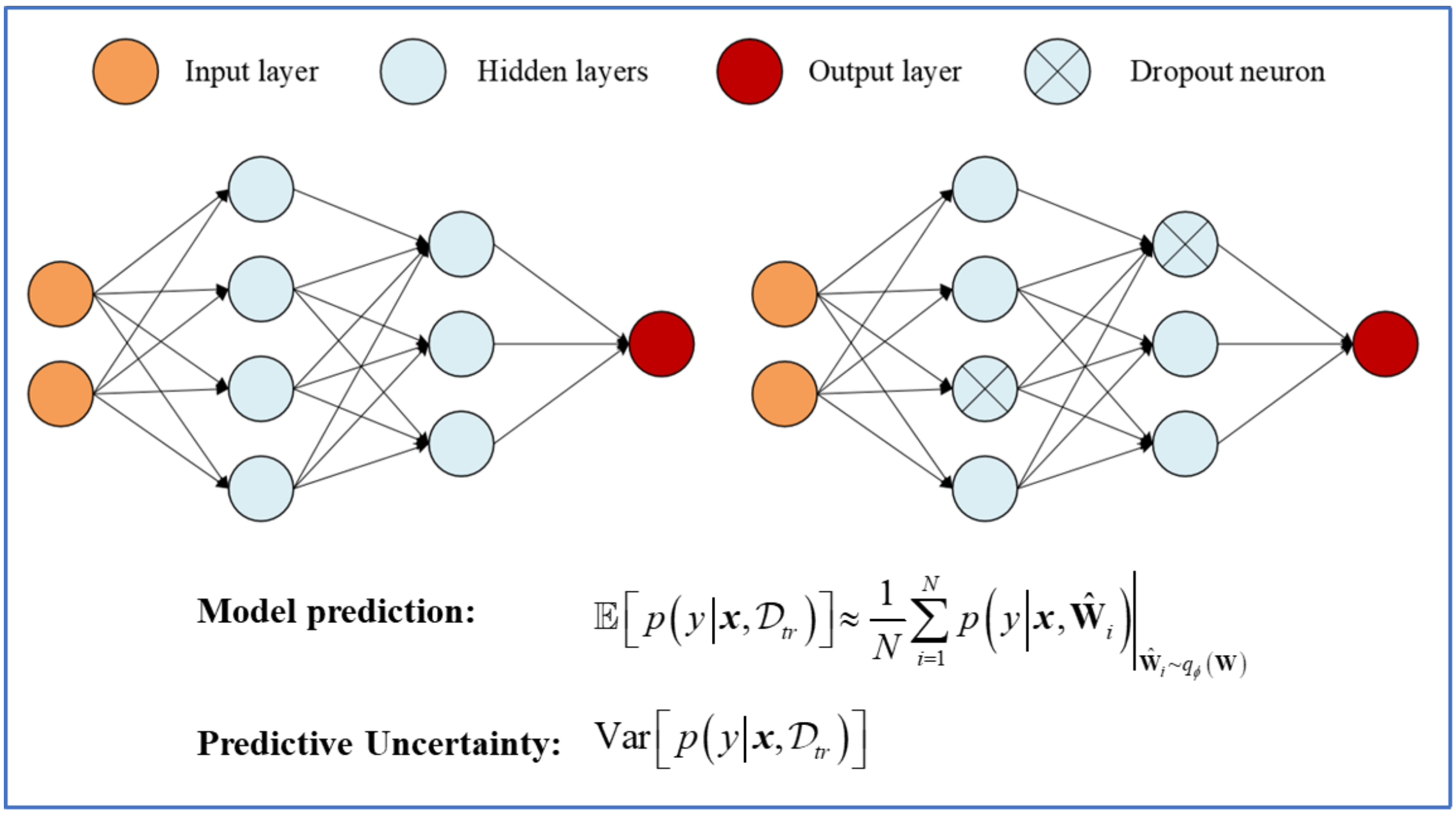

Figure 7. A fully connected neural network (left) and the same network with dropout in a particular training step (right).

#### 4.1.4.2 Fully connected Bayesian neural networks

The fully connected neural network is the most fundamental DNN architecture, which consists of a series of fully connected layers and can be directly converted to a BNN by treating the network parameters as uncertain variables, as shown in Fig. 6. Owing to their implementation convenience and uncertainty quantification capability, fully connected Bayesian neural networks (FCBNNs) have found extensive applications in structural dynamics [373-376]. However, despite of these successful applications, a limitation of FCBNNs is that the fully connected layers could not properly capture the complex temporal and spatial information in the observed data, which motivates the integration of BNNs with more sophisticated NN architectures.

#### 4.1.4.3 Bayesian convolutional neural networks

Convolutional neural networks (CNNs) employ a mathematical operation known as convolution to replace general matrix multiplication in fully connected layers in at least one of the network layers for processing data with a known, grid-like topology [377], which typically incorporate five types of layers: the input layer, convolutional layer(s), pooling layer(s), fully connected layer(s), and the output layer. A convolution layer convolves the input matrix with a convolution kernel at each spatial position, resulting in a series of dot products that preserve spatial information in the input, followed by which a pooling layer simply treats the output of the convolution layer and reduces its dimensionality. By recursively combining convolution and pooling layers, CNNs effectively transform the raw input matrix into a feature map and automatically extract important features [377]. To mitigate overfitting of CNNs when dealing with small datasets, the Bayesian convolutional neural network (BCNN) is developed by integrating the principles of BNNs into the CNN architecture. Traditionally, BCNNs merely utilize probability distributions over the parameters of the fully connected layers for uncertainty quantification, as shown in Fig. 8, which is equivalent to applying a finite deterministic transformation to the raw inputs before feeding them into FCBNNs [16]. This approach, known for its flexibility and ease of implementation, has found applications in structural dynamics for tasks such as seismic response prediction [378] and structural damage classification [379].

However, a significant drawback of this type of BCNN lies in its ignorance of uncertainties associated with the parameters of convolutional layers, especially the kernel weights. Meanwhile, it is crucial to consider the uncertainty in convolutional layers for a more comprehensive estimation of the overall predictive uncertainty of CNNs. To address this issue, Shridhar et al. [380] proposed a variant of Bayes by Backprop that can be implemented on convolutional layers, which use a Gaussian variational posterior with coupled mean and variance to approximate the true posterior of kernel weights. Subsequently, the output of the convolutional layer is formulated using the reparameterization trick. During the training phase, two sequential convolutional operations are conducted to update the mean and variance coefficients of the variational posterior, respectively. This approach provides a fully Bayesian perspective on CNNs within the framework of Bayes by Backprop, enabling more reasonable uncertainty estimates for BCNNs. Alternatively, for Monte Carlo dropout, Gal et al. [16, 369] introduced an approach that assigns a prior distribution over each kernel and approximately integrates each kernels-patch pair with Bernoulli variational distributions, which is equivalent to applying dropout after each convolution layer before pooling. Consequently, this method accounts for uncertainty arising not only from the fully connected layers but also the convolution layers involved in feature extraction. Subsequently, the posterior predictive distribution can be estimated by conducting dropout during the testing phase. This approach offers greater interpretability of uncertainty in CNNs compared to using Bayesian inference solely on fully connected layers. Moreover, it is more computationally efficient than Bayes by Backprop-based BCNNs, resulting in a higher prevalence of its applications in structural dynamics for training the network and estimating the predictive uncertainty [381, 382].

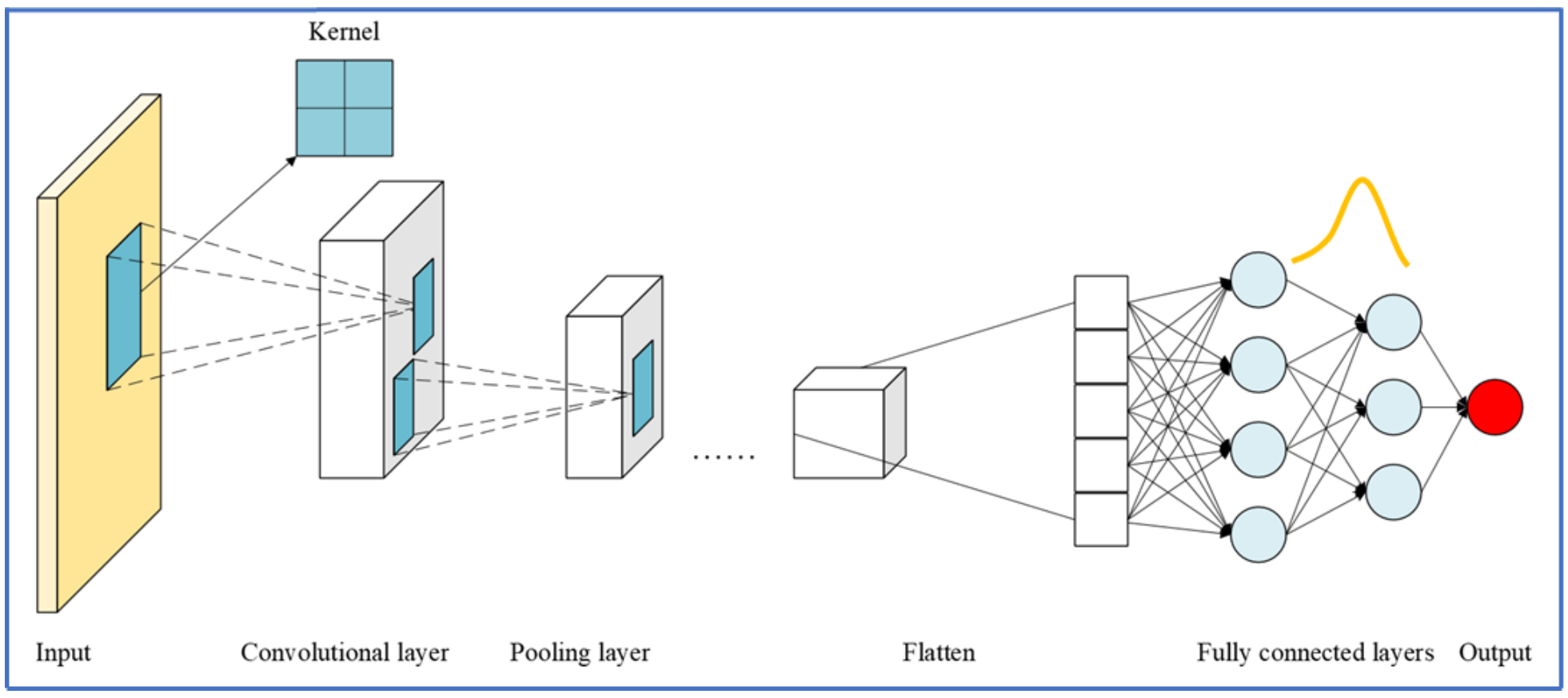

Figure 8. The architecture of a typical Bayesian convolutional neural network.

#### 4.1.4.4 Bayesian recurrent neural networks

Recurrent neural networks (RNNs) are specifically designed for modelling sequential data, such as time series dynamic responses, which possess internal memory to preserve information from previous inputs in the network's internal states, thereby influencing the network output [377], as depicted in Fig. 9. To implement Bayes by Backprop on RNNs and their variants, Fortunato et al. [383] introduced novel framework that truncates the sequences in each minibatch of training data, with the initial state of a sequence set to be the last state of its ancestral sequence. Subsequently, standard Bayes by Backprop can be adopted to train the RNN by sampling network parameters for each minibatch of training data, while the parameters of the variational posterior remain fixed throughout each sequence within the current minibatch. Due to the flexibility and stability, the Bayesian recurrent neural network (BRNN) with Bayes by Backprop has been modified and applied in structural dynamics for tasks such as dynamic response analysis [384] and residual useful life (RUL) prediction [385].

In terms of applying Monte Carlo dropout in RNNs, Gal [16] introduced a new dropout variant that randomly masks (zeros) rows of each weight matrix through all time steps, which is equivalent to dropping the same network units and randomly discarding inputs, outputs, and recurrent connections in an RNN at each time step. This method can also be generalized to variants of RNN, such as gated recurrent units (GRUs) and long-short term memories (LSTMs), by using different dropout masks for different gates [16]. This type of BRNN provides computational efficiency and has been applied in structural dynamics for dynamic response analysis [386].

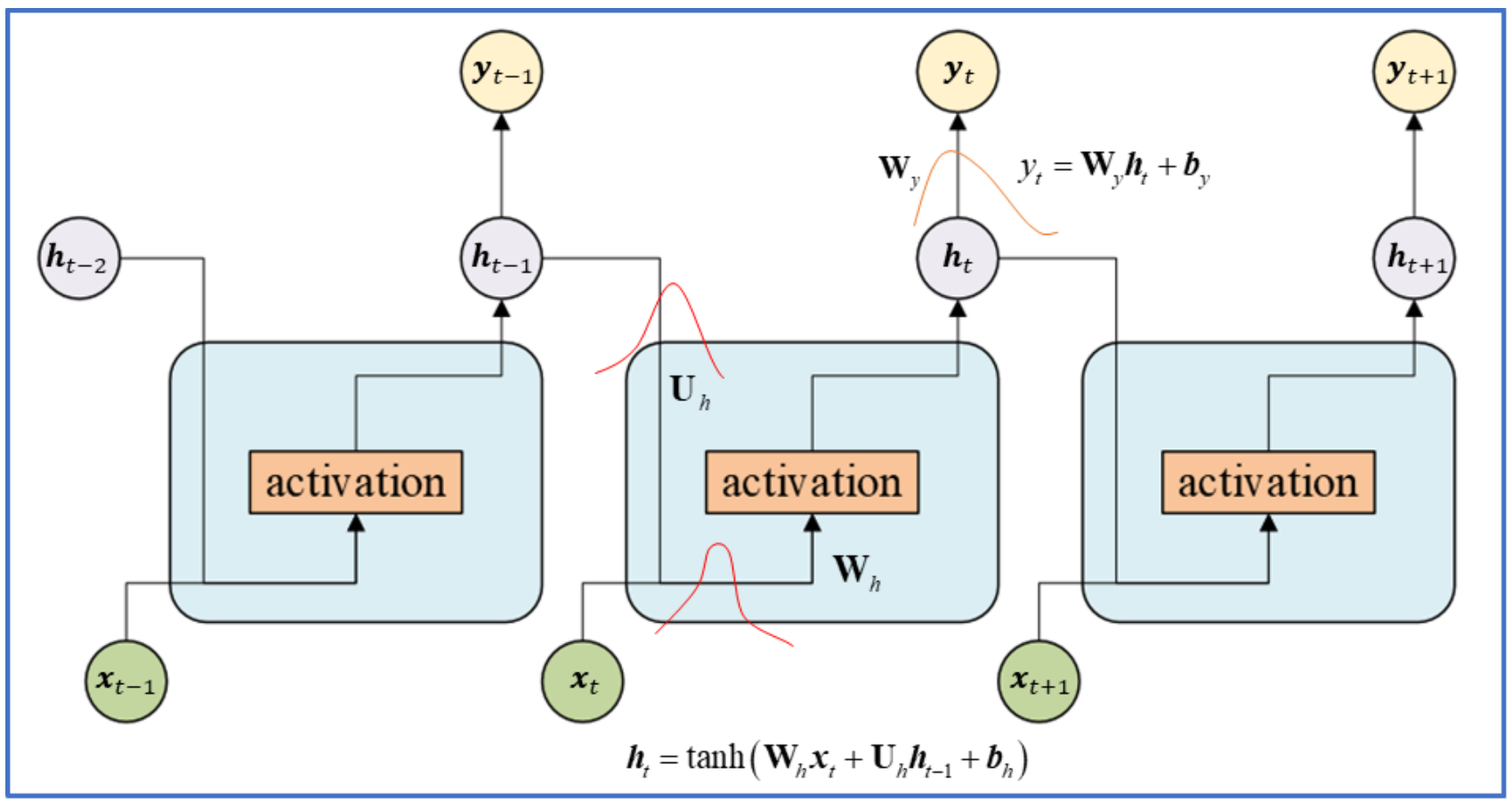

Figure 9. The architecture of the BRNN.

## 4.2 Applications to Structural Dynamics

Section 4.1 provides a concise summary of four Bayesian approaches for data-centric statistical models commonly used in structural dynamics. This section will delve into their detailed applications, focusing primarily on four key areas: structural damage diagnosis, model updating, response prediction, and reliability analysis. A summary of these applications is provided in Table 1.

Table 1. Selection of literature on the applications of Bayesian inference for data-centric statistical models in structural dynamics.

| Approaches | Applications | References |
|---|---|---|
| DPMM | damage diagnosis | [339, 346, 387] |
| | reliability analysis | [388] |
| GP | damage diagnosis | [389] |
| | model updating | [139] |
| | response prediction | [355, 390] |
| | reliability analysis | [391, 392] |
| BDLM | damage diagnosis | [290, 393] |
| | model updating | [394, 395] |
| | response prediction | [359, 360, 362-364] |
| | reliability analysis | [361] |
| BNN | damage diagnosis | [374, 375, 379, 381, 382, 385, 396-399] |
| | model updating | [394, 395] |
| | response prediction | [373, 378, 384, 386, 400-403] |
| | reliability analysis | [376, 404-406] |

### 4.2.1 Data-driven structural damage diagnosis

In contrast to model-driven damage diagnosis, data-driven approaches directly construct data-centric statistical models based on the training data and utilize pattern recognition methods to assign measured data from the monitoring phase to relevant diagnostic class labels [1]. Given that structural damage diagnosis is inherently hierarchical [285], the applications of Bayesian approaches for data-centric statistical models in this area can also be organized according to this hierarchy. The DPMM is typically used for fundamental tasks, predominately damage detection and sometimes damage localization, due to its unsupervised nature. This characteristic enhances its applicability as it does not rely on labeled training data. However, it has limited capacity for more in-depth damage diagnosis, such as damage classification and quantification. In contrast, GPs and BNNs can achieve higher levels of damage diagnosis tasks by utilizing the information of labeled training data, but their application is often constrained by the scarcity or even inaccessibility of well-annotated training data from various damage scenarios. Additionally, it is worth mentioning that, unlike the other three approaches, BDLMs are rarely applied to damage diagnosis as they are originally designed for modeling time series and thus are mainly used for response prediction in the context of structural dynamics [359, 362, 363]. However, a simplified form of BDLMs, which is referred to as Bayesian linear regression models (BLRMs) and assumes time-invariant regression coefficients, has found applications in structural damage detection due to its relative simplicity and interpretability [407, 408].

In terms of BNP mixture models, a pioneering work in this field was proposed by Rogers et al. [339], where a DPMM clustering-based method was introduced for damage detection and semi-supervised damage diagnosis under the influence of EOVs. This method exhibits superior performance compared to traditional Mahalanobis squared distance (MSD)-based method due to effectively addressing the epistemic uncertainties arising from the structure and parameters of the data-centric statistical model used to represent the normal operating conditions. Additionally, there are also studies utilizing DPMMs as density estimators to model the normal condition distribution of the structural features for damage detection, aiming at addressing the effects of various environmental and operational conditions [346, 347, 409]. However, existing works rely on MCMC sampling-based methods to approximate the posterior distribution, which can result in unaffordable computational costs and difficulties in convergence diagnosis, especially in large-scale damage diagnosis problems with high-dimensional structural features and tremendous response measurements [53]. To address this issue, Mei et al. [387] proposed a streaming VI-enhanced DPMM for online structural damage detection, which combines the uncertainty handling capability of DPMM with the efficiency of VI. In addition to DPMM-based methods, Ni et al. [407, 408] proposed a series of structural damage detection approaches based on BLRMs and physically interpretable anomaly indices. These methods have been further enhanced through the use of kernel functions to capture nonlinear relationships and sparse Bayesian learning techniques to improve model efficiency and robustness [290], as well as by replacing BLRMs with BDLMs to account for changes in regression coefficients for real-time damage assessment and uncertainty quantification [393].

Compared to DPMMs and BDLMs, GPs are more commonly used for a wider range of damage diagnosis tasks due to their supervised nature, which enables them to address a greater variety of applications. Liu et al. [350] proposed a crack detection method for composite sandwich structures with probabilistic interpretation of diagnostics based on GP classification, which exhibits good performance in terms of crack detection accuracy and quantifies the uncertainties through the predictive variance efficiently, illustrating the feasibility and advantages of GPs in damage diagnosis. Teimouri et al. [357] proposed a GP-based framework for damage localization and quantification of multi-layer composite airfoil structures, which exhibits better robustness against data noise than NN-based methods with a significantly reduced the cost of data preprocessing, owing to the kernelization of GP. In addition, this work also evaluates the effect of different covariance functions on the predictive performance of the GP-based framework, which provides further insights towards hyperparameter selection when using GPs for structural damage diagnosis. Considering the generalization capability of damage diagnosis methods, Marcus et al. [351] introduced a damage quantification method by integrating transfer component analysis with a GP regression model, which exhibits satisfying damage detection and quantification accuracy and robustness when transferring from a numerical model with labels in damaged states to an experimental test without labels for label consistency cases. Owing to their flexibility, efficiency, and uncertainty quantification capability resulting from the nonparametric nature, GPs have been widely and successfully applied in structural damage diagnosis for complex tasks. However, improving computational efficiency and optimizing hyperparameter selection in GP-based methods remains an ongoing area of research to support their broader adoption in structural dynamics.

BNNs, though introduced later than the other three approaches, have gained increasing attention in recent years. They combine the feature extraction and nonlinear mapping strengths of traditional NNs while mitigating overfitting issues in scenarios with limited training data, thus are well-suited for damage diagnosis. In terms of FCBNN, a pioneering work was proposed by Hoskere et al. [374], where they utilized a FCBNN with Monte Carlo dropout to predict the depth of bearing gaps based on strain measurements for damage quantification. A case study on the miter gates at the Greenup locks and dam demonstrates that this method learns a robust mapping from the observable quantity to the damage extent with explicit uncertainty quantification, which helps effective decision-making for the maintenance of lock gates. Considering the deterioration of wooden railway ties, He et al. [376] proposed a deterioration interval estimation framework based on the FCBNN with Monte Carlo dropout, which utilizes the FCBNN to predict the deterioration rate given weather and historical detection datasets. Subsequently, the cover width deviation and error are introduced as a metric to evaluate the prediction interval of deterioration rate. Numerical experiments demonstrate that the BNN-based framework outperforms some traditional methods in terms of both accuracy and robustness, which facilitates a more informed maintenance procedure of railway ties. For the application of BCNNs in damage diagnosis, Sajedi and Liang [379, 382] conducted a series of studies focusing on damage detection, localization, and classification, with predictive uncertainty quantified through variations in softmax probabilities and the entropy of dropout model predictions. Case studies involving crack detection, local damage identification, and bridge component damage diagnosis demonstrated that BCNNs produce reliable results with explicit uncertainty quantification, enabling more informed decision-making. For BRNNs, it is less widely applied to damage diagnosis. A pioneering work was proposed by Caceres et al. [385], where a BRNN trained using Bayes by Backprop with the Flipout method was used for RUL prognostics. This method outperforms some deterministic DL models as well as the MC dropout-based BRNN, especially when dealing with more complex scenarios involving multiple operating conditions and fault modes, due to considering both epistemic and aleatoric uncertainties.

### 4.2.2 Model updating with data-centric statistical models

Bayesian statistical models have found extensive applications in structural engineering, broadly divided into two main categories. The first category uses Bayesian models, including Gaussian Process Regression (GPR) and BNN, as efficient alternatives to costly physical forward solvers. While these models serve a similar purpose to other surrogate models, they offer distinct advantages, such as reducing overfitting and quantifying output uncertainties. The second category utilizes generative models for structural parameter inference, where the posterior distributions of unknown variables can be approximated by expressive generative models.

In the early works of Khodaparast et al. [410], Erdogan et al. [411], and Wan and Ren [139], the GP model is utilized as a surrogate model representing the relationship between structural parameters and responses. Wan and Ren [412] proposed a residual-based GP framework for finite element model updating, where the GP is adopted to characterize the relationship between the residual and the structural parameters. Zhou and Tang [413] adopted the multi-response Gaussian process strategy which can emulate frequency responses at multiple sensor locations, avoiding training multiple single-response GPs, and enabling them to capture inherent correlation of frequency responses at different locations. He et al. [414] utilized the Gaussian process model to construct the mapping between the selected parameters and fusion features in model updating and an adaptive sampling strategy was developed to reduce the computation cost. Yoshidaa et al. [415] proposed a Bayesian updating method using GPR surrogate model with multiple random fields, and a new 'learning function' is developed to find the location of large values of the posterior PDF and avoids those that have been visited. In addition to GP, BNN has also been adopted for surrogate modelling as it can alleviate overfitting with limited data and the learning process can be more robust when presented within the Bayesian framework. Yin and Zhu [394] proposed a tailor-made algorithm for architecture selection of Bayesian neural network in model updating, where the hidden neuron number and custom transfer functions in both hidden and output layers are simultaneously considered. Zhang et al. [395] took a Bayesian neural network as the surrogate model for probabilistic model updating. The modal data are inputs and structural parameters to be updated are outputs, enabling uncertainty quantification of the structural parameters.

The second category of applications is founded on generative models. Zeng et al. [416] investigated a likelihood-free Bayesian inference method, named BayesFlow, for probabilistic damage inference through model updating, where the BayesFlow approximates the posterior distributions of structural parameters by jointly training a conditional invertible neutral network and a summary network. Li et al. [417] applied the variational autoencoder to Bayesian model updating, and replaced the original decoder with a polynomial chaos expansion surrogate model to reduce the computational cost involved in repeated model evaluations. Itoi et al. [418] utilized the surrogate unimodal encoders of a multimodal variational autoencoder for Bayesian structural model updating, facilitating an approximation of the likelihood when tackling a small number of observations. Lee et al. [419] proposed a novel latent space-based method for stochastic model updating that leverages limited data to effectively quantify uncertainties in engineering applications, which relies on an amortized variational autoencoder, whereby circumventing probability estimations at each iteration of MCMC.

### 4.2.3 Response prediction with data-centric statistical models

Predicting the response of a dynamical system given a specific input is one of the most intuitive applications of data-centric statistical modeling in structural dynamics. However, this task intrinsically involves significant uncertainties from various sources, such as measurement noise, the random nature of stochastic vibration, the high nonlinearity of complex dynamical systems, and environmental variabilities, making Bayesian approaches promising in this area for more accurate and reliable results [2]. Due to their capabilities in state-space modeling and uncertainty treatment, BDLMs are widely used in response prediction as surrogate models to represent real-world physical systems. A first attempt to employ BDLMs for response prediction was presented by Solhjell [420], where several BDLM-based methods were applied to predict the strain of a bridge, demonstrating the feasibility of BDLMs in response prediction. Zhang and his colleagues [359, 360, 421] proposed a series of works utilizing BDLMs to predict dynamic responses such as displacement and strain, illustrating that BDLMs exhibit better forecasting accuracy and robustness compared to some deterministic approaches. However, despite these achievements, BDLMs inevitably face challenges when dealing with real-world structural systems with high nonlinearity due to their inherent linear assumptions. This limitation has led to increased attention on GPs and BNNs for response prediction, owing to their superior capabilities in nonlinear modeling and uncertainty treatment. Considering the advantages of GPs in predicting nonlinear time series and quantifying uncertainty, Zhang et al. [422] introduced a dynamic GPR model to predict the international roughness index for flexible pavements, which outperformed traditional BDLMs in accuracy and uncertainty estimates. Wang et al. [423] proposed a GP-based method to predict the responses of nonlinear structural dynamic models by utilizing a constrained GP regression to represent empirically inferred state-dependent parameters. Numerical simulations emphasize the significant role of prior knowledge in enhancing predictive performance. Leveraging the modeling capability of GPs, Wan and Ni [424, 425] proposed a series of GP-based response prediction and reconstruction methods, which achieve accurate and robust results with improved efficiency compared to traditional methods. In terms of BNN-based response prediction, Wang et al. [378] proposed a BCNN framework for seismic response prediction of three-dimensional structures. This framework treats the weights of fully connected layers in CNNs as uncertain variables and employs Bayes by Backprop for training. A numerical simulation illustrated that the BCNN effectively captures the randomness of structural responses and is robust against input noise. Kim et al. [400] proposed a BCNN framework with probabilistic fully connected layers trained through MC dropout for seismic response prediction, where a new loss function was designed to account for data uncertainty. Additionally, Li et al. [384] developed a BNN framework for response prediction that combines CNN and LSTM architectures. This framework was trained using Bayes by Backprop and treated selected network parameters as uncertain variables to estimate the uncertainty in vehicle-induced bridge response. These works highlight the effectiveness and potential of BNNs in delivering robust response predictions and handling uncertainties arising from factors such as training dataset size variability, vehicle speed variance, and measurement noise.

### 4.2.4 Reliability analysis with data-centric statistical models

In structural reliability analysis, the Gaussian process models are mainly utilized as surrogate models of physical solvers. In order to reduce the computational cost involved in the large number of evaluations of the performance functions, Su et al. [426] applied the GP model to reliability analysis such that the implicit performance and its derivatives can be approximated by the trained GP using explicit formulations. Su et al. [427] proposed a dynamic Gaussian process regression surrogate model based on Monte Carlo Simulation for reliability analysis of complex structures, and the most probable point was quickly predicted using Monte

Carlo sample technique. Zhou and Peng [428] combined kernel principal component analysis (KPCA)-based nonlinear dimension reduction and the GPR surrogate model, leading to an optimal KPCA-based subspace. Then, the KPCA-GPR model is combined with the active learning-based sampling strategy and the Monte Carlo simulation for reliability analysis. Qian et al. [391] employed a multiple response GP model to construct the surrogate model of a multi-output structural system, where the correlation among different failure modes is characterized. Li et al. [429] proposed a novel system reliability analysis method based on a hybrid of multivariate GP and univariate GP models where multivariate GP models are constructed over the groups of highly interdependent components and univariate GP models are built over the components which are relatively independent of one another. Saida and Nishio [392] developed a transfer learning GPR surrogate model to reduce the computational cost of structural reliability analysis with the assumption that the input-output relationship of the source analysis is similar to that of the target analysis. Espoeys et al. [430] reviewed and compared the reliability analysis techniques based on multifidelity GPs, and evaluated these techniques from two aspects including the construction of the multifidelity surrogate models and the enrichment criterion. In the work of Menz et al. [431], the sensitivity of the failure probability estimator to uncertainties generated by the Gaussian process and the sampling strategy was analyzed, enabling the control of the whole error associated to the failure probability estimate.

## 5 Challenges and Future Outlook

Despite the exhaustive efforts and impressive achievements analyzed in previous sections, the development of Bayesian approaches in structural dynamics remains an active research topic with some open issues and research gaps to be addressed in the future. This section outlines several challenges related to the desirable properties that practical Bayesian approaches for SI and SHM should bear in mind, along with some potential directions for future work.

## 5.1 Enhancing efficiency of Bayesian approaches in structural dynamics

Bayesian inference approaches usually require a large number of numerical simulations. In each simulation, the model response should be numerically evaluated by resorting to FE packages such as ANSYS and ABAQUS. In all iterative parameter updating methods, each iteration requires an FE analysis for the given set of updated parameters. If the structure of interest is composed of a large number of FEs (which is generally the case), the large number of computations involved in repeated FE runs can rule out many approaches due to the expense of carrying out an exhaustive number of runs. To mitigate these challenges, the following approaches are promising to address the computational issue:

- **Surrogate model:** Complex numerical simulations for predicting are replaced by cheap and fast surrogate models [432] built using an experiment design strategy to overcome the computational challenges of repeated likelihood evaluations and the difficulty of interfacing different software environments. Metamodeling in computational mechanics faces significant challenges in balancing accuracy with computational efficiency, particularly when handling high-dimensional input spaces and complex nonlinear behaviors. The "curse of dimensionality" exponentially increases the data requirements for constructing reliable surrogates, while dynamic systems (e.g., fatigue crack propagation) demand adaptive frameworks to maintain predictive validity over time.

Additionally, disentangling surrogate approximation errors from inherent physical model uncertainties remains a critical barrier to robust uncertainty quantification, often leading to biased Bayesian updates or reliability assessments.

- **Reduced order model**: Reduced order models are simplified representations of complex, high-fidelity FE models, which are designed to reduce computational costs in numerical simulations of structural models with a large number of DOFs while preserving essential system behavior and accuracy. In structural dynamics, commonly applied reduced order model techniques include component mode synthesis [433], adaptive meta-modeling [434], and substructure-based techniques [435]. However, reduced-order models struggle to preserve essential physics during dimensionality reduction, especially in multi-physics or geometrically nonlinear scenarios, where oversimplification risks losing localized critical behaviors. While techniques like component mode synthesis improve efficiency, their limited generalization across varying boundary conditions or parametric regimes undermines practical utility. Shared challenges include the lack of standardized validation metrics for engineering decision-making and integration barriers with legacy simulation software, hindering real-time applications.

## 5.2 Advanced inference strategies

Currently, Bayesian approaches for structural dynamics mainly rely on MCMC sampling or VI to estimate the posterior distribution, owing to their flexibility and approximating capability. Although the effectiveness of MCMC sampling is widely recognized, it’s also known for its computational inefficiency, primarily due to the tedious sampling process required to maintain the stationarity of the Markov chain. Conversely, while VI offers greater efficiency, its performance is significantly influenced by the selection of the variational distribution. For example, the mean-field assumption inevitably limits the approximation capacity of the variational distribution, as it ignores the correlations among high-dimensional variables. Moreover, a Gaussian variational distribution cannot appropriately captures the multi-modal characteristics of the true posterior [43]. With these challenges in mind, advanced inference schemes for estimating the posterior distribution still require further efforts to address the intrinsic limitations of MCMC and VI. A promising direction could be the integration of VI with sampling techniques to improve the estimation of posterior distributions [436, 437].

Parallel computing strategies, notably distributed task decomposition [151], exploit high-performance architectures to accelerate stochastic sampling. For example, TMCMC exemplifies inherent parallelism through stage-wise independent chains [321], outperforming conventional MCMC in scalability. However, TMCMC's efficiency degrades in high-dimensional spaces due to resampling bottlenecks and communication overhead in distributed implementations. Emerging solutions integrate hardware-aware optimizations, such as GPU-accelerated gradient computations and tensorized sampling kernels, to enhance throughput while preserving statistical rigor. Future advancements may couple these approaches with quantum-inspired annealing techniques and federated learning paradigms to address dimensionality challenges in large-scale infrastructure systems.

## 5.3 Non-Gaussian noise assumption for prediction error

In the field of structural dynamics, the Gaussian white noise is commonly used to capture the uncertainty from the inherent noise in response measurements. This assumption facilitates formulating the posterior prediction distribution and sometimes enables analytical analysis of forward and inverse propagations of uncertainty within the physical or ML models. However, a simple Gaussian assumption may not always be appropriate to capture the properties of data noise due to its intricate mechanism and various sources [259, 438]. This indicates a demand for a more general and flexible approach to model the noise, thus allowing for a more comprehensive uncertainty estimation. The GP framework proposed by Kennedy and O'Hagan (KOH) [439] has been widely utilized to characterize prediction errors in model calibration. However, the KOH methodology exhibits critical limitations: (1) non-identifiability arising from intricate interdependencies between prediction errors and model outputs, which compromises the accuracy of both physical parameter estimation and GP hyperparameter calibration; and (2) computational intractability in Bayesian inference of high-dimensional posterior distributions, particularly when modeling non-stationary errors requiring complex covariance structures. These challenges are exacerbated by the curse of dimensionality inherent to non-stationary prediction error model formulations, where the proliferation of kernel parameters renders sampling prohibitively expensive for large-scale systems.

## 5.4 Prior selection for Bayesian approaches in structural dynamic analysis

In a broad sense, assessing the likelihood of an event involves combining observed data with prior knowledge, the latter typically encoded as a prior PDF. The choice of this prior can significantly influence the shape of the resulting posterior distribution, particularly in cases involving limited data or when the data itself carries insufficient information. In such scenarios, prior misspecification can hinder decision-making by substantially underestimating or overestimating key quantities of interest. While the selection of an appropriate prior distribution has a profound impact on the performance of Bayesian methods [440], it remains a challenging and relatively underdeveloped area in Bayesian structural dynamic analysis. To reduce subjectivity and facilitate computation, uniform or Gaussian priors are commonly employed, especially in high-dimensional settings. However, despite their simplicity, these conventional priors often fail to adequately represent the complex, and sometimes multi-modal, nature of posterior distributions, particularly when both physical and statistical parameters are inferred jointly, as is increasingly common in modern approaches incorporating machine learning. Moreover, in data-scarce regimes, posterior estimates can become highly sensitive to prior assumptions, leading to biased or unstable inferences. These issues underscore the need for more principled prior selection strategies that balance flexibility with physical realism. One such approach was pioneered by Coolen [441] through the introduction of imprecise probability priors. This framework has since been extended by others (e.g., [442]) through the specification of upper and lower bounds on the prior distribution, generating a set of posterior distributions under certain ad hoc likelihood formulations. Nevertheless, due to the conceptual complexity and high computational cost, particularly in applications involving expensive, physics-based models, these methods are often impractical. Promising alternatives include hierarchical Bayesian modeling, which introduces structured uncertainty via hyperpriors; ensemble-based priors [443], which aim to capture variability across model instances; and physics-informed priors [155], which incorporate domain knowledge to improve identifiability and interpretability.

## 5.5 Online learning approaches

Typically, structural dynamic analysis tasks involve a long-term process that could span the entire service life of the monitored structure. Therefore, it is desirable for the Bayesian methods applied in this field to make predictions and estimate uncertainty in an online manner as time progresses [387]. Meanwhile, Bayesian approaches are inherently advantageous in online inference as the information of previous data can be stored in the posterior as each step, which is then used as the prior for the subsequent step [339, 444]. However, as only one sample is obtained by the model at each step, online learning is more vulnerable to uncertainty arising from noisy data and insufficient knowledge [443]. Moreover, online learning requires the model to estimate the posterior predictive distribution with the acquisition of each new sample, resulting in higher computational costs compared to offline methods due to the repeated inference. This computational challenge is especially serious for Bayesian methods based on MCMC sampling, as these sampling approaches require reassessing all of the previous data to estimate the new posterior [339]. Consequently, there is an ongoing demand for more efficient and stable Bayesian methods suitable for online structural dynamic analysis.

## 5.6 Bayesian inference for physics-informed machine learning in structural dynamics

Despite the growing success and widespread adoption of data-oriented statistical modeling in structural dynamics, its application still faces challenges, primarily due to the scarcity of high-quality training data and the lack of physical interpretability. In response, physics-informed machine learning (PIML) techniques, especially physics-informed neural networks (PINNs), have attracted increasing attention in recent years as a promising solution to improve the performance and applicability of ML in structural dynamic analysis by incorporating information based on physical mechanisms [445-453]. These approaches offer the potential to significantly enhance the modeling and prediction of structural dynamics, yet inevitably introduce additional uncertainties from both the ML models and the underlying physical models [454, 455]. Consequently, integrating Bayesian into these PIML-based methods for structural dynamic analysis holds great promise as a direction for future research efforts, as evidenced by pioneering works such as [456, 457]. It is worth pointing out that integration of physical laws with probabilistic deep learning may induce pathological loss landscapes that destabilize both Bayesian inference and gradient-based training. Furthermore, the computational cost of Bayesian marginalization over high-dimensional neural network parameters clashes with the iterative nature of solving time-dependent PDEs, rendering traditional MCMC or variational methods impractical for large-scale structures. Current implementations also lack rigorous mechanisms to disentangle model-form errors (e.g., neural network approximation gaps) from inherent physical uncertainties (e.g., material randomness), undermining the reliability of posterior predictions. Addressing these issues requires rethinking architecture design, inference schemes, and adaptive training strategies tailored to dynamical systems.

## 5.7 Generative models for Bayesian structural dynamic analysis

Generative models are a class of machine learning models that aim to understand and capture the underlying distribution of a dataset to estimate probability densities and generate new, similar data. They possess strong fitting capability, efficient training modes and exceptional advantage in handling high-dimensional data. In Bayesian inference, sampling-based techniques like MCMC methods are used for approximation when exact inference is intractable, whereas they confront with the issues of high cost and slow convergence for high-dimensional problems. Generative models, particularly those trained with variational methods (such as variational autoencoders), act as efficient alternatives by approximating the posterior distribution directly, often with lower computational cost [458, 459]. The generative models have found some applications [187, 418, 419, 460-462] in structural dynamic analysis but still deserve further attention. Key directions include developing hybrid architectures that synergize variational inference with adversarial training for efficient multi-modal posterior estimation, embedding uncertainty-aware mechanisms for joint epistemic-aleatoric quantification, and optimizing hardware-accelerated frameworks to enable real-time digital twin applications. Addressing data scarcity through physics-guided few-shot learning and establishing validation protocols grounded in mechanical principles will be critical for bridging theoretical advances with engineering trustworthiness.

# 6 Conclusion

Bayesian learning has emerged as a prevalent and promising approach in structural dynamics, providing a systematic probabilistic framework and comprehensive uncertainty treatment for analyzing complex dynamical systems. This review highlights the development of Bayesian learning in structural dynamics over the past three decades, showcasing the foundational principles, advanced methodologies, and diverse applications that have significantly contributed to the field. By examining three approaches to estimating the posterior distribution, i.e., Laplace approximation, stochastic sampling, and variational inference, this work delves into Bayesian learning for both physical and data-centric statistical models, underscoring its versatility in addressing a wide range of challenges, from system identification and model updating to damage detection and reliability assessment. These applications illustrate the pivotal role of Bayesian methods in enhancing comprehension of dynamical systems and decision-making processes. Moving forward, the paper address current research gaps by proposing ways to enhance Bayesian inference efficiency through techniques such as surrogate models and parallel computation. Furthermore, it explores advanced methods for estimating posterior distributions, addresses non-Gaussian and non-stationary modeling errors, and envisions the forefront of online Bayesian learning amid evolving trends in machine learning. Different from previous research, this study offers a thorough examination of both traditional and cutting-edge Bayesian methods, along with their applications to various tasks in structural dynamics. It underscores the transformative impact of Bayesian approaches, guiding researchers toward new advancements and opportunities while facilitating the careful selection and refinement of suitable methods for addressing various challenges in structural dynamics.

**Acknowledgement**

This research has been supported by the Science and Technology Development Fund, Macau SAR (File no.: 101/2021/A2, 0038/2024/RIB1, 001/SKL/2024), the Research Committee of University of Macau (File no.: MYRG-GRG2024-00119-IOTSC), the State Key Laboratory of Internet of Things for Smart City (University of Macau) Open Research Project (SKL-IoTSC(UM)/ORP02/2025) and Guangdong-Hong Kong-Macau Joint Laboratory Program (Project No.: 2020B1212030009).